\documentclass[a4paper,11pt,reqno]{article}
\usepackage{epsfig}
\usepackage{hyperref,amsmath,mathrsfs}
\usepackage{graphicx}
\usepackage{parskip}
\usepackage{pgf,tikz,tikz-cd,pgfplots,physics}
\usepackage{amsfonts}
\usepackage{pdflscape}
\usepackage{amssymb}
\usepackage[bbgreekl]{mathbbol}
\usepackage{bbm}

\usepackage{a4wide}

\usepackage{multirow} 
\usepackage{color} 
\definecolor{rougef}{rgb}{0.7,0,0}
\definecolor{vertf}{rgb}{0,0.6,0}
\definecolor{bleuf}{rgb}{0,0,0.9}

\newcommand{\be}{\begin{equation}}
\newcommand{\ee}{\end{equation}}
\newcommand{\bea}{\begin{eqnarray}}
\newcommand{\eea}{\end{eqnarray}}

\newcommand{\eq}[1]{(\ref{#1})}

\def\y{\eta}
\def\a{\alpha} \def\ad{\dot{\a}} \def\ua{{\underline \a}}

\def\b{\beta}  \def\bd{\dot{\b}} 
\def\c{\gamma} 
\def\C{\Gamma}
\def\d{\delta} 
\def\D{\Delta}
\def\e{\epsilon}

\def\k{\kappa}

\def\l{\lambda}
\def\L{\Lambda}

\def\r{\rho}
\def\s{\sigma}
\def\S{\Sigma}

\def\y{\eta}

\def\O{\Omega}
\def\o{\omega}

\def\sb{{\bar\s}}

\def\cP{{\cal P}}

\def\yb{{\bar y}}
\def\zb{{\bar z}}

\def\ft#1#2{{\textstyle{{\scriptstyle #1}
\over {\scriptstyle #2}}}} 

\def\mso{\mathfrak{so}}

\def\msl{\mathfrak{sl}}
\def\msp{\mathfrak{sp}}

\def\Real{{\mathbb R}}
\def\Comp{{\mathbb C}}

\def\ket#1{|#1\rangle}
\def\bra#1{\langle#1|}
\numberwithin{equation}{section}
\usepackage[utf8]{inputenc}
\begin{document}

\begin{center}
\thispagestyle{empty}

{\LARGE Fractional Spins, Unfolding, and Holography:}\\[10pt] {\LARGE II. 4D Higher Spin Gravity and 3D Conformal Dual}


\vskip .6cm

{Felipe Diaz${}^{a,}$\footnote{\href{mailto:f.diazmartinez@uandresbello.edu }{\texttt{f.diazmartinez@uandresbello.edu}}}, Carlo Iazeolla${}^{b,}$\footnote{\href{mailto:c.iazeolla@gmail.com  }{\texttt{c.iazeolla@gmail.com }}}, and Per Sundell${}^{c,d,}$\footnote{
\href{mailto:per.anders.sundell@gmail.com }{\texttt{per.anders.sundell@gmail.com}}}}
\vskip .2cm
$^{a}$ \textit{\small  Departamento de Ciencias F\'isicas, Universidad Andres Bello, Sazi\'e 2212, Santiago, Chile}\\
$^{b}$\textit{\small Dipartimento di Scienze Ingegneristiche, G. Marconi University -- Via Plinio 44, 00193, Roma, Italy \& Sezione INFN Roma “Tor Vergata” -- Via della Ricerca Scientifica 1, 00133, Roma, Italy}\\
${}^{c}$\textit{\small Instituto de Ciencias Exactas y Naturales, Universidad Arturo Prat, Playa Brava 3265, 1111346 Iquique, Chile}\\
${}^{d}$\textit{\small
Facultad de Ciencias, Universidad Arturo Prat, Avenida Arturo Prat Chacón 2120, 1110939 Iquique, Chile}
\end{center}
\vskip 1cm

\paragraph{Abstract:}

This paper completes the analysis initiated in the companion work arXiv:2403.02283 --- referred to as Paper I --- by showing how Vasiliev's 4D higher-spin gravity (HSG) and 3D coloured conformal matter fields coupled to conformal higher-spin gauge fields and colour gauge fields (coloured conformal HSG, or CCHSG) emerge as consistent reductions of a common parent model.
The latter is a Frobenius-Chern-Simons model with superconnection valued in a fractional-spin extension of Vasiliev's higher-spin algebra, and was defined and studied in Paper I.
Here, we i) realize HSG as a subcase of a more general 4D reduction, describing HSG coupled to coloured, fractional-spin matter, which we refer to as 4D fractional-spin gravity; ii) study the CCHSG model, in particular exhibiting the crucial role played by novel colour gauge fields in coupling conformal matter to conformal HSG, thereby completing the models due to Vasiliev and Nilsson; iii) extract conformal currents and composite coloured sources in an expansion of the CCHSG model around 3D Minkowskian leaves; and iv) compare our results with Vasiliev's holography proposal of arXiv:1203.5554.
The common origin of HSG and CCHSG, besides relating the two models directly, is the starting point for establishing the holographic correspondence between the two models via overlap conditions, to be presented separately.

\newpage

\clearpage
\setcounter{page}{1}

\tableofcontents

\section{Introduction}

This paper is the second in a series of works, initiated in \cite{paperI}, developing an approach to higher-spin holography based on a natural extension of the AKSZ formalism \cite{Alexandrov:1995kv} applied to the Frobenius--Chern--Simons (FCS) formulation \cite{Boulanger:2015kfa, Bonezzi:2016ttk} of Vasiliev's 4D higher-spin gravity (HSG) \cite{Vasiliev90,properties,more,Vasiliev03} (see \cite{review99,Bekaert:2005vh,Didenko:2014dwa} for reviews). The fundamental idea is that the dual bulk and boundary systems share a common origin as two reductions, with distinct field contents, of a single parent field equation, given by a flatness condition on a superconnection. The parent model can, in turn, be obtained as a boundary field configuration of an underlying, multi-dimensional AKSZ sigma model.  
This common embedding provides a rationale for deriving holographic relations from multi-dimensional AKSZ partition functions on cylinders with dual boundary conditions to be reported on in a third component of the aforementioned series \cite{OLC}.   

The first paper \cite{paperI}, referred henceforth to as Paper I, outlines our approach to holography\footnote{We refer the reader to the Introduction of Paper I for a more detailed framing of our results within the context of holographic dualities involving HSG in the bulk and conformal fields at the boundary, and for related references.} based on dual operator algebras arising at boundaries of topological parent models upon deformation quantization of spaces of classical boundary field configurations, referred to as \emph{defects}, characterized by holonomies, reduced structure groups and two-form cohomology elements.
The parent model is itself formulated in terms of horizontal differential forms on a fibered non-commutative manifold referred to as \emph{correspondence space} (see \cite{paperI} for more details); that is, its classical, fundamental fields belong to an operator algebra capturing a projection of a pair of first-quantized conformal particles, or string partons.
The resulting parent, boundary field equation is a flatness condition on a superconnection $X$ valued in a graded algebra ${\rm mat}_{1|1}$, viz., 
\be X = \left[\begin{array}{c|c}\mathbb{A}&\mathbb{B}\\\hline\widetilde{\mathbb{B}}&\widetilde{\mathbb{A}}\end{array}\right]
\ ,
\ee
where $(\mathbb{A},\widetilde{\mathbb{A}};\mathbb{B},\widetilde{\mathbb{B}})$ are factorizable, horizontal forms of degrees $(1,1,0,2)$ , i.e., they are symbols of operators that can be expanded in a basis of fibre zero-forms with coefficients being forms on the base.
The choice of operator algebra off-shell is the non-commutative analogue of imposing boundary conditions in an ordinary commuting field theory.
Lorentzian holography in asymptotically, locally anti-de Sitter backgrounds is captured by symbols of inhomogenous, holomorphic metaplectic group algebras (which are complex extensions of the real metaplectic group giving rise to positive and negative energy modes \cite{meta,Iazeolla:2022dal,paperI,paper0}).
The two defects containing the classical moduli spaces of HSG and its holographic dual arise by breaking the symmetry under exchange of the two partons and fixing a holonomy group given by subgroup of the real, metaplectic group acting on the first parton and a $U(N,N)$ group acting on the second parton, leading to the \emph{fractional-spin algebra} $\boldsymbol{\cal FS}$ of \cite{Boulanger:2013naa}.
This algebra thus consists of an infinite-dimensional external algebra and one finite-dimensional internal algebra, leading to the schematic decomposition 
\be \left. (\mathbb{A}, \ \widetilde{\mathbb{A}}, \ \mathbb{B},\ \widetilde{\mathbb{B}}) \ \right\downarrow_{\boldsymbol{{\cal FS}}} \ \sim \ \left[\begin{array}{c|c}\ket{{\rm ext}}\bra{{\rm ext}} &\ket{{\rm ext}}\bra{{\rm int}}\\\hline\ket{{\rm int}}\bra{{\rm ext}}\ &\ket{{\rm int}}\bra{{\rm int}}\end{array}\right]  \ ,\label{FSdec}\ee
suppressing holomorphic Wigner--Ville maps defined in Paper I \cite{paperI}, and where the external algebra is taken to be the group algebra of special holomorphic metaplectic group elements (see Section 2 in \cite{paperI} and \cite{meta}) representing endomorphisms of the conformal or non-compact singleton module (see Appendix \ref{App:emb} and \cite{paperI} for their definition); while the internal algebra is the algebra of endomorphisms of a hermitian colour module, in general with split signature $(N,N)$. The off-diagonal blocks of \eq{FSdec} correspond to intertwining sectors, transforming under the external symmetry algebra on one side and under the internal one on the other side.  

As we shall see in this paper, a prominent role in the definition of the two dual reductions is played by the expectation value of the dynamical two-form $\widetilde{\mathbb{B}}$, which determines the unbroken structure group of the reduction. The relevant two-form deformation for HSG was determined (in the present form) in \cite{more} and encoded as the expectation value for the dynamical two-form of the FCS model in \cite{Boulanger:2015kfa}. Two of the main results of Paper I are the realization that the HSG two-form expectation value, coupled with the fractional-spin algebra expansion of the parent superconnection $X$, actually leads to a more general 4D theory, that can, in turn, be truncated to 4D HSG; and the identification of a cohomologically non-trivial two-form for a 3D candidate holographic dual of 4D HSG, breaking down the structure group to the expected residual group $SL(2,\Real)_{\rm Lor}\times O(1,1)_{\rm Dil}$. We also evaluated a parent second Chern class on the two reductions, thereby assessing their non-triviality and the compatibility of the two models as holographic duals.

In the present paper, we complete the analysis undertaken in Paper I by showing in detail that, as anticipated in \cite{paperI}, the 3D dual reduction describes a \emph{conformal HSG plus coloured matter}, for which we shall use the acronym CCHSG --- i.e., a system describing coloured conformal matter fields coupled to topological conformal higher-spin gauge fields and colour gauge fields. This theory can thus be considered a fully non-linear completion (at least within the spirit of the AKSZ program, defined in \cite{paperI}) of the 3D matter-coupled \emph{conformal HSG} (CHSG) models considered by Vasiliev \cite{Misha} and Nilsson \cite{Nilsson:2013tva,Nilsson}, obtained by adding topological gauge fields of a colour gauge group. In particular, we analyse the CCHSG model perturbatively in an expansion around a foliated vacuum with 3D Minkowski leaves (the foliation parameter playing the role of a scale dimension, in analogy with the radial coordinate of the AdS Poincar\'e patch) and we compare in detail our results at first non-trivial order with those obtained in \cite{Misha}.

The embedding of 4D HSG and 3D CCHSG into the same parent model depends on the interplay between the graded-algebra decomposition and the fractional-spin algebra decomposition of $X$. Indeed, as a result of the first decomposition, the flatness condition $dX + X\star X = 0$ of the parent theory decomposes into curvature conditions on the one-form connections, dictating that the curvatures be sourced by composite operators built by the zero-form and two-form components; and ``mixed'' covariant constancy conditions on the latter two,
\begin{align}\label{1.3}
d\mathbb{A}+\mathbb{A}\star \mathbb{A}+\mathbb{B}\star \widetilde{\mathbb{B}}= 0\ ,\qquad d\widetilde {\mathbb{A}}+\widetilde{\mathbb{A}}\star \widetilde{\mathbb{A}}+\widetilde{\mathbb{B}}\star {\mathbb{B}}= 0\ ,\\ \label{1.4}
d\mathbb{B}+\mathbb{A}\star \mathbb{B}-\mathbb{B}\star \widetilde{\mathbb{A}}= 0\ ,\qquad d\widetilde {\mathbb{B}}+\widetilde{\mathbb{A}}\star \widetilde{\mathbb{B}}-\widetilde{\mathbb{B}}\star {\mathbb{A}}= 0\ ,
\end{align}
Within this parent scheme, (CC)HSG models are then embedded as consistent truncations that turn on only certain subsectors of the full fractional-spin algebra, with the above-mentioned expectation values for the two-form $\widetilde{\mathbb{B}}$, which leave different unbroken structure group. 4D HSG corresponds to turning on only the purely external ($\ket{{\rm ext}}\bra{{\rm ext}}$) sectors of every superconnection component (identifying the two one-form connections, $\mathbb{A}=\widetilde{\mathbb{A}}$), and an $SL(2,\Comp)$-invariant expectation value for the two-form. The 3D CCHSG reduction is instead obtained by turning on both the pure bimodule components ($\ket{{\rm ext}}\bra{{\rm ext}}$ and ($\ket{{\rm int}}\bra{{\rm int}}$)) of the one-form connections (again taken to be identical) and the hybrid, intertwining ones ($\ket{{\rm ext}}\bra{{\rm int}}$ and ($\ket{{\rm int}}\bra{{\rm ext}}$) for the zero-form and the two-form; the latter's expectation value only preserving an $SL(2,\Real)_{{\rm Lor}}\times O(1,1)_{{\rm Dil}}$-symmetry. Such choices change the meaning of the curvature constraints of the two reductions: on the HSG reduction, having only ``pure'' external components allows for the identification of the source term $\mathbb{B}\star \widetilde{\mathbb{B}}$ with the curvature components that are left free to fluctuate (i.e., Weyl tensors); while on the CCHSG reduction, in which both the zero-form and the two-form have a hybrid bimodule structure that can accommodate a conformal coloured 3D scalar field, the composite sources $\mathbb{B}\star \widetilde{\mathbb{B}}$ and $\widetilde{\mathbb{B}}\star\mathbb{B}$ naturally read as sesquilinear constructs in the 3D scalar ---  respectively providing pure bimodule sources of type $\ket{{\rm ext}}\bra{{\rm ext}}$ for the external gauge field, to be interpreted in terms of colour-singlet conformal currents, and of type $\ket{{\rm int}}\bra{{\rm int}}$ for the internal one, giving rise to colour generators. 

Moreover, the more general 4D system that gives Vasiliev's HSG upon truncation is obtained by leaving one-form connections as well as zero-form in $X$ with all non-vanishing entries, and only the hybrid components of the two-form are set to zero. The two-form still preserves $SL(2,\Comp)$, and the resulting system can be described as HSG coupled to coloured, fractional-spin matter fields, in their turn coupled to an internal colour sector required for integrability. In this paper we study the linearization of this 4D \emph{fractional-spin gravity} (FSG) model around $AdS_4$, leaving a more thorough exploration of its properties to a future work. 

\subsection{Main results of this paper}

More in detail, in this paper:

\begin{itemize}

\item We write down a fully non-linear unfolded CCHSG system with $U(N,N)$-coloured matter fields that admits a consistent truncation to $U(N)$ colour group\footnote{The CCHSG system is formulated in a correspondence space using the same formalism that was used to write Vasiliev's equations for HSG \cite{properties,review99,Didenko:2014dwa,Bekaert:2005vh}. As for the latter, it is by now well known that the extension of the base of the correspondence space with non-commutative $Z$ variables leads to a compact generating system whose homotopy contraction introduces a method to control field redefinitions; on the flip side, the determination of the right gauge choice in $Z$-space leading to ``minimally non-local'' spacetime vertices is at the moment a research topic (see \cite{Vasiliev:2017cae,Gelfond:2018vmi,Didenko:2018fgx,Didenko:2019xzz,Gelfond:2019tac,COMST,Didenko:2020bxd,Vasiliev:2022med} for progress in this direction). However, we stress that while the elimination of $Z$ variables and the extraction of pure spacetime vertices is part of the standard approach to extracting holographic data  (see \cite{GiombiYin,GiombiYin2,Boulanger:2015ova,Didenko:2017lsn,Sezgin:2017jgm}), these steps are bypassed in our proposed AKSZ approach, which is based on invariant functionals for parent fields computed on the full correspondence space.
These functionals factor out the data entering the specification of gauges and ordering schemes in $Z$-space used in the Vasiliev spin-local approach; for a summary of approaches to HSG, see Section 1.1 in \cite{paperI}.}.  

\item In the leading order, to be studied here, the CCHSG system encodes conformal currents into a colour-singlet, sesquilinear construct in 3D matter fields, explicitly written in terms of star products. The currents are coupled to colourless, topological CHSG  gauge fields and built from matter fields forming a colour $U(N)$ vector multiplet. The latter couple to topological colour gauge fields via non-local sources (see below). 

\item Our fully non-linear, coordinate-independent CCHSG system is written in terms of a 3D matter scalar encoded into a master zero-form with hybrid bimodule structure, schematically $C\sim \ket{\rm{conformal}}\bra{\rm{colour}}$, i.e., expanded over non-polynomial functions of the non-commutative fibre coordinates corresponding to metaplectic group elements (or rescaled limits thereof \cite{meta,Iazeolla:2022dal}) that realize a conformal left-module and a colour right-module. This hybrid structure is fundamental to achieving regular current sources without any need for a star-product regularization. Indeed, the core element of the conformal current master field is $C\star \overline C$, and the presence of colour states in $C$ avoids the direct clash of conformal singleton/anti-singleton states, which gives rise to a divergent star-product (that, in turn, calls for a realization of singleton state projectors by means of a specific integral presentation in order for such divergence to be tamed) \cite{2011,2017,COMST,corfu19,meta}. Colour states, coming from the fractional-spin algebra expansion, help to smooth the star product $C\star\overline C$ (see Sections \ref{sec:currents} and \ref{sec:roleV} for a more detailed discussion, and Appendix \ref{App:B} for a proof), giving rise --- upon further star product with the distinctive $Z$-space two-form factor of CCHSG --- to regular source terms that can be identified as conformal current generating functions, without any need for a star-product regularization. Such regular source terms then drive the perturbative expansion of our CCHSG system.

\item 4D HSG and 3D CCHSG are obtained as two consistent reductions of the same parent system descending from an FCS variational principle. Besides relating them directly, this opens the way to (second) quantization of both theories and to a new rationale for the holographic duality. 

\item For the reasons above summarized, and explained in greater detail in Paper I, the 3D CCHSG system is thus a candidate holographic dual for 4D HSG (to be verified via overlap conditions). As such, it verifies certain aspects of Vasiliev's refined HS/CFT conjecture \cite{Misha}, namely that i) the duality actually takes place on any 3D leaf of $AdS_4$ (in a Fefferman--Graham expansion of 4D HSG); and ii) the 3D dual of 4D HSG is not simply a free CFT, but in general a CFT coupled to a CHSG. On the other hand, consistency of the non-linear 4D HSG and 3D CCHSG reductions requires the dynamical two-form of the parent model to be built using two distinct two-forms on $Z$-space, respectively denoted by $I_{\Comp}$ and $I_\Real$. As a result, in \cite{Misha}, the linearized 4D HSG equations, featuring $I_{\Comp}$, were directly pulled back on a 3D leaf, leading to a decoupling of currents from gauge fields in the case of Type A and B models (which extends to second order within the spin-local approach to 4D HSG \cite{Didenko:2017lsn}). On the other hand, in our AKSZ approach, 3D CCHSG arises as an independent unfolded system, featuring $I_\Real$, which does not allow any decoupling\footnote{We find it reasonable to expect that the 3D CCHSG model reproduces the correlation functions of the $O(N)$ model to the leading order in $1/N$ where both theories can be treated as free quantum theories (using the Hubbard--Stratonovich transformation in the strongly coupled fixed point), in turn corresponding to the classical HSG model in the bulk.
Moreover, beyond this order, the coupling to topological HS fields may introduce double-trace deformations of the type proposed by 
\cite{Sezgin:2002rt} as opposed to those proposed in \cite{KP}, leaving the issue of which type of quantizations of 4D HSG these deformations correspond to. We propose that the 3D CCHSG model corresponds to a quantization of 4D HSG using the AKSZ approach, which is now under study.}; for comments on a possible generalized reduction Ansatz that may accommodate such decoupling, see Conclusions. Extending the 3D system by a fourth direction dual to the dilation operator, the resulting (3+1)D system contains a (free-theory) renormalization group equation. For a related reason (see Section \ref{App:D}), the conserved conformal currents are embedded differently into our model and into that of \cite{Misha}, leading to generating functions that only agree on the boundary. On any other 3D leaf, our conserved currents keep the same form, whereas those of  \cite{Misha} receive higher-derivative corrections (that do not ruin conservation) weighted by appropriate powers of the radial/foliation coordinate, under the assumption that the bulk Weyl zero-form can be factorized in terms of 3D matter fields. A detailed comparison with the results of \cite{Misha} is given in Section \ref{App:D}. 

\item The other novelty of our system is that its matter fields are coupled to extra, topological, colour gauge fields, and we show via perturbative analysis the role of the latter and why they are crucial to complete matter-coupled CHSG to fully non-linear level. The CCHSG system candidate dual to HSG naturally incorporates a minimally coupled $U(N)$ gauge field $V$ --- in this sense resembling an ordinary colour gauge field. However, due to their fractional-spin algebra origin, the colour gauge group generators are not fundamental generators that are introduced independently, but are composite in terms of the scalar matter fields. This feature gives rise to a source term for $V$, a two-form along the non-commutative base directions of the correspondence space, given by a non-local construct of the scalar field; which distinguishes it from an ordinary 3D colour Chern-Simons gauge field \cite{Giombi:2011kc}, sourced by a local current. This composite origin also implies that the colour generators are actually realized in terms of the same oscillators that build current and CHSG gauge field generating functions, and therefore the interaction of the colour gauge fields with the latter is controlled by a single star-product algebra. In this sense, we can think of the colour degrees of freedom in our system as dynamically realized Chan-Paton factors.

\end{itemize}

Following the approach introduced in a number of papers \cite{fibre,2011,2017,BTZ,COMST,corfu19} and more formally spelled out in \cite{meta,Iazeolla:2022dal,paperI,paper0}, we realize endomorphisms $\ket{m}\bra{n}$ of hermitian modules via symbols of homolorphic metaplectic group elements --- Gaussian functions of the oscillators that realize the algebra generators. The star product among such elements, which are eigenfunctions carrying definite quantum numbers under left/right action of symmetry generators, implements the product of group elements in a given ordering and reproduces the inner bra-ket product among states defined by said quantum numbers. This enables us to implement Dirac-style bra-ket-like computations without any need of introducing states as separate entities from symmetry algebras. Even the hybrid bimodule elements can be realized as star products of  Gaussian elements representing pure bimodule elements, schematically $\ket{{\rm ext}}\bra{{\rm int}}=\ket{{\rm ext}}\bra{{\rm ext}}\star \ket{{\rm int}}\bra{{\rm int}}$. 
On the other hand, the star product computations can handle more general operator algebras not necessarily realized in terms of Hilbert spaces, as the group algebra of the holomorphic metaplectic group $Mp(4;\mathbb{C})$, whose product rules are defined by analytic continuation; see \cite{meta,Iazeolla:2022dal,paperI} for further details. We also refer the reader to \cite{paper0} for a more complete description of our operator bases. 

Finally, we stress that this paper --- and in general our proposed formulation of HS/CFT correspondence combining bulk and boundary theories into a single, parent AKSZ sigma model --- is part of an in-development approach to HSG \cite{Engquist:2005yt,fibre,2011,Sezgin:2011hq,Boulanger:2011dd,Colombo:2012jx,Boulanger:2015kfa,Bonezzi:2016ttk,2017,BTZ,COMST,corfu19,meta,Iazeolla:2022dal} that exploits the compact, geometric formulation of the Vasiliev system in correspondence space. Two key ideas underpinning the ensuing AKSZ approach to HSG and CCHSG, as outlined in Paper I, are: i) to treat the entire master fields as the fundamental variables of the second-quantized formulation, rather than just as tools to extract component fields; and ii) to define the master fields as elements of a well-defined operator algebra, rather than as formally defined generating functions. In this framework, spacetime fields of given spin represent well-defined data only in asymptotic spacetime regions where appropriate boundary conditions hold; and physical quantities are instead extracted via gauge-invariant functionals of the master fields defined by traces on the full correspondence space.
Correspondingly, as we have anticipated throughout this Introduction, our approach only requires well-defined operator products and traces (to build observables), and thus relies on fibre algebras beyond the Weyl algebra, accommodating different boundary conditions and thus different sectors of the theory. 
The computation of boundary correlators from higher-spin invariant functionals has so far been tested at the leading order  \cite{Colombo:2010fu,Colombo:2012jx,Didenko:2012tv}.
However, \cite{COMST} (see also \cite{corfu19,meta}) contains a proposal for extending this test to subleading orders via a Fefferman-Graham-like scheme for computing asymptotic master field configurations in the full correspondence space (including a nontrivial $Z$-space dependence) by perturbatively adjusting gauge functions \emph{and} integration constants (containing the local data of the master fields) via the imposition of boundary conditions. 
The resulting perturbative corrections to the gauge-invariant observables of the unfolded system, induced by those of the master fields, are sensitive to the bulk interactions and may determine non-trivial corrections to the free CFT current correlation functions. 

\subsection{Outline of the paper}

The plan of the paper is as follows: 

\noindent We use results, notation and nomenclature of Paper I, the most relevant of which will be recalled for the reader's convenience. 

\noindent Section \ref{sec:recap} provides a brief recapitulation of the basic features of the parent model and of the structure groups of the two defects, recalling some of the notations of Paper I that will be used in the following Sections.

\noindent Section \ref{sec:FSG} provides the FSG model, describing the coupling of 4D fractional-spin fields, representing dynamical domain walls, to Vasiliev's HSG and additional, topological colour gauge fields, and initiates its linearization around $AdS_4$. We also show how to further truncate to Vasiliev's HSG.

\noindent Section \ref{sec:CHSG} spells out the CCHSG model, i.e. 3D, conformal, coloured matter fields coupled to conformal HSG and colour gauge fields, in the case of conformal scalars.
In a perturbative expansion around ${\rm Mink}_3$, it is shown that the matter fields source the conformal HSG fields via primary, conformal currents.
Assuming boundary conditions on conformal ${\rm Mink}_3$, the matter fields are shown to source the colour gauge fields via a composite cocycle in twistor space.
As a step towards geometrizing conformal-symmetry breaking, the CCHSG model is re-formulated on ${\rm Mink}_3\times\Real$. 
Finally, we comment on the role of the colour states and colour gauge fields in the model.

\noindent Section \ref{App:D} is devoted to a detailed comparison of our results with those obtained in \cite{Misha}.

\noindent Finally, in Section \ref{sec:conclusions}, we conclude and provide an outlook to future works.

\noindent Appendix \ref{App:emb} contains our conventions for realizing $\mso(2,3)\cong \msp(4;\Real)$ using $SL(2;\Comp)$- and $O(1,1)\times SL(2;\Real)$-covariant oscillators, and the internal matrix algebra using a finite-dimensional Fock space.

\noindent Appendix \ref{App:B} concerns the metaplectic nature of the oscillator realization of the Hermitian singleton module with conformal ground states, and its role in the computation of quadratic, conformal currents starting from the two-form in twistor space.

\section{Parent field equations}\label{sec:recap}

The parent model is an AKSZ sigma model of the type introduced in \cite{BarnichGrigoriev,Grigoriev} with a non-commutative source and target (see also \cite{Boulanger:2011dd}), i.e., its configuration space is a BV manifold consisting of morphisms from a source differential graded associative algebra (DGA) to a target DGA, equipped with a master action built from the source and target DGA operations (including traces).
The source DGA is represented using symbols given by holomorphic differential forms on a non-commutative, holomorphic, differential Poisson manifold $\widehat{\boldsymbol{M}}$ with boundary $\boldsymbol{M}$, and has differential $d$ and product $\star$ given by deformations of the de Rham differential and wedge product, respectively, and contains a set of projections, referred to as virtual source DGAs, induced by fibrations with sections, and equipped with graded trace operations given by regularized integration measures.
The graded target manifold is in itself a DGA with monary and binary operations $m_1$ and $m_2$, hence equipped with a canonical Q-structure, viz., $\vec Q Z= m_1(Z)+m_2(Z,Z)$, where $Z$ is a target coordinate of total degree $1-{\rm deg}(m_2)$; we assume that ${\rm deg}(m_2)=0$ and write $m_1(Z)\equiv m(Z)$ and $m_2(Z,Z)\equiv Z\star Z$.
Taking target is also a graded cotangent bundle over a sub-DGA acting on the fibre.

The non-commutative structures of the target and the boundaries of the source arise from an underlying first-quantized system \cite{Engquist:2005yt,Barnich:2006pc,BarnichGrigoriev,Grigoriev,Arias:2016agc} given by a set of partons on a fibered, non-commutative, holomorphic, differential Poisson manifold $\boldsymbol{Y}\to {\boldsymbol{C}}\to {\boldsymbol{M}}$ with holomorphic, symplectic fibre, referred to as a correspondence space, equipped with a holomorphic Hermitian conjugation $\dagger$ composed out of the Hermitian conjugation operation $\dagger_\Comp$ of the underlying complex manifold and a structure-preserving involution $r$ that exchanges holomorphic and anti-holomorphic coordinates, i.e., $\dagger=r^\ast\circ \dagger_\Comp$.
The correspondence space is assumed to carry the action of a discrete group ${\cal K}$ of structure-preserving maps used in imposing twisted boundary conditions as particle segments are glued together into boundaries of the two-dimensional sources of the sigma model used to quantize ${\boldsymbol{C}}$.
The complex nature of $\boldsymbol{C}$ implies that ${\cal K}$ can be chosen such that the group algebra elements representing ${\cal K}$ form a discrete group connecting various boundary conditions on horizontal forms on $\boldsymbol{C}$.
The first-quantization scheme is also assumed to activate boundary conditions describing how subsets of partons are attached to a base point in ${\boldsymbol{C}}$ through partial fibrations which activates projector algebras.

Restricting the first-quantized system to a projected two-parton system with projector algebra ${\rm mat}_2$ and gauge group $\Real$ with ghost system generating a three-graded ${\rm mat}
_{1|1}$ \cite{Grigoriev} yields a total DGA 
$\boldsymbol{\cal E}^H_{\rm hor}({\cal K}\times 
{\boldsymbol{C}};\boldsymbol{\cal N})$ with structure group $H$, consisting of horizontal, holomorphic, differential forms on $\boldsymbol{C}$ valued in the graded algebra $\boldsymbol{\cal N}={\rm mat}_{1|1}\star {\rm mat}_2$ glued together using transition functions from $H$ as described in \cite{paperI}.
Letting the base and fibre of the target be coordinatized by $X$ and $P$, respectively, giving rise to conjugate AKSZ superfields of the non-commutative AKSZ sigma model on $\widehat{\boldsymbol{M}}$ and imposing boundary conditions on $P$, the resulting boundary field equation, viz.,
\begin{align}\label{2.1}
dX+X\star X=0\ ,\qquad X\in \boldsymbol{\cal E}^H_{\rm hor}({\cal K}\times \boldsymbol{C};\boldsymbol{\cal N})\ ,\qquad {\rm deg}_{\boldsymbol{\cal E}}(X)=1\ ,   
\end{align}
can be interpreted as a deformation of the BRST operator of the first-quantized system on $\boldsymbol{C}$ \cite{Grigoriev}; as for the master action, two models arise depending on the dimension of $\widehat{\boldsymbol{M}}$ mod 2 \cite{OLC}. 

Thus, moduli spaces of first-quantized parton geometries are identified as spaces of classical, flat superconnections $X$, referred to as defects, providing boundary conditions for a multi-dimensional, second-quantized partition function \cite{paperI,OLC}.
The superconnection consists of operators with symbols given by , viz.,
\begin{align}
X=\left[\begin{array}{c|c}\mathbb{A}&\mathbb{B}\\\hline\widetilde{\mathbb{B}}&\widetilde{\mathbb{A}}\end{array}\right]=\left[\begin{array}{cc|cc} A&\Psi&B&\Sigma\\\overline{\Psi}&U&\overline{\Sigma}&M\\\hline\widetilde{B}&\widetilde{\Sigma}&\widetilde{A}&\widetilde{\Psi}\\\widetilde{\overline{\Sigma}}&\widetilde{M}&\widetilde{\overline{\Psi}}&\widetilde{U}\end{array}\right]\ ,\qquad {\rm deg}_{\boldsymbol{C}}(\mathbb{A},\widetilde{\mathbb{A}},\mathbb{B},\widetilde{\mathbb{B}})=(1,1,0,2)\ ,
\end{align} 
obeying Eqs. \eqref{1.3} and \eqref{1.4}, where the components are horizontal forms on $\boldsymbol{C}$ obeying the reality conditions
\begin{eqnarray} &A^\dagger=-\widetilde A\ ,\qquad U^\dagger=-\widetilde{U}\ ,\qquad \Psi^\dagger=-\widetilde{\overline{\Psi}}\ ,\qquad \widetilde{\Psi}^\dagger=-\overline{\Psi}\ ,&\\ &B^\dagger=B\ ,\qquad M^\dagger=M\ ,\qquad \Sigma^\dagger=\overline{\Sigma}\ ,&\\& \widetilde{B}^\dagger=-\widetilde{B}\ ,\qquad \widetilde{M}^\dagger=-\widetilde{M}\ ,\qquad
\widetilde{\Sigma}^\dagger=-\widetilde{\overline{\Sigma}}\ .&\end{eqnarray}
The defects, which we label by $\Xi$, are characterized by holonomies, reduced structure groups, two-form cohomologies and choices of real integration domains inside $\boldsymbol{C}$, referred to as chiral domains. 
In particular, the FSG ($\Xi=\Comp$) and CCHSG ($\Xi=\Real$) defects have reduced structure groups 
\begin{align}
H_\Comp=\frac12(1+\pi^\ast)(H)\stackrel{\#}{\longrightarrow} SL(2;\mathbb{C})_{\rm Lor}\times U(N,N)\ ,\\
H_\Real=\frac12(1+\pi^\ast_{\!\mathscr{P}})(H)\stackrel{\#}{\longrightarrow}  SL(2;\mathbb{R})_{\rm Lor}\times O(1,1)_{\rm Dil}\times U(N,N)\ ,
\end{align} 
using notation introduced in Paper I and with properties explained there, including the closed and central two-form cohomology elements $I_\Comp$ and $I^\pm_\Real$ appearing in the consistent reductions; the further reductions indicated above, to be discussed in Sections \ref{sec:LorHSG} and \ref{sec:LorCCHSG}, select structure groups that can be maintained at the level of homotopy contraction down to component formulations on commutative spacetime manifolds.

\section{Fractional-spin gravity and higher-spin gravity defects}\label{sec:FSG}

In this Section, we give the classically consistent reduction of the parent field equation \eq{2.1} to 4D FSG describing the coupling of coloured singletons to Vasiliev's 4D HSG  and topological, colour gauge fields.
From Paper I, we recall that on the HSG defect the complex correspondence space has the structure of a fibration $\boldsymbol{C}^{(\Comp)}_{12}\to \boldsymbol{X}^{(\Comp)}_4$ over a K\"ahler manifold $\boldsymbol{X}^{(\Comp)}_4$ of complex dimension four equipped with commuting, holomorphically real coordinates, with non-commutative, holomorphic, symplectic fibres given by fibre bundles $\boldsymbol{Y}\to \boldsymbol{T}^{(\Comp)}_8\to \boldsymbol{Z}^{(\Comp)}_4$ (whose bundle structure may thus vary over $\boldsymbol{X}^{(\Comp)}_4$).
The natural coordinates on $\boldsymbol{Y}$ are holomorphically complex doublets $(y^\alpha,\bar y^{\dot\alpha})$ obeying
\be [y^\alpha,y^\beta]_\star=2i\epsilon^{\alpha\beta}\ ,\qquad [\yb^{\ad},\yb^{\bd}]_\star=2i\epsilon^{\ad\bd}\ , \qquad \yb^{\dot\alpha}:=(y^\alpha)^\dagger\ ,
\ee
and on which the holomorphic, symplectic involutions $\pi_y$ and $\bar\pi_{\yb}$ act as
\begin{align}\label{piyyb}
(y^\alpha,\bar y^{\dot\alpha})\circ \pi_y:=(-y^\alpha,\bar y^{\dot\alpha})\ ,\qquad (y^\alpha,\bar y^{\dot\alpha})\circ \bar\pi_{\bar y}:=(y^\alpha,-\bar y^{\dot\alpha})\ .
\end{align}  
Analogously, the non-commutative base directions are coordinatized by holomorphic, symplectic variables $(z^\alpha,\bar z^{\dot\alpha})$ obeying
\be [z^\alpha,z^\beta]_\star=-2i\epsilon^{\alpha\beta}\ ,\qquad [\zb^{\ad},\zb^{\bd}]_\star=-2i\epsilon^{\ad\bd}\ , \qquad \zb^{\dot\alpha}:=-(z^\alpha)^\dagger\ ,\label{realz}
\ee
and on which the symplectic involutions $\pi_z$ and $\bar\pi_{\zb}$ act as
\begin{align}\label{pizzb}
(z^\alpha,\bar z^{\dot\alpha})\circ \pi_z:=(-z^\alpha,\bar z^{\dot\alpha})\ ,\qquad (z^\alpha,\bar z^{\dot\alpha})\circ \bar\pi_{\bar z}:=(z^\alpha,-\bar z^{\dot\alpha})\ .
\end{align}  
The space $\boldsymbol{T}^{(\Comp)}_8$ is equipped with the complex, cohomologically non-trivial, horizontal two-form 
\begin{align}
I_\Comp=-\frac{i}4  dz^2 \kappa_y\star \kappa_z\star k\ ,\qquad k=i k_y\star k_z\ , \label{IC}
\end{align}
making use of inner and outer Klein operators on the fibre and base. The inner Klein operators $\kappa_y$ and $\bar\kappa_{\bar y}$ implement the $\pi_y$ and $\bar\pi_{\yb}$ maps via their adjoint action, are unimodular, and are given in Weyl order by the analytic delta functions
\begin{align}
\kappa_y:=2\pi\delta^2_\Comp(y)\ ,\qquad \bar\kappa_{\bar y}:=2\pi\delta^2_\Comp(\bar y)\ ,\qquad (\kappa_y)^\dagger=\bar\kappa_{\bar y}\ ,\label{kappay}
\end{align}
defined on the two-sheeted Riemann surface of the square-root function, viz., 
\begin{align}
\delta^2_\Comp(M y)=\frac{1}{{\rm det}(M)}\delta^2_\Comp(y)\ ,\qquad \delta^2_\Comp(\bar M\bar y)=\frac{1}{{\rm det}(\bar{M})}\delta^2_\Comp(\bar y)\ ,\label{deltaC}
\end{align}
for $M,\bar M\in GL(2;\Comp)$; for details, see \cite{meta}.  The Klein operators $\k_z$ and $\bar\kappa_{\zb}$ are defined analogously. See Section 3.1 in Paper I \cite{paperI} for more details on the outer Klein operators.

\subsection{Bifundamental FSG defect}

On the FSG defect, the DGA $\boldsymbol{\cal E}^H_{\rm hor}({\cal K}\times 
{\boldsymbol{C}};\boldsymbol{\cal N})$ contains a one-parameter family of cohomologically nontrivial and central two-forms 
\begin{align}\label{3.15}
\mathbb{I}^{(\theta_0)}_{\Comp}:=b \,\mathbb{I}_{\Comp}-\bar b \,\overline{\mathbb{I}}_{\Comp}\ ,\quad \mathbb{I}_{\Comp}=I_\Comp\star\left[\begin{array}{cc}
  {\rm Id}_{{\cal S}(\xi_0)}   &0  \\
   0  & {\rm Id}_{{\cal C}(N,N)}
\end{array}\right]\ ,\quad  b=e^{i\theta_0}\ ,\quad \theta_0\in [0,\pi[\ ,
\end{align}
using the identity on the fractional-spin algebra (see Section 5.3 of Paper I \cite{paperI}).
The reduction 
\begin{align}  \widetilde{\mathbb{B}}= \mathbb{I}^{(\theta_0)}_{\Comp}\star \widetilde{\mathbb{C}}\equiv {I}^{(\theta_0)}_{\Comp}\star \widetilde{\mathbb{C}}\ ,\qquad \widetilde{\mathbb{C}}^\dagger =\widetilde{\mathbb{C}}\ ,\qquad {\rm deg}_{\boldsymbol{{\cal B}}}\,\widetilde{\mathbb{C}}=0\ ,\qquad {I}^{(\theta_0)}_{\Comp}=bI_\Comp-\bar b\bar{I}_\Comp\ ,
\end{align}
is classically consistent modulo the integrable equations of motion
\begin{align}
d\mathbb{A}+\mathbb{A}\star \mathbb{A}+\mathbb{B}\star \widetilde{\mathbb{C}}\star {I}^{(\theta_0)}_{\Comp}= 0\ ,\qquad d\widetilde {\mathbb{A}}+\widetilde{\mathbb{A}}\star \widetilde{\mathbb{A}}+\widetilde{\mathbb{C}}\star {\mathbb{B}}\star I^{(\theta_0)}= 0\ ,\\
d\mathbb{B}+\mathbb{A}\star \mathbb{B}-\mathbb{B}\star \widetilde{\mathbb{A}}= 0\ ,\qquad d\widetilde {\mathbb{C}}+\widetilde{\mathbb{A}}\star \widetilde{\mathbb{C}}-\widetilde{\mathbb{C}}\star {\mathbb{A}}= 0\ ;
\end{align}
the connection of $H_\Comp$ consists of $\frac12(1+\pi^\ast)(\mathbb{A},\widetilde{\mathbb{A}})$.

\subsection{Broken FSG defect}

The further reduction
\begin{align} \widetilde{\mathbb{A}}={}&\mathbb{A}=\left[\begin{array}{cc} A&\Psi\\ \overline \Psi&U\end{array}\right]\ ,\qquad A^\dagger=-A\ ,\quad \Psi^\dagger=-\overline\Psi\ ,\quad U^\dagger=-U\ ,\\
\mathbb{B}\star \widetilde{\mathbb{C}}={}&\widetilde{\mathbb{C}}\star \mathbb{B} =:\mathbb{C}:=\left[\begin{array}{cc} C&\Xi\\ \overline \Xi& N\end{array}\right]\ ,\qquad C^\dagger=C\ ,\quad \Xi^\dagger=-\overline\Xi\ ,\quad N^\dagger=N\ ,
\end{align}
is classically consistent modulo the integrable equations of motion
\begin{align}
d\mathbb{A}+\mathbb{A}\star \mathbb{A}+\mathbb{C}\star I^{(\theta_0)}_{\Comp}= 0\ ,\qquad d\mathbb{C}+\mathbb{A}\star \mathbb{C}-\mathbb{C}\star {\mathbb{A}}= 0\ ,    
\end{align}
describing a coupling of coloured singletons $(\Psi,\Xi)$ \cite{Bergshoeff:1988jm} to Vasiliev's pure HSG master fields $(A,C)$ and an internal colour sector $(U,N)$.
Taking $\mathbb{A}$ and $\mathbb{\Phi}:=\mathbb{C}\star k$ to be $\boldsymbol{\cal K}$-independent, the further consistent truncation
\begin{equation}
d\mathbb{A}+\mathbb{A}\star \mathbb{A}+\mathbb{\Phi}\star {J}^{(\theta_0)}_{\Comp}= 0\ ,\qquad d\mathbb{\Phi}+\mathbb{A}\star \mathbb{\Phi}-\mathbb{\Phi}\star \pi^\ast({\mathbb{A}})= 0\ , 
\end{equation}
where $\pi:=\pi_y\circ \pi_z$ and ${J}^{(\theta_0)}_{\Comp}:=k\star{I}^{(\theta_0)}_{\Comp}$, referred to as the unaugmented model.

\subsection{Pure HSG defects}

Setting coloured singletons and colour gauge fields to zero, i.e., $\Psi=0 =\Xi$, $U=0=N$, yields Vasiliev's pure 4D HSG \cite{properties,review99,Bekaert:2005vh,Didenko:2014dwa}, viz., 
\be
dA+A\star A + C\star I^{(\theta_0)}_{\Comp}= 0 \ ,\qquad dC+A\star C-C\star A= 0\ ,
\ee
with master fields given by operators in the Hermitian, metaplectic singleton ${\cal S}(\xi_0)$, and unaugmented truncation 
\begin{equation}
    dA+A\star A + \Phi\star J^{(\theta_0)}_{\Comp} =0 \ , \qquad d\Phi+A\star \Phi-\Phi\star \pi^\ast(A)=0\ ,  \label{HSG}
\end{equation}
where the twisted-adjoint Weyl zero-form $\Phi:=C\star k$, which consists of one real, spin-$s$ Weyl tensor for $s=0,1,2\dots$ upon homotopy contraction of $\boldsymbol{Z}$ followed by linearization around locally anti-de Sitter spacetimes \cite{more,review99,Didenko:2014dwa,Didenko:2015cwv}.

\subsection{Manifest Lorentz covariance} \label{sec:LorHSG}

As outlined in Paper I, classical defects can be constructed in various bases for the horizontal forms on $\boldsymbol{T}^{(\Comp)}_8$, with the AKSZ formalism operating at the level of metaplectic group algebras, which are strict operator algebras referring to background polarizations. 
To make contact with the Fronsdal formalism, the superconnection $X$ is instead reduced to by contracting $\boldsymbol{Z}^{(\Comp)}_4$ before imposing any boundary conditions on $\boldsymbol{X}^{(\Comp)}_4$ by working in a basis of Lorentz tensors, which yields a manifestly $SL(2,\Comp)$-covariant albeit formally defined homotopy associative algebra (alias $A_\infty$-algebra).
The manifest Lorentz covariance is achieved \cite{review99,Colombo:2010fu} by embedding a bona fide $SL(2,\Comp)_{\rm Lor}$ connection $(\omega^{\alpha\beta},\bar\omega^{\dot\alpha\dot\beta})$ on $\boldsymbol{X}^{(\Comp)}_4$ into the projection $A_{\boldsymbol{X}}$ of $A$ onto $\boldsymbol{X}^{(\Comp)}_4$ via the redefinition
\begin{align}\label{4.11}
A_{\boldsymbol{X}}=:A'+\frac{i}4\left(\omega^{\alpha\beta} M^{({\rm Lor})}_{\alpha\beta}+\bar\omega^{\dot\alpha\dot\beta}\overline{M}^{({\rm Lor})}_{\dot\alpha\dot\beta}\right)\ ,\qquad M^{({\rm Lor})}_{\alpha\beta}:=M^{(0)}_{\alpha\beta}+M^{(S)}_{\alpha\beta}\ ,
\end{align}
where $A'$ is a one-form on $\boldsymbol{X}^{(\Comp)}_4$, and the Lorentz generators
\begin{align}
M^{(0)}_{\alpha\beta}:={}&y_\alpha y_\beta-z_\alpha z_\beta\ ,\qquad[M^{(0)}_{\alpha\beta},M^{(0)}_{\gamma\delta}]_\star =4i\epsilon_{(\beta|(\gamma} M^{(0)}_{\delta)|\alpha)}\ ,\\ M^{(S)}_{\alpha\beta}:={}&S_{(\alpha}\star S_{\beta)}\ ,\qquad\quad [M^{(S)}_{\alpha\beta},M^{(S)}_{\gamma\delta}]_\star =-4i\epsilon_{(\beta|(\gamma} M^{(S)}_{\delta)|\alpha)}\ ,
\end{align}
where the latter are formed out of the Wigner-deformed oscillators
\begin{align}
S_\alpha:={}& z_\alpha - 2i A_\alpha\ ,\qquad \overline{S}_{\dot\alpha}:=-(S_{\alpha})^\dagger=\bar z_{\dot\alpha} -2i \overline{A}_{\dot\alpha}\ ,\\
[S_\alpha,S_\beta]_\star ={}&-2i\epsilon_{\alpha\beta}(1+\Phi\star \kappa)\ ,\qquad  S_\alpha\star \Phi+e^{i\theta_0}\Phi\star \pi(S_\alpha)=0\ ;
\end{align}
it follows that the equations of motion take the manifestly Lorentz covariant form
\begin{eqnarray}
&\nabla A'+A'\star A'+ \frac{i}4\left(r^{\alpha\beta} M^{({\rm Lor})}_{\alpha\beta}+\bar r^{\dot\alpha\dot\beta}\overline{M}^{({\rm Lor})}_{\dot\alpha\dot\beta}\right)=0\ ,&\\&
\nabla \Phi+A'\star \Phi-\Phi\star \pi^\ast(A')=0\ ,\qquad 
\nabla S_\alpha+A'\star  S_\alpha- S_\alpha\star A'=0\ ,&
\end{eqnarray}
where the Lorentz-covariantized exterior derivatives on $\boldsymbol{X}^{(\Comp)}_4$ are given by
\begin{align}
\nabla A':={}&d_{\boldsymbol{X}}A'+ [\omega^{(0)},A']_\star\ ,\quad r^{\alpha\beta}=d\omega^{\alpha\beta}-\omega^{\alpha\gamma}\wedge \omega_\gamma{}^\beta\ ,\\  \nabla \Phi:={}&d_{\boldsymbol{X}}\Phi+ [\omega^{(0)},\Phi]_\star \ ,\quad \nabla S_\alpha:=d_{\boldsymbol{X}} S_\alpha-\omega_\alpha{}^\beta S_\beta+[\omega^{(0)},S_\alpha]_\star\ ,
\end{align}
using $\omega^{(0)}:=\frac{i}4\left(\omega^{\alpha\beta} M^{(0)}_{\alpha\beta}+\bar\omega^{\dot\alpha\dot\beta}\overline{M}^{(0)}_{\dot\alpha\dot\beta}\right)$.

\paragraph{4D fractional-spin gravity.}Towards a Lorentz covariantization of FSG, it is natural to combine \eqref{4.11} with an embedding of the Lorentz connection into the projection $U_{\boldsymbol{X}}$ of the internal one-form $U$ onto $\boldsymbol{X}$ using a set of Lorentz generators that act faithfully on forms valued in ${\cal C}$  viz.,
\begin{align}
U_{\boldsymbol{X}}=:{}&U'+\frac{i}4\left(\omega^{\alpha\beta} M^{({\rm Lor},{\cal C})}_{\alpha\beta}+\bar\omega^{\dot\alpha\dot\beta}\overline{M}^{({\rm Lor},{\cal C})}_{\dot\alpha\dot\beta}\right)\ ,
\\
M^{({\rm Lor},{\cal C})}_{\alpha\beta}:={}&M^{(0,{\cal C})}_{\alpha\beta}+M^{(T)}_{\alpha\beta}\ ,\quad M^{(0,{\cal C})}_{\alpha\beta}:=-z_\alpha z_\beta\ ,\quad M^{(T)}_{\alpha\beta}:=T_{(\alpha}\star T_{\beta)}\ ,\\
T_\alpha:={}& z_\alpha - 2i U_\alpha\ ,\qquad \overline{T}_{\dot\alpha}:=-(T_{\alpha})^\dagger=\bar z_{\dot\alpha} -2i \overline{U}_{\dot\alpha}\ ;&
\end{align}
it follows that the intertwiners $S_\alpha \star \Sigma-\Sigma\star T_\alpha$ are proportional to $\Psi$, leading to cancellations.
Unlike the master fields of HSG, which decompose into towers of finite-dimensional $SL(2,\Comp)$-irreps, the singleton-valued master fields of FSG transform in infinite-dimensional $SL(2,\Comp)$-irreps, for which is it possible to choose several reference states; we hope to return to this interesting issue in a separate work.

\subsection{Linearization and COMST.} 

The FSG model admits locally anti-de Sitter vacua 
\be {A}^{(0)}=\Omega\ ,\qquad d\Omega+\Omega\star \Omega=0\ ,\ee
where $\Omega$ is an $\mso(2,3)$-valued one-form.
Linearization of the unaugmented model around these vacua yields 
\bea & dA^{(1)}+\Omega\star A^{(1)}+A^{(1)}\star\Omega+ \Phi^{(1)}\star J^{(\theta_0)}_\Comp= 0\ ,\qquad d\Phi^{(1)}+\Omega\star \Phi^{(1)}-\Phi^{(1)}\star \pi^\ast(\Omega)= 0\ ,& \label{125}\\
& (d+\Omega)\star \Psi^{(1)}+ \Xi^{(1)}\star k\star J^{(\theta_0)}_\Comp = 0\ ,\qquad (d+\Omega)\star \Xi^{(1)}\star k= 0\ ,&\label{126} \\
& dU^{(1)}+ N^{(1)}\star k\star J^{(\theta_0)}_\Comp = 0\ ,\qquad d N^{(1)}\star k= 0\ ,&\label{127}\eea
containing cocycles on $\boldsymbol{Z}^{(\Comp)}_4$.
Following the gauge function method, the system is first homotopy contracted on $\boldsymbol{X}^{(\Comp)}_4$ using a vacuum gauge function, and then on $\boldsymbol{Z}^{(\Comp)}_4$ using deformed oscillators.
Alternatively, aiming at asymptotic expansions on locally constantly curved backgrounds on $\boldsymbol{X}^{(\Comp)}_4$, the system is first homotopy contracted on $\boldsymbol{Z}^{(\Comp)}_4$ using a homotopy contraction $d^{\ast}_{\boldsymbol{Z}}$ of the DGA $\Omega(\boldsymbol{Z}^{(\Comp)}_4)$ \cite{more,review99,Bekaert:2005vh,Didenko:2014dwa,Didenko:2015cwv}, viz.,
\begin{align}
{\rm Id}_{\Omega\left(\boldsymbol{Z}^{(\Comp)}_4\right)} = {\rm pr}_{H\left(d_{\boldsymbol{Z}}\right)} + {\rm Im}(d^{\ast}_{\boldsymbol{Z}})+{\rm Im}(d_{\boldsymbol{Z}})\ ,\qquad d_{\boldsymbol{Z}} d^{\ast}_{\boldsymbol{Z}}=1\ ,
\end{align}
i.e., an $f\in \Omega(\boldsymbol{Z}^{(\Comp)}_4)$ can be written as 
\begin{align}
f=\check f+  d^{\ast}_{\boldsymbol{Z}} g+ d_{\boldsymbol{Z}} h\ ,\qquad d_{\boldsymbol{Z}} f=g\ ,\qquad \check f\in H\left(d_{\boldsymbol{Z}}\right)\ .
\end{align}
The contraction of Eq. \eqref{125} on $\boldsymbol{Z}^{(\Comp)}_4$ yields 
\be A^{(1)}= \check A^{(1)}- d^{\ast}_{\boldsymbol{Z}}(\Phi^{(1)}\star J^{(\theta_0)}_\Comp)\ ,\qquad \Phi^{(1)}= \check \Phi^{(1)}\ ,\ee
leading to a projected Cartan integrable system on $\boldsymbol{X}^{(\Comp)}_4$, viz.,
\begin{align}
d_{\boldsymbol{X}} \check A^{(1)}+\Omega\star \check A^{(1)}+\check A^{(1)}\star\Omega+\Sigma^{\check A}(e,e;\check \Phi^{(1)})= 0\ ,\label{COMSTHSG1}\\
d_{\boldsymbol{X}}\check \Phi^{(1)}+\Omega\star \check \Phi^{(1)}-\check \Phi^{(1)}\star \pi^\ast(\Omega)= 0\ , \label{COMSTHSG2}    
\end{align}
where $e$ is the background frame field, and 
\begin{align}
\Sigma^{\check A}(e,e;\check \Phi^{(1)}) = \frac{ib}{4}\,e^{\a\ad} e_{\a}{}^{\bd}\partial_{\yb^{\ad}}\partial_{\yb^{\bd}} \left. \check \Phi^{(1)}\right|_{y=0}+\frac{i\bar b}{4}\,e^{\a\ad} e^{\b}{}_{\ad}\partial_{y^{\a}}\partial_{y^{\b}} \left. \check \Phi^{(1)}\right|_{\yb=0} \label{ChEil}    
\end{align} 
is a manifestly, background Lorentz-covariant cocycle on $\boldsymbol{X}^{(\Comp)}_4$ sourcing $\check A^{(1)}$ by $\check \Phi^{(1)}$. 
Under the assumption that $e$ is an invertible frame field, Eqs. \eq{COMSTHSG1} and \eq{COMSTHSG2} constitute a linearized unfolded system that is equivalent to the equations of motion of a set of real Fronsdal fields of spins $s=0,1,2,\dots$ \cite{Misha1987,MV,more,review99,Bekaert:2005vh,Didenko:2014dwa}, which is referred to in the literature as the Central On-Mass-Shell Theorem (COMST).
Contracting Eq. \eq{126} on $\boldsymbol{Z}^{(\Comp)}_4$ yields a projected system
\be d_{\boldsymbol{X}} \check \Psi^{(1)}+\Omega\star \check \Psi^{(1)}+\Sigma^\Psi(\Omega,\Omega;\check \Sigma^{(1)})= 0\ ,\qquad d_{\boldsymbol{X}}\check \Sigma^{(1)}+\Omega\star \check \Sigma^{(1)}= 0\ ,\ee
propagating singletons valued in ${\cal C}(N,N)$; the system's Lorentz covariance properties will be analysed elsewhere.

\section{Coloured conformal higher-spin gravity defect}\label{sec:CHSG}

In this Section, we give the classically consistent reduction of the parent field equation \eq{2.1} to a fully non-linear formulation of 3D, $U(N,N)$-coloured, complex conformal matter fields coupled to conformal HSG and internal, colour gauge fields in terms of horizontal forms on the correspondence space, which provides a fully non-linear completion of the perturbatively defined models constructed by Vasiliev \cite{Misha} and Nilsson \cite{Nilsson}. 

In brief, the reduced model admits a manifestly locally $SL(2,\Real)\times \Real$-covariant formulation and a perturbatively defined homotopy contraction to a Cartan integrable system on $\boldsymbol{X}^{(\Real)}_4=\boldsymbol{X}_3\times \Real$. 
On a fixed, 3D leaf, the first two sub-leading orders of its perturbative expansion around locally, conformally flat spacetime yields an unfolded system of linearized matter fields sourcing the fluctuations in the higher-spin gauge fields by sesquilinear, colour-singlet, primary spacetime currents --- constituting the Central On Mass-Shell Theorem (COMST) for CHSG plus matter.
The extension of the system to $\boldsymbol{X}^{(\Real)}_4$ by adding a background dilation gauge field and one-form components along the foliation direction does not produce any additional cocycles at second order.
As for the colour gauge fields, they are set to zero in the background, and their quadratic fluctuations remain pure gauge on $\boldsymbol{X}^{(\Real)}_4$.
Instead, its projection to $\boldsymbol{Z}^{(\Real)}_4$ is sourced by a sesquilinear, spacetime non-local, higher-spin invariant, colour-adjoint construct, which is finite as the matter fields are expanded harmonically over momentum eigenstates in terms of $L^2$-normalizable wave-functions (see Eqs. \eqref{mom}-\eqref{coherent} and Paper I \cite{paperI}) with Hermitian form \eqref{2.124}. Interestingly, such colour source construct is interpretable as a non-abelian statistics connection for an anyon on a two-dimensional, Lagrangian submanifold of $\boldsymbol{Z}^{(\Real)}_4$.  

We recall, from Paper I, that the natural fibre coordinates on the CCHSG defects (due to the manifest $SL(2;\Real)\times \Real\subset Sp(4;\Comp)$ symmetry of the conformal basis of $\mathfrak{sph}(4;\Real)$ defined in App. \ref{App:emb}) split into the  conjugate pair of holomorphically real doublets $y^{\xi,\alpha}$, $\xi=\pm$, normalized such that
\begin{align}
[y^{\xi,\alpha},y^{\xi',\beta}]_\star=2i\delta^{\xi,\xi'}\epsilon^{\alpha\beta}\ ,\qquad (y^{\xi,\alpha})^\dagger=y^{\xi,\alpha}\ ,\label{3.9}
\end{align}
where the conformal weights are determined by
\begin{align}
[D,y^\xi_\alpha]_\star:=\frac{i\xi}2 y^\xi_\alpha\ ,\qquad D:=\frac14 y^+ y^- ,
\end{align}
where in the last equation we used the NW-SE convention for implicit spinor indices, in which $y^+ y^- \equiv y^{+\a} y^-_\a$ (see Appendix \ref{App:emb}).

Analogously, on the non-commutative base submanifold $\boldsymbol{Z}^{(\Real)}_4$, we introduce canonical coordinates $z^\xi_\alpha$ obeying 
\begin{align}
[z^\xi_\alpha,z^{\xi'}_\beta]_\star= -2i\epsilon_{\alpha\beta}\d^{\xi,-\xi'}
\ ,\qquad (z^\xi_\alpha)^\dagger=-z^\xi_\alpha\ , 
\end{align}
such that 
\begin{align}
[D^{(z)},z^\xi_\alpha]_\star:=\frac{i\xi}2 y^\xi_\alpha\ ,\qquad D^{(z)}:=-\frac14 z^+ z^- \ .
\end{align}
$\boldsymbol{Z}^{(\Real)}_4$ is equipped with the real, comohologically non-trivial two-forms 
\begin{align}
I^\pm_{\Real}=j_{z^\pm}\ ,\qquad j_{z^\pm}:=\frac{i}8 dz^{\pm\alpha}\wedge dz^\pm_\alpha \, \tilde\kappa_{z^\pm}\ ,\quad \tilde\kappa_{z^\pm}:=4\pi\delta^2_\Comp(z^\pm)\ ,\label{IR} 
\end{align}
which are built from manifestly $SL(2,\mathbb{R})_{\rm Lor}\times O(1,1)_{\rm Dil}$-covariant twisted projectors \cite{paperI}.

\subsection{Coloured conformal HSG reduction} \label{sec:CCHSGred}

The reductions 
\begin{align}
\mathbb{A}=\widetilde{\mathbb{A}}= \left[\begin{array}{cc}W&0\\ 0&V\end{array}\right]\equiv \mathbb{W} \ ,\qquad \mathbb{B}= \left[\begin{array}{cc}0&C\\ \overline{C}&0\end{array}\right]\equiv \mathbb{C}\ ,\qquad \widetilde{\mathbb{B}}=I^{\xi}_{\Real}\star \mathbb{C}\ ,\qquad \xi=\pm\ ,\label{CCHSGansatz}
\end{align} 
where $I^\xi_\Real$ are the cohomology elements on $S^2_\xi$ given in \eqref{IR}, and $(W,V,C)$ are horizontal forms of degrees $(1,1,0)$ on the correspondence space obeying 
\begin{align}
W^\dagger=-W\ ,\qquad V^\dagger=-V\ ,\qquad C^\dagger=\overline{C}\ ,
\end{align} 
are classically consistent truncations of the flat superconnection modulo the  equations of motion
\begin{align}\label{5.3}
d\mathbb{W}+\mathbb{W}\star\mathbb{W}+\mathbb{C}\star \mathbb{C}\star I^\xi_\Real= 0\ ,\qquad d\mathbb{C}+\mathbb{W}\star\mathbb{C}-\mathbb{C}\star \mathbb{W}= 0\ ,
\end{align} 
or, in components,
\begin{align}
dW+W\star W + C\star \overline{C} \star I^\xi_{\Real} ={}&0 \ , \label{fullCHSG1}\\dC+W\star C-C\star V={}&0\ , \label{fullCHSG2} \\
dV+V\star V+\overline{C}\star C \star I^\xi_{\Real}={}&0\ ,\label{fullCHSG3} 
\end{align}
provided that $(W,V,C)$ are projected from $S^2_+\times S^2_-$ to $S^2_\xi$, i.e.,
\begin{align}\label{4.12}
 (W,V,C)={\rm Pr}_{\xi}^\ast(W,V,C)\ ,\qquad {\rm Pr}_{\xi}:=\sigma_\xi\circ \pi_\xi\ ,     
\end{align}
where $\pi_\xi : S^2_+\times S^2_-\to S^2_\xi$ is a fibration with section $\sigma_\xi: S^2_\xi\to S^2_+\times S^2_-$; indeed, from 
\begin{align}
{\rm Pr}_{\xi}^\ast(I^\xi_\Real)=I^\xi_\Real\ ,   
\end{align}
it follows that \eqref{5.3} is equivalent to its ${\rm Pr}_{\xi}^\ast$-projection, which is integrable in view of 
\begin{align}
[{\rm Pr}_{\xi}^\ast(\psi_1),{\rm Pr}_{\xi}^\ast(\psi_2)]_\star=0\ ,\qquad \psi_1,\psi_2\in {\rm Pr}_{\xi}^\ast\left(\Omega(S^2_+\times S^2_-)\right)\ .
\end{align}
Before assigning any vacuum expectation value to the one-forms, the $\xi=\pm$ reductions are equivalent. 
For a given $\xi$, \eqref{5.3} admits two inequivalent locally, conformally flat spacetime vacua depending on whether the frame field is taken to gauge i) $y^{-\xi}_\alpha y^{-\xi}_\beta$, which leads to a perturbative description in terms of conformal matter fields; or ii) $y^{\xi}_\alpha y^{\xi}_\beta$, which leads to a perturbative description in terms of dual, conformal matter fields \cite{BMVI}.
In what follows, we consider case (i), leaving case (ii) for a separate work.

\subsection{Manifest Lorentz and dilation covariance} \label{sec:LorCCHSG}

The CCHSG equations of motion can be written on manifestly $SL(2,\Real)\times \Real$-covariant form by embedding a bona fide $SL(2,\Real)\times \Real$ connection $\omega^{\alpha,\beta}=\omega^{\alpha\beta}+\epsilon^{\alpha\beta}\omega$ on $\boldsymbol{X}^{(\Real)}_4$ into the projections $W_{\boldsymbol{X}}$ and $V_{\boldsymbol{X}}$ of $W$ and $V$ onto  $\boldsymbol{X}^{(\Real)}_4$, viz.,
\begin{align}
W_{\boldsymbol{X}}=:W'+\frac{i}4 \omega^{\alpha,\beta} M^{({\cal S})}_{\alpha,\beta}\ ,\qquad
V_{\boldsymbol{X}}=:V'+\frac{i}4 \omega^{\alpha,\beta} M^{({\cal C})}_{\alpha,\beta}\ ,
\end{align}
where the generators
\begin{align}
M^{({\cal S})}_{\alpha,\beta}:={}&M^{(0,{\cal S})}_{\alpha,\beta}+M^{(S)}_{\alpha,\beta}\ ,\qquad M^{(0,{\cal S})}_{\alpha,\beta}:=y^+_\alpha y^-_\beta-z^+_\alpha \star z^-_\beta\ ,\qquad M^{(S)}_{\alpha,\beta}:= S^+_\alpha\star S^-_\beta\ ,\\
M^{({\cal C})}_{\alpha,\beta}:={}&M^{(0,{\cal C})}_{\alpha,\beta}+M^{(T)}_{\alpha,\beta}\ ,\qquad M^{(0,{\cal C})}_{\alpha,\beta}:=-z^+_\alpha \star z^-_\beta\ ,\qquad M^{(T)}_{\alpha,\beta}:= S^+_\alpha\star S^-_\beta\ ,
\end{align}
are built from deformed oscillators
\begin{align}
S^\pm_\alpha=z^\pm_\alpha-2i W^\pm_\alpha\ ,\qquad  T^\pm_\alpha=z^\pm_\alpha-2i V^\pm_\alpha\ ,  
\end{align}
with $W^-_\alpha=0=V^-_\alpha$, obeying
\begin{align}
S^\pm_\alpha \star C=C\star T^\pm_\alpha\ ,\qquad T^\pm_\alpha \star \overline{C}=\overline{C}\star S^\pm_\alpha\ ,   
\end{align}
\begin{align}
[S^+_\alpha,S^+_\beta]_\star ={}&-2i\epsilon_{\alpha\beta} C\star \overline{C}\star \delta^2_\Comp(z^-)\ ,\qquad 
[S^+_\alpha,S^-_\beta]_\star =-2i\epsilon_{\alpha\beta}\ , \qquad   [S^-_\alpha,S^-_\beta]_\star =0\ ,\\
[T^+_\alpha,T^+_\beta]_\star ={}&-2i\epsilon_{\alpha\beta} \overline{C}\star C\star \delta^2_\Comp(z^-)\ ,\qquad 
[T^+_\alpha,T^-_\beta]_\star =-2i\epsilon_{\alpha\beta}\ , \qquad   [T^-_\alpha,T^-_\beta]_\star =0\ ,
\end{align}
from which it follows that $[S^+_{\alpha}\star S^-_{\beta},S^+_\gamma]_\star=2i\epsilon_{\gamma\beta}S^+_{\alpha}$ and $[S^+_{\alpha}\star S^-_{\beta},S^-_\gamma]_\star=2i\epsilon_{\gamma\alpha}S^+_{\beta}$ idem $T^\xi_{\gamma}$. 
The remaining equations of motion take the manifestly $SL(2,\Real)\times \Real$-covariant form
\begin{align}
D W'+W'\star W'+r^{\alpha,\beta} M^{({\cal S})}_{\alpha,\beta}=0\ ,\qquad D V'+V'\star V'+r^{\alpha,\beta} M^{({\cal C})}_{\alpha,\beta}=0\ ,\\ 
DC+W'\star C-C\star V'=0\ ,\qquad D S^\pm_\alpha +[W',S^\pm_\alpha]_\star=0\ ,\qquad D T^\pm_\alpha +[V',T^\pm_\alpha]_\star=0\ ,
\end{align}
using $SL(2,\Real)\times \Real$-covariantized derivatives on $\boldsymbol{X}^{(\Real)}_4$ given by
\bea
& D W':=d_{\boldsymbol{X}}A'+ [\omega^{(0,{\cal S})},W']_\star\ ,\qquad D V':=d_{\boldsymbol{X}}V'+ [\omega^{(0,{\cal C})},V']_\star\ ,&\\
&r^{\alpha,\beta}=d\omega^{\alpha,\beta}-\omega^{\alpha,\gamma}\wedge \omega_{\gamma,}{}^\beta\ ,&\\ & D C:=d_{\boldsymbol{X}}C+ \omega^{(0,{\cal S})}\star C-\overline{C}\star  \omega^{(0,{\cal C})}\ ,&\\
&D S^\pm_\alpha:=d_{\boldsymbol{X}} S^\pm_\alpha-\omega_{\alpha,}{}^\beta S^\pm_\beta+[\omega^{(0,{\cal S})},S_\alpha]_\star\ ,\quad D T^\pm_\alpha:=d_{\boldsymbol{X}} T^\pm_\alpha-\omega_{\alpha,}{}^\beta T^\pm_\beta+[\omega^{(0,{\cal C})},T^\pm_\alpha]_\star\ ,&
\eea
using $\omega^{(0,{\cal S})}:=\frac{i}4\omega^{\alpha,\beta} M^{(0,{\cal S})}_{\alpha,\beta}$ and $\omega^{(0,{\cal C})}:=\frac{i}4\omega^{\alpha,\beta} M^{(0,{\cal C})}_{\alpha,\beta}$.

\subsection{Minkowski vacuum} \label{sec:mink}

In this and the following sections, we restrict the CCHSG defect to the leaf $\boldsymbol{X}_3\times \{1\}$.
The solution space to
\be C^{(0)}=0=V^{(0)}\ ,\qquad W^{(0)}=\Omega_x\ ,\qquad d\Omega_x +\Omega_x\star \Omega_x=0\ ,\label{Minkvac}\ee
with
\be \Omega_x = i \left(e^{m} T_{m} + \frac12\, \omega^{mn} M_{mn}\right)= -i \left(e^{\alpha\beta} T_{\alpha\beta} + \omega^{\alpha\beta} M_{\alpha\beta}\right)\in \mathfrak{iso}(1,2)\ ,\label{Omegasp}\ee
(see Appendix \ref{App:emb} for our spinor basis conventions) assuming that $e^m$ is an invertible frame field, contains locally conformally 3D flat spacetimes. 

The Minkowski translation generator $T_m$ can be conventionally chosen to be either the $D$-raising or the $D$-lowering operator. In the following we choose $T_m$ as the $D$-raising operator (and the special conformal transformation generators $K_m$ as the $D$-lowering ones, resulting in the conformal algebra as given in \eq{dt}-\eq{mt}), which corresponds to the oscillator realization \eq{PKapp} and breaks the symmetry between conformal singleton and anti-singleton basis \eq{confsing} by selecting, as Poincar\'e-invariant vacuum, the singleton highest-weight state $\ket{(-i/2)}$. In other words, this choice for the background connection $\O_x$ implies that a standard 3D conformal scalar be expanded over states $(y^-)^n_{\a_1...\a_n}\star\ket{(-i/2)}$, which in turn breaks the polarization symmetry in the reduction of the CCHSG two-form $\widetilde{{\mathbb{B}}}$ in \eq{CCHSGansatz}: indeed, as we shall see in \eq{5.32}, for the term $C\star\overline C\star I_{\Real}$ to be a regular source term for the gauge fields in $W$, at the first non-trivial order in \eq{fullCHSG1}, it is necessary that $I_\Real = I_\Real^-$ for $C$ encoding a 3D scalar fluctuation\footnote{On the other hand, within the realization \eq{PKapp} for the translation generator in $\O_x$, the scalar expanded on the conformal anti-singleton states $(y^+)^n_{\a_1...\a_n}\star\ket{(i/2)}$ actually acquires the interpretation of unfolded dual scalar, which, as it can be shown, encodes the singular, fundamental solution to the Klein-Gordon equation (see \cite{corfu19} for the corresponding bulk master field). Interestingly, one can also consider the possibility of expanding $\widetilde{{\mathbb{B}}}$ on both the standard and the dual scalar, thus including both $I^+_\Real$ and $I^-_\Real$ in the CCHSG reduction, giving rise to a doubling of conformal currents and creating room for a relative phase between the two terms similar to the $b$ phase in the two-form $I_{\Comp}$ of HSG. We defer the study of this system to a future work.}. 
Of course, one could as well use the completely parallel formulation with $T_m$ realized as the $D$-lowering operator and thus expand around the Poincar\'e-invariant vacuum $\ket{(i/2)}$, which choice selects $I^{+}_{\Real}$ in $\widetilde{{\mathbb{B}}}$.

\subsection{Propagating fields and conformal currents} \label{sec:currents}

In what follows, we derive the 3D central on mass-shell theorem stating that linearized, conformal HSG fields on locally, conformally, 3D flat spacetimes couple to conformal scalars (and spinors) on their mass shells via conformal currents.

Expanding in the zero-form around the vacuum \eq{Minkvac}, yields 
\begin{align}
    &dC^{(1)} +\Omega_x\star C^{(1)}=0\ ,\qquad d\overline C^{(1)} -\overline C^{(1)}\star \Omega_x =0\ ,&\label{C1}\\
    &d W^{(2)}+  {\Omega}_x\star W^{(2)}+W^{(2)}\star \Omega_x + C^{(1)}\star \overline{C}^{(1)}\star I_{\Real}=0\ ,&\label{W2}\\
    &d V^{(2)}+ \overline C^{(1)}\star {C}^{(1)}\star I_{\Real}=0\ .&\label{V2}
\end{align}

\paragraph{Zero-form master field.} A set of unfolded, coloured, conformal, free, bosonic scalar fields can be encoded into the linearized intertwining zero-form master field $C$ obeying \eq{C1} by assuming the expansion\footnote{We recall that, in  equations such as \eq{expC}, for notational simplicity we are omitting the Wigner-Ville map, which represents the associative algebra of endomorphism of some Hermitian module in terms of Gaussian symbols of operators (including distributions) on $\boldsymbol{Y}$. Thus, \eq{expC} should more properly be written as $C^{(1)}=  c^{(1)\hat I}\star \varphi_{WV}(\ket{(-i/2)}\bra{e_{\hat I}})$, and in the following we shall simply identify, with a slight abuse of notation, the abstract endomorphisms with their symbol realization, e.g., $ 4\exp(\pm iy^+y^-)\equiv 4\exp(\pm 4iD)=\ket{(\pm i/2)}\bra{(\pm i/2)}$. See Appendix \ref{App:B} for a concrete realization of the intertwining bimodule $\ket{(-i/2)}\bra{e_{\hat I}}$ in
terms of $Y$ oscillators.}
\be C^{(1)}= c^{(1)\hat I}\star \ket{(-i/2)}\bra{e_{\hat I}}\ ,\label{expC}\ee
with Hermitian conjugate 
\be \overline C^{(1)}= \ket{e^{\hat I}}\bra{(i/2)}\star\, \bar c^{(1)}_{\hat I}\ , \label{expCb}\ee
where: i) $\ket{(-i/2)}$ is a highest-weight state of ${\cal T}^-(-i/2)$ obeying
\be y^+_\alpha\star \ket{(-i/2)}=0\ ,\qquad \left(D+\frac{i}2\right)\star \ket{(-i/2)}=0\ ,\label{HW}\ee
and $\bra{(i/2)}$ its Hermitian conjugate, satisfying
\be  \bra{(i/2)}\star y^+_\a =0\ ,\qquad \bra{(i/2)}\star \left(D-\frac{i}2\right)=0\ ,\label{HWconj}\ee
(see also Paper I \cite{paperI} and Appendix \ref{App:emb}); ii) $\ket{ e^{\hat I}}\equiv \ket{ e^{ I}_\epsilon}$, $\epsilon=\pm$, $I=1,\dots, N$, span a finite-dimensional Hermitian space of signature $(N,N)$, with normalization
\bea & \langle e^{\epsilon,I}|e^{\epsilon'}_J\rangle =\delta_{\epsilon,\epsilon'}\delta^I_J\ ,&\label{orthocol}\\
&\pi^\ast\bar \pi^\ast(\mid (-i/2)\rangle\langle e^\epsilon_I\mid)=\epsilon\mid (-i/2)\rangle\langle e^\epsilon_I\mid\ ,&\eea
where $\bra{e_I^\epsilon}:= \epsilon (\ket{e^I_\epsilon})^\dagger$; iii) the $SL(2,R)$-covariant generating functions
\be c^{(1)\hat I}=\sum_{n=0}^\infty \frac{1}{n!}\,\phi^{\hat I,m(n)} K_{m_1}...K_{m_n} \ ,
\qquad \bar c_{\hat I}^{(1)}= \sum_{n=0}^\infty \frac{1}{n!} \,\bar \phi_{\hat I}{}^{m(n)}  K_{m_1}...K_{m_n}\ ;\label{130}\ee
where $\phi^{\hat I,m(n)}$ and $\bar \phi_{\hat I}{}^{m(n)}$ are traceless zero-forms coefficients\footnote{We use the shorthand notation $T_{m(n)}$ to denote a tensor with $n$ totally symmetrized indices $T_{m_1...m_n} =
T_{(m_1...m_n)}$, both for vector and spinor indices. Repeated non-contracted indices are also to be understood as totally symmetrized, $S_{\a(2)}T_{\a(2)}=S_{\a\a}T_{\a\a} :=
S_{(\a_1 \a_2} T_{\a_3 \a_4)}$.} on ${\boldsymbol X}_3$, $\eta^{m_1m_2}\phi^{\hat I}_{m_1m_2m_3...m_n}=0$; and iv) the adjoint, colour matrix elements
\be U_{\hat I}{}^{\hat J}:=\bra{(i/2)} \star \bar c^{(1)}_{\hat I}\star c^{(1)\hat J}\star \ket{(-i/2)}\ ,\label{Ucol}
\ee
are finite\footnote{The introduction of fermionic, spinor fields requires an extension of the oscillator algebra by a pair of fermionic oscillators of same type used in the ${\cal N}=1$ supersymmetric extension of Vasiliev's bosonic HSG; we defer this construction to a future work.\label{Fermfoot}}. 

As can be seen using the commutation relations \eq{dt}-\eq{mt}, the condition \eq{HW} and that $M_{mn}\star \ket{(-i/2)}=0$, Eq. \eq{C1}, together with \eq{expC} and \eq{130}, are equivalent to the unfolded chain of equations 
\be \nabla\phi^{\hat I}_{m(n)} - i \, e^p\,\phi^{\hat I}_{m(n)p} = 0 \ , \label{unfeq} \ee
(while $\bar\phi_{\hat I}^{m(n)}$ satisfy the complex conjugate equations) where $\nabla$ is the flat Lorentz-covariant derivative on ${\boldsymbol X}_3$, which imply that $\phi^{\hat I}$ obeys the massless Klein-Gordon equation (see \cite{review99,Bekaert:2005vh} and references therein). 

Any hybrid bimodule like \eq{expC} can be written as a product of two pure bimodules, one involving conformal states and another one involving colour states. For instance, one can take $\bra{e_{\hat I}}=\bra{e^+_0}\star f_{\hat I}(a)$ using a compact colour reference state $\bra{e^+_0}$, and use
\be \ket{(-i/2)}\bra{e^+_0}=\ket{(-i/2)}\bra{(\mp i/2)}\star \ket{e^+_0}\bra{e^+_0}\ ,\label{hybr}\ee
which have the same left/right conformal/colour eigenvalues, and possess the same left/right polarizations, i.e.
\be y^+_\a \star \ket{(-i/2)}\bra{e^+_0} = 0 =  \ket{(-i/2)}\bra{e^+_0} \star a^{\dagger i}  \ ,\label{pol}\ee
where $a^{\dagger i}= \gamma^{i\a} y^-_\a$ are compact creation operators (see Appendix \ref{App:emb} for a realization of compact and non-compact oscillators). Conditions \eq{pol} are an integrable system of two linear first-order partial differential equations, admitting a unique solution up to an overall normalization absorbed into $c^{(1)I}$ (as we shall show in detail in a paper in preparation \cite{paper0}). Thus, the master field corresponding to \eq{expC} is unique, and this gives us the freedom of using one or another concrete realization \eq{hybr} to our convenience \cite{paperI}. See also Appendix \ref{App:B} for an explicit computation of the hybrid element in $C^{(1)}$ via star product of a conformal-state and a colour-state projector.

\paragraph{Conformal currents.} The system can then be integrated using the normal ordering scheme (with respect to the combinations $Y\pm Z$) in which 
\be (f\star g)(y^\sigma,z^\sigma)= 
\int \frac{d^2 a^+d^2 a^- d^2b^+d^2b^-}{(2\pi)^4} \,e^{i(b^+a ^-+b^- a^+)}
f(y^\sigma+a^\sigma,z^\sigma+a^\sigma) g(y^\sigma+b^\sigma,z^\sigma-b^\sigma)\ .\label{starpm}\ee
Let us begin by analysing the source term $C^{(1)}\star\overline{C}^{(1)}\star I_{\mathbb{R}}$. First, using \eq{expC} and \eq{orthocol},
\be  C^{(1)}\star\overline{C}^{(1)}  =   c^{(1) {\hat I}}\star\ket{(-i/2)}\bra{(i/2)}\star \bar{c}_ {\hat I}^{(1)}  \ ,\label{CCbar}\ee
where the real twisted projector\footnote{Following the terminology of \cite{2017,COMST,meta}, we refer to a Gaussian fibre element with opposite left and right eigenvalues of a $\pi$-odd isometry generator (here $D$) as \emph{twisted projector}, as it can be obtained from a projector via the one-sided action of $\k_y$, e.g. $\ket{(-i/2)}\bra{(i/2)}\sim \ket{(-i/2)}\bra{(-i/2)}\star\k_y$.}  $\ket{(-i/2)}\bra{(i/2)}$ has the Weyl-ordering symbol
\be  \ket{(-i/2)}\bra{(i/2)} 
\ = \ 4\pi \,\delta^2_\Comp(y^+)\ .\label{realtwist}\ee
We refer the reader to Appendix \ref{App:B} for an explicit star-product computation of $C^{(1)}\star \overline C^{(1)}$ leading to the result \eq{realtwist} (thus reproducing the bra-ket computation, up to an overall constant real factor that we can absorb in the definition of $c^{(1)}$). What is paramount is that the star product with the delta function included in $I_{\Real}$ (see the definitions \eq{IR}) gives rise to a regular element: as explained in Section \ref{sec:mink}, this selects
\be I_\Real = I^-_\Real \ , \ee
since
\be \tilde{\kappa}_{z^-}\star \ket{(-i/2)}\bra{(i/2)} = 4\exp(iy^+ z^-) \ ,\label{5.32}\ee
whereas $\tilde \kappa_{z^+}\star \ket{(-i/2)}\bra{(i/2)}$ diverges. Thus, the gauge field curvature has a regular source term given by 
\be C^{(1)}\star\overline{C}^{(1)}\star I_{\mathbb{R}}= \frac{i}{2} \,dz^{-\a}\wedge dz^-_\a \,c^{(1) {\hat I}}\star \exp(iy^+ z^-) \star \bar{c}_ {\hat I}^{(1)} \ . \ee
In order to compute the remaining star products involving the scalar master fields \eq{130}, it is useful to rewrite the latter as
\bea c^{(1) {\hat I}}(x,Y)  &=& \sum_{n=0}^\infty \frac{1}{(2n)!}\,\phi^{ {\hat I},\a(2n)}(x)\,y^-_{\a_1}\ldots y^-_{\a_{2n}} \nonumber\\ &=&  \sum_{n=0}^\infty \frac{(-1)^n}{ (2n)!}\,\phi^{ {\hat I},\a(2n)}(x) \left.\left(\frac{\partial^2}{\partial u^{+}\partial u^{+}}\right)_{\a(2n)}^n \exp(iu^+ y^-)\right|_{u^+=0} \ ,\label{phider}\eea
\bea \bar c^{(1)}_ {\hat I}(x,Y)  &=& \sum_{n=0}^\infty \frac{1}{(2n)!}\,\bar \phi^{\a(2n)}_ {\hat I}(x)\,y^-_{\a_1}\ldots y^-_{\a_{2n}} \nonumber\\ &=&  \sum_{n=0}^\infty \frac{(-1)^n}{(2n)!}\,\bar \phi_{ {\hat I}}^{\a(2n)}(x) \left.\left(\frac{\partial^2}{\partial v^{+}\partial v^{+}}\right)_{\a(2n)}^n \exp(iv^+ y^-)\right|_{v^+=0} \ , \label{phibarder}\eea
in view of the oscillator realization \eq{PKapp}.
The normal-ordering symbol of the generating functional
\be J'(z^-,y^+,y^-;u^+,v^+):=\exp (iu^+ y^-) \star \exp(iy^+z^-) \star \exp (iv^+ y^-)\ ,\ee
can be obtained by computing the star products with the help of \eq{starpm}, and is given by
\be J'(z^-,y^+,y^-;u^+,v^+)=\exp i\left[y^+z^-+(u^++v^+)y^-+(v^+-u^+)z^-\right]\ .\label{J'} \ee
In terms of these building blocks we can then rewrite
\be C^{(1)}\star \overline C^{(1)} \star I_{\Real} =  \frac{i}{2}\, dz^{-\a}\wedge dz^-_\a\,{\cal J}\left[J'\right]\ ,\label{CCbIJ}\ee
where 
\bea  &{\cal J}\left[J'\right]&:= c^{(1) {\hat I}}\star \exp(iy^+ z^-) \star \bar{c}_ {\hat I}^{(1)}\nonumber\\
&=&\sum_{m,n}\frac{(-1)^{m+n}}{(2m)!\,(2n)!}\,\phi^{ {\hat I},\a(2m)}\bar\phi_ {\hat I}{}^{\b(2n)} \left.\left(\frac{\partial^2}{\partial u^{+}\partial u^{+}}\right)_{\a(2m)}^m\left(\frac{\partial^2}{\partial v^{+}\partial v^{+}}\right)_{\b(2n)}^n J'\,\right|_{u^+=0=v^+}  \label{calJ}\eea 
is a linear map acting on the spinor sources $(u^+,v^+)$ and commuting with the DGA operations .
We can now turn to analysing Eq. \eq{W2}, i.e.,
\be  dW^{(2)} +\{\Omega_x,W^{(2)}\}_\star +\frac{i}{2}\, dz^{-\a}\wedge dz^-_\a\,  {\cal J}\left(J'\right)=0\ ,\label{147}\ee
on ${\boldsymbol X}_3\times {\boldsymbol Z}^{(\Real)}_4$. 

In accordance with the projection \eq{4.12}, the source terms are purely along $z^-$, and hence the components of \eq{147} that have at least one component along $z^+$ together with the gauge symmetries associated to $W^{(2)}$ allow to gauge fix $W^{(2)}_+= dz^{+\a} W_{\a}^{(2)}=0 $ and to declare the remaining component independent of $z^+$, i.e., $W^{(2)}=W^{(2)}_x+W^{(2)}_- := dx^m W_m^{(2)}+dz^{-\a}W^{(2)}_{\a}=W^{(2)}(x,z^-, y^+, y^-)$. Let us denote with $d_x$ and $\partial_-$ the components  of the total exterior differential along ${\boldsymbol X}_3$ and along $S^2_- \subset {\boldsymbol Z}^{(\Real)}_4$, respectively.   

Then, the components of Eq. \eq{147} that have 
at least one component along $S^2_-$ 
have the form
\be \partial_- f  =  g \ee
with $f$ and $g$ differential forms on ${\boldsymbol X}_3\times S^2_-$, and have general solution 
\be f = \partial^\ast_- g + \partial_- h+c \ ,  \ee
where $\partial^\ast_-$ is a resolution operator, providing the particular solution, $h$ is a gauge function (or form) and $c$ is an element of the $\partial_-$-cohomology $H(\partial_-)\subset \Omega(S^2_-)$. In Vasiliev gauge $\imath_{\vec E_-}W^{(2)}_- = 0$ (in the terminology of \cite{2017,review,COMST}), with $\vec E_- := z^{-\a}\partial_{z^{-\a}}$, and using a resolution operator with standard homotopy contraction along $\vec E_-$ \cite{review99,Didenko:2015cwv,COMST} 
\be \partial^\ast_- g(dx,dz^-;x,z^-,y^+,y^-)  = \imath_{\vec E_-} \int_0^1\frac{dt}{t}\,g(dx,tdz^-;x,tz^-,y^+,y^-) \ ,\label{hom}\ee
with no cohomological part associated to $ W^{(2)}_-\in \Omega^1(S^2_-)$, the equation
\be \partial_- W^{(2)}_- = -\frac{i}{2} \,dz^{-\a}\wedge dz^-_\a\,{\cal J}\left[J'\right] \ee
is solved as
\be W^{(2)}_- = i\,dz^-\,z^- \int_0^1dt\,t\,{\cal J}\left[J'\right](x,tz^-,y^+,y^-) \ .  \label{W2-}\ee
The spacetime component $W^{(2)}_x$ is then determined from the ``mixed'' component equation
\be \partial_-W^{(2)}_x+d_x  W^{(2)}_- +\{\Omega_x,W^{(2)}_-\}_\star = 0 \ee
as
\be W^{(2)}_x = \imath_{\vec E_-} \int_0^1\frac{dt}{t}\left(d_x  W^{(2)}_- +\{\Omega_x,W^{(2)}_-\}_\star\right)(dx,tdz^-;x,tz^-,y^+,y^-)+w_x^{(2)}(x,y^+,y^-)\ , \label{W2x}\ee
where $w^{(2)}_x\in H^0(\partial_-)$ is a spacetime one-form, which is the generating function for the 3D CHSG gauge fields, and \eq{W2-} implies that the term $d_x  W^{(2)}_-$ does not contribute, since $\imath_{\vec E_-} W^{(2)}_-=0$. Likewise, in $\{\Omega_x,W^{(2)}_-\}_\star$ only the terms that are not in ${\rm ker}(\imath_{\vec E_-})$ give a non-trivial contribution to $W^{(2)}_x$. Taking into account \eq{Omegasp} and the realization \eq{PKapp}-\eq{MDapp}, the latter are
\bea\{\Omega_x,W^{(2)}_-\}_\star &=& -dx^m\wedge dz^{-\a}\left[e_{m\a}{}^{\b}\int_0^1 dt\,t\,\partial_{y^{-\b}}{\cal J}\left[J'\right](x,tz^-,y^+,y^-) \right.\nonumber\\
&& \left.+\o_{m\a}{}^\b\int_0^1 dt\,t\,\partial_{y^{+\b}}{\cal J}\left[J'\right](x,tz^-,y^+,y^-) \right]+{\rm irrelevant}\ ,\eea
while the irrelevant terms are in ${\rm ker}(\imath_{\vec E_-})$. Substituting in \eq{W2x}, one obtains
\be W_x^{(2)} = w_x^{(2)}(x,y^+,y^-) + \widetilde w_x^{(2)}(x,z^-,y^+,y^-)\ , \ee
where  
\be \widetilde w_x^{(2)}(x,z^-,y^+,y^-) := -z^{-\a}\int_0^1 dt'\int_0^1 dt\,t\, \left[e_\a{}^\b\partial_{y^{-\b}} + \o_\a{}^\b\partial_{y^{+\b}}\right]{\cal J}\left[J'\right](x,t'tz^-,y^+,y^-)  \ ,\ee
is the particular solution.
It is then possible to substitute in the pure spacetime component of \eq{147} 
\be d_x W_x^{(2)}+\{\O_x,W_x^{(2)}\}_\star = 0\label{147xx}\ee
which, projected on $z^-=0$, gives the spacetime equation for the gauge fields contained in $w_x^{(2)}=W_x^{(2)}|_{z^-=0}$. Thus, 
\be \left.\{\Omega_x,W^{(2)}_x\}_\star\right|_{z^-=0} = \{\Omega_x,w_x^{(2)}\}_\star + \left.\{\Omega_x,\widetilde w_x^{(2)}\}_\star\right|_{z^-=0} \ .\ee
Now, $\partial_{y^{+\b}}J'\propto z^-_\b J'$, and thus the $\omega$-term in $\widetilde w_x^{(2)}$ is bilinear in $z^-$. But $\{\Omega_x,\widetilde w_x^{(2)}\}_\star$ does not contain any double $z^-$-contraction, which implies that all terms containing the Lorentz connection disappear in the $z^-=0$ projection. On the other hand,
\be -\frac{i}{2}\left.\{e^{\a\b}y^+_\a y^+_\b,\widetilde w_x^{(2)}\}_\star\right|_{z^-=0} = i \,e^{\a\gamma}e_\gamma{}^\b\,\partial_{y^{-\a}} \partial_{y^{-\b}}  \left.{\cal J}\left[J'\right]\right|_{z^-=0}(x,y^-)  \ ,  \ee
and thus \eq{147xx} at $z^-=0$ reads
\be dw_x^{(2)}+\{\Omega_x,w_x^{(2)}\}_\star = i \,e^{\a\gamma}e_\gamma{}^\b\,\partial_{y^{-\a}} \partial_{y^{-\b}}  \left.{\cal J}\left[J'\right]\right|_{z^-=0}\ ,  \label{COMST}\ee
a linearized Cartan integrable system on ${\boldsymbol X}_3$ where the cocycle on the r.h.s. couples the gauge fields contained in $w^{(2)}_x$ to a generating function $\left.{\cal J}\left[J'\right]\right|_{z^-=0}$ of conformal scalar currents of all integer spins\footnote{The additional ``asymmetric'' current found in \cite{Misha} is not obtained in our purely bosonic 3D theory, as it is built out of a sesquilinear construct in the 3D fermion field. We will return to the full CCHSG model including fermionic excitations in a future work, see Footnote \ref{Fermfoot}.}. Indeed, recalling the definitions \eq{calJ}, \eq{J'}, \eq{phider}, \eq{phibarder}, $\left.{\cal J}\left[J'\right]\right|_{z^-=0}$ reads
\bea  \left.{\cal J}\left[J'\right]\right|_{z^-=0}
= \left. \sum_{m,n}\frac{(-1)^{m+n}}{(2m)!\,(2n)!}\,\phi^{ {\hat I},\a(2m)}\bar\phi_ {\hat I}{}^{\b(2n)} \left(\partial^2_{u^+}\right)_{\a(2m)}^m\left(\partial^2_{v^+}\right)_{\b(2n)}^n   \,e^{i(u^++v^+)y^-}\right|_{u^+=0=v^+} \ ,  \label{calJ0}\eea 
and, expanding in $y^-$ with the same normalization used for $\phi^ {\hat I}$ and $\bar\phi_ {\hat I}$, 
\be \left.{\cal J}\left[J'\right]\right|_{z^-=0}= \sum_{s=0}^\infty \frac{1}{s!}\,{\cal J}_{s}^{m(s)}(x) K_{m_1}...K_{m_s}  = \sum_{s=0}^\infty \frac{1}{(2s)!}\,{\cal J}_{s}^{\a(2s)}(x) y^-_{\a_1}...y^-_{\a_{2s}}\ ,\label{Jexp}\ee
and noting that, for every fixed order $2s$ in $y^-$, the only terms surviving the $u^+=0=v^+$ projection are those for which $m+n=s$, it is easy to see that the spin-$s$ coefficient of \eq{calJ0} is the spin-$s$ conserved current 
\be {\cal J}_{s}^{\a(2s)} = \sum_{k=0}^s {2s \choose 2k}\phi^{ {\hat I},\a(2k)}\bar\phi_ {\hat I}^{\a(2(s-k))} = i^s\sum_{k=0}^s {2s \choose 2k}(-1)^k\,\nabla^{\a(2k)}\phi^ {\hat I}\,\nabla^{\a(2(s-k))} \bar \phi_ {\hat I} \ ,\label{4.59} \ee
that is, with vector indices,
\bea {\cal J}_{s,m(s)} 
= i^s\sum_{k=0}^s {2s \choose 2k}(-1)^k\,\nabla_{\{m(k)}\phi^ {\hat I}\, \nabla_{m(s-k)\}}\bar\phi_ {\hat I} 
\ , \label{currents}\eea
where the on-shell equality holds by virtue of the unfolded equations \eq{unfeq} and curly brackets enclosing indices denote symmetric (on-shell-)traceless projection. The relative coefficients of the various terms in  \eq{currents} indeed reproduce conformal currents in $D=3$ \cite{Anselmi,Konstein:2000bi,Giombirev}. The spin-$0$ coefficient ${\cal J}_{0}=\bar\phi_ {\hat I}\,\phi^ {\hat I}$ is a mass term,  and is cut off from \eq{COMST} by the second $y^-$-derivative acting on $\left.{\cal J}\left[J'\right]\right|_{z^-=0}$. This way, at $y^-=0$ in \eq{COMST} the spin-$1$ gauge field in $w^{(2)}$ is sourced by the spin-$1$ conformal current 
\bea
{\cal J}_{1,m}=-i(\bar \phi_ {\hat I} \, \nabla_m\phi^ {\hat I}-\phi^ {\hat I}\,\nabla_m\bar \phi_ {\hat I})\ ;\eea
at order $(y^{-})^2$ \eq{COMST} glues the spin-$2$ gauge field in $w^{(2)}$ to the spin-$2$ conformal current 
\bea {\cal J}_{2,mn}=-\left[\bar \phi_ {\hat I} \,\nabla^2_{mn}\phi^ {\hat I}-6\left(\nabla_{(m}\bar\phi_ {\hat I}\,\nabla_{n)}\phi^ {\hat I}-\frac13\eta_{mn}\partial \bar\phi_ {\hat I}\cdot \nabla \phi^ {\hat I}\right)+\phi^ {\hat I}\,\nabla^2_{mn}\bar\phi_ {\hat I}\right]\ ;\eea
and so on.

\subsection{CCHSG on conformal Minkowskian leaves} \label{sec:rhocurrents}

The perturbative expansion of the CCHSG system above treated can be generalized to take place on a one-parameter family of conformal Minkowskian vacua on $\boldsymbol{X}^{(\Real)}_4$, given by 
\be C^{(0)}=0=V^{(0)}\ ,\qquad W^{(0)}=\Omega\in \mathfrak{iso}(1,2)\oplus \mathfrak{so}(1,1)_D\ ,\label{MinkxRvac}\ee
where $\Omega$ satisfies
\begin{align}
    \qquad d\Omega +\Omega\star \Omega=0\ .
\end{align}
In such foliation topology, $\Omega({\boldsymbol X}_3\times \Real)\equiv \bigoplus_{m,n}\Omega_{(m,n)}({\boldsymbol X}_3\times \Real)$, where $\Omega_{(m,n)}({\boldsymbol X}_3\times \Real)$ consists of $m$-forms on ${\boldsymbol X}_3$ and $n$-forms on $\Real$.
We now decompose  
\bea
   \left. d\right|_{{\boldsymbol X}_3\times \Real}=d_x+d_\rho\ ,\qquad & d_x:\Omega_{(m,n)}({\boldsymbol X}_3\times \Real)\to \Omega_{(m+1,n)}({\boldsymbol X}_3\times \Real) \ ,& \nonumber  \\
   & d_\rho:\Omega_{(m,n)}({\boldsymbol X}_3\times \Real)\to \Omega_{(m,n+1)}({\boldsymbol X}_3\times \Real)\ ,&
\eea
and
\be \Omega = \widetilde \O_x+\O_\r\ ,\qquad \widetilde \O_x\in \Omega_{(1,0)}({\boldsymbol X}_3\times \Real)\ ,\qquad  \O_\r\in \Omega_{(0,1)}({\boldsymbol X}_3\times \Real)\ ,\label{twocp}\ee
and we take
\begin{align}
\displaystyle \widetilde \Omega_x ={}& i \left(\frac{1}{\r}\, e^{m} T_{m} + \frac12\,  \omega^{mn} M_{mn}\right)= -i \left(\frac{1}{\r} \,e^{\alpha\beta}T_{\alpha\beta} + \omega^{\alpha\beta} M_{\alpha\beta}\right)\ , \label{Omegasprho} \\
\displaystyle \O_\r ={}& -i \frac{d\rho}{\r}D\ ,\label{Orho}    
\end{align}  
where $\rho$ is a conformal factor, i.e., a strictly positive coordinate function on $\Real$ such and $\rho\to 0^+$ in one asymptotic region of $\Real$. Finally, Minkowskian leaves arise by assuming that $e^m$ is an invertible frame field. 

Thought of from a 4D perspective, and with the holographic correspondence with 4D HSG in mind, these ${\rm Mink}_3\times \Real$ vacua correspond to foliations of $AdS_4$ in Poincar\'e coordinates, which shares the same background connection $\O$, $\r$ being identified with the radial coordinate, leading to the metric
\be ds^2_{AdS_4} = \frac{dx^2+d\r^2}{\r^2} \ .\ee

\paragraph{Propagating fields and conformal currents on Minkowskian leaves.} 

We now turn to upgrading the perturbative analysis of the CCHSG system to the first non-trivial order on the Minkowskian leaves. Expanding in the zero-form around the vacuum \eq{MinkxRvac} now yields 
\begin{align}
    &dC^{(1)} +\Omega\star C^{(1)}=0\ ,\qquad d\overline C^{(1)} -\overline C^{(1)}\star \Omega =0\ ,&\label{C1xR}\\
    &d W^{(2)}+  {\Omega}\star W^{(2)}+W^{(2)}\star \Omega + C^{(1)}\star \overline{C}^{(1)}\star I_{\Real}=0\ ,&\label{W2xR}\\
    &d V^{(2)}+ \overline C^{(1)}\star {C}^{(1)}\star I_{\Real}=0\ ,&\label{V2xR}
\end{align}
A propagating massless scalar solution of \eq{C1xR} is now encoded in the linearized intertwining zero-form master fields
\be C^{(1)}= \sqrt{\r}\,c^{(1) {\hat I}}\star \ket{ (-i/2)}\bra{e^+_ {\hat I}}\ , \qquad \overline C^{(1)}=\sqrt{\r} \,\ket{ e^{+, {\hat I}}}\bra{ (i/2)}\star\, \bar c^{(1)}_ {\hat I}\ , \label{expCbis}\ee
where now  
\be c^{(1) {\hat I}}=\sum_{n=0}^\infty \frac{\rho^n}{n!}\,\phi^{ {\hat I},m(n)} K_{m_1}...K_{m_n} \ ,
\qquad \bar c_ {\hat I}^{(1)}= \sum_{n=0}^\infty \frac{\rho^n}{n!} \,\bar \phi_{ {\hat I}}{}^{m(n)}  K_{m_1}...K_{m_n}\ ,\label{130rho}\ee
and $\phi^{ {\hat I},m(n)}$ and $\bar \phi_{ {\hat I}}{}^{m(n)}$ are zero-forms on ${\boldsymbol X}_3$. The $x$-components of \eq{C1xR} imply that $\phi^{ {\hat I},m(n)}$ satisfy the unfolded chain of equations 
\be \nabla_x\phi^ {\hat I}_{m(n)} - i \, e^p\,\phi^ {\hat I}_{m(n)p} = 0  \ . \label{unfeqbis} \ee
Thus, on any Minkowskian leaf at fixed $\r$, $\phi^ {\hat I}$ obeys the massless Klein-Gordon equation. Note that the factors of $\r$ in \eq{expCbis}-\eq{130rho} precisely account for the conformal weights of the state $K_{m_1}...K_{m_n}\star\ket{ (-i/2)}\bra{e_ {\hat I}}$ and of its hermitian conjugate, i.e., the expansions \eq{130rho} identically solve the $\r$-components of \eq{C1xR}.

The system \eq{C1xR}-\eq{V2xR} can be integrated as the one on the leaf at $\r=1$ examined before.  
The source term $C^{(1)}\star\overline{C}^{(1)}\star I_{\mathbb{R}}$ is now dressed with one overall power of $\r$, 
\be C^{(1)}\star\overline{C}^{(1)}\star I_{\mathbb{R}}= \frac{i}{2} \,dz^{-\a}\wedge dz^-_\a \,\r\,c^{(1) {\hat I}}\star \exp(iy^+ z^-) \star \bar{c}_ {\hat I}^{(1)} \ , \ee
following \eq{expCbis}, while we absorb the powers of $\r$ of the expansions \eq{130rho} into rescaled oscillators
\be q^\pm_\a := \sqrt{\r}\,y^\pm_\a  \ , \qquad p^{\pm}_\a := \frac{1}{\sqrt{\r}}\,y^\pm_\a \label{pq}\ee
as
\bea c^{(1) {\hat I}}  &=& \sum_{n=0}^\infty \frac{1}{ (2n)!}\,\phi^{ {\hat I},\a(2n)}(x)\,q^-_{\a_1}\ldots q^-_{\a_{2n}} \nonumber\\ &=&  \sum_{n=0}^\infty \frac{(-1)^n}{(2n)!}\,\phi^{ {\hat I},\a(2n)}(x) \left.\left(\partial^2_{u^+}\right)_{\a(2n)}^n \exp(iu^+ q^-)\right|_{u^+=0} \ ,\label{phiderw}\eea
\bea \bar c^{(1)}_ {\hat I}  &=& \sum_{n=0}^\infty \frac{1}{(2n)!}\,\bar \phi^{\a(2n)}_ {\hat I}(x)\,q^-_{\a_1}\ldots q^-_{\a_{2n}} \nonumber\\ &=&  \sum_{n=0}^\infty \frac{(-1)^n}{(2n)!}\,\bar \phi_{ {\hat I}}^{\a(2n)}(x) \left.\left(\partial^2_{v^+}\right)_{\a(2n)}^n \exp(iv^+ q^-)\right|_{v^+=0} \ . \label{phibarderw} \eea
The oscillator dependence of the generating functional for conserved currents is now contained in
\bea \widetilde J'&:=&\exp (iu^+ q^-) \star \exp(iy^+z^-) \star \exp (iv^+ q^-)\nonumber\\
&=& \exp i\left[y^+z^-+(u^++v^+)q^-+\sqrt{\r}(v^+-u^+)z^-\right]\ .\eea
Thus, \eq{W2xR} becomes 
\be  dW^{(2)} +\{\Omega,W^{(2)}\}_\star +\frac{i}{2}\,\r\, dz^{-\a}\wedge dz^-_\a\,  {\cal J}\left(\widetilde J'\right)=0\ ,\label{dW2rho}\ee
where 
\bea  &{\cal J}\left[\widetilde J'\right]&:= c^{(1) {\hat I}}\star \exp(iy^+ z^-) \star \bar{c}_ {\hat I}^{(1)}\nonumber\\
&=&\sum_{m,n}\frac{(-1)^{m+n}}{(2m)!\,(2n)!}\,\phi^{ {\hat I},\a(2m)}\bar\phi_ {\hat I}{}^{\b(2n)} \left.\left(\partial^2_{u^+}\right)_{\a(2m)}^m\left(\partial^2_{v^+}\right)_{\b(2n)}^n \widetilde J'\,\right|_{u^+=0=v^+}  \label{calJrho}\eea 
and all differential forms live on ${\boldsymbol X}_3\times \Real \times {\boldsymbol Z}$. 

As before, we can gauge fix $W_+=0$ and trivialize the $z^+$-dependence of all fields. Then, the integration of the $z^-z^-$-component and of the $z^-x$-component of \eq{dW2rho} in Vasiliev gauge proceed as for the $\r=1$ case treated before, respectively giving  
\be W^{(2)}_- = i\r\,dz^-\,z^- \int_0^1dt\,t\,{\cal J}\left.\left[\widetilde J'\right]\right|_{z^-\to tz^-} \ , \label{W2-rho}\ee
and
\be W_x^{(2)} = w^{(2)}_x + \widetilde w^{(2)}_x \ , \label{Wx2rho}\ee
with 
\be \widetilde w^{(2)}_x := -\,z^{-\a}\int_0^1 dt'\int_0^1 dt\,t\, \left[e_\a{}^\b\partial_{y^{-\b}} +\r \,\o_\a{}^\b\partial_{y^{+\b}}\right]{\cal J}\left.\left[\widetilde J'\right]\right|_{z^-\to t'tz^-} \label{tw2x}\ee
where now $w^{(2)}_x=w^{(2)}_x(x,\r,y^+,y^-)\in H^{0}(\partial_-)$ and, in computing \eq{tw2x}, the form of \eq{twocp} and \eq{Omegasprho} has been taken into account. Note in particular that, due to the $\r$-rescalings, the term containing the Dreibein, which is the one which will survive in the final equations gluing currents to gauge fields, turns out to be $\r$-independent. Finally, the pure spacetime component of \eq{dW2rho} imposes  
\be d_x w^{(2)}_x+\{\widetilde \Omega_x,w^{(2)}_x\}_\star = \frac{i}{\r} \,e^{\a\gamma}e_\gamma{}^\b\,\partial_{y^{-\a}} \partial_{y^{-\b}}  \left.{\cal J}\left[\widetilde J'\right]\right|_{z^-=0} \ .  \label{COMSTrho1}\ee
The remaining dependence on $\r$ is in fact reabsorbed into the rescaling \eq{pq} of the oscillators (which in turn is determined by the conformal weight of the 3D scalar field states). Expanding also the gauge master field and background connection in terms of rescaled oscillators and choosing $(q^-_\alpha,p^+_\alpha)$ as the new independent variables, i.e., with
\bea  & \displaystyle \widetilde \O_x(x,y^+,y^-)=-\frac{i}{2} \left(\frac{1}{\r} \,e^{\alpha\beta}y^+_\alpha y^+_\beta + \omega^{\alpha\beta} y^+_\alpha y^-_\beta\right)=-\frac{i}{2} \left( e^{\alpha\beta}p^+_\alpha p^+_\beta + \omega^{\alpha\beta} p^+_\alpha q^-_\beta\right)=\O_x(x,q^-,p^+)\ ,& \nonumber\\
& w^{(2)}_x=w^{(2)}_x(x,\r,q^-,p^+) & \eea 
(in such a way that gauge fields are expanded in terms of a pair of oscillators whose commutation relations are $\r$-independent like $y^\pm$ \cite{Misha}), we can finally write 
\be d_x w^{(2)}_x+\{\Omega_x,w^{(2)}_x\}_\star =i\,e^{\a\gamma}e_\gamma{}^\b\,\partial_{q^{-\a}} \partial_{q^{-\b}}  \left.{\cal J}\left[\widetilde J'\right]\right|_{z^-=0}\ ,  \label{COMSTrho2}\ee
which verifies that the central on-mass-shell theorem of CHSG theory coupled to conformal scalars is indeed scale-independent. The conserved current generating function now maintains the same form as \eq{calJ0} except for the scaling accompanying the oscillators, 
\bea  \left.{\cal J}\left[\widetilde J'\right]\right|_{z^-=0}
= \left. \sum_{m,n}\frac{(-1)^{m+n}}{(2m)!\,(2n)!}\,\phi^{ {\hat I},\a(2m)}\bar\phi_ {\hat I}{}^{\b(2n)} \left(\partial^2_{u^+}\right)_{\a(2m)}^m\left(\partial^2_{v^+}\right)_{\b(2n)}^n   \,e^{i(u^++v^+)q^-}\right|_{u^+=0=v^+} \ ,  \label{calJ0rho}\eea 
which means that, expanding in $y^-$ with the same normalizations used for $c^ {\hat I}$ and $\bar c_ {\hat I}$, 
\be \left.{\cal J}\left[\widetilde J'\right]\right|_{z^-=0}= \sum_{s=0}^\infty \frac{\r^s}{s!}\,{\cal J}_{s}^{m(s)}(x) K_{m_1}...K_{m_s}  = \sum_{s=0}^\infty \frac{\r^s}{(2s)!}\,{\cal J}_{s}^{\a(2s)}(x) y^-_{\a_1}...y^-_{\a_{2s}}\ ,\label{Jexprho}\ee
the conserved currents keep the same form on any leaf,  
\bea {\cal J}_{s,m(s)} 
= i^s\sum_{k=0}^s {2s \choose 2k}(-1)^k\,\nabla_{\{m(k)}\phi^ {\hat I}\, \nabla_{m(s-k)\}}\bar\phi_ {\hat I} 
\ . \label{currents2}\eea
Let us now consider the additional components of \eq{dW2rho}, i.e., those that have one component along the ``radial'' direction $\r$. First, from 
\be  \partial_- W^{(2)}_\r+d_\r W^{(2)}_- +\{\O_\r,W^{(2)}_-\}_\star=0 \ ,\ee
using \eq{W2-rho}, one obtains
\be W^{(2)}_\r =w_\r^{(2)} + \imath_{\vec E_-} \int_0^1\frac{dt}{t}\left.\left(d_\r  W^{(2)}_- +\{\Omega_\r,W^{(2)}_-\}_\star\right)\right|_{z^-\to tz^-, \ dz^- \to tdz^-} \ , \ee
with $w_\r^{(2)}= w_\r^{(2)}(x,\r,y^+,y^-)\in H^0(\partial_-)$. 
Once more, the form of $W^{(2)}_-$ \eq{W2-rho} implies that the term $d_\r  W^{(2)}_-$ does not contribute, since $\imath_{\vec E_-} W^{(2)}_-=0$. Likewise, taking into account the definition of $\O_\r$ \eq{Orho} and  that
\bea [D,z^-_\a f(z^-,y^+,y^-)]_\star &=& \frac{1}{2}\left[iz^-_\a (y^+ \partial_{y^+} -y^-\partial_{y^-})f + z^-_\a \partial_{z^-}\partial_{y^+}f-\partial_{y^{+\a}}f\right]\ ,\eea
it is immediately evident that (as $z^{-\a}z^-_\a=0$)
\be \imath_{\vec E_-}\{\Omega_\r,W^{(2)}_-\}_\star \propto \imath_{\vec E_-}\,d\r\, dz^- \partial_{y^+} \int_0^1 dt\,t\,{\cal J}\left.\left[\widetilde J'\right]\right|_{z^-\to tz^-} \ ,\label{5.95}\ee
and since $\partial_{y^{+\a}}\widetilde J'\propto z^-_\a \widetilde J'$ the r.h.s. of \eq{5.95} also vanishes, thus leaving
\be W^{(2)}_\r =w_\r^{(2)} \ .\label{Ww} \ee
This $z^\pm$-independent one-form connection along $\r$ is constrained by the $x\rho$-component of \eq{dW2rho},
\be  d_x w_\r^{(2)}+d_\r W^{(2)}_x +\{\O_x,w_\r^{(2)}\}_\star +\{\O_\r,W^{(2)}_x\}_\star=0  \ ,\label{lasteq}\ee
which determines it in terms of $W^{(2)}_x$ \eq{Wx2rho}, in its turn determined via \eq{tw2x}-\eq{COMSTrho2} in terms of the currents in ${\cal J}\left[\widetilde J'\right]$. Denoting $D_x g := d_x g  + \O_x\star g - (-1)^{{\rm deg}(g)} g\star \O_x$, analogously for $D_\r$, and considering that, for the same reasons leading to \eq{Ww},
\be \left.\{\O_\r,W^{(2)}_x\}_\star\right|_{z^-=0} = \left. \{\O_\r,w^{(2)}_x + \widetilde w^{(2)}_x\}_\star \right|_{z^-=0} = \left. \{\O_\r,w^{(2)}_x\}_\star \right|_{z^-=0}\ ,   \ee
on the physical surface $z^-=0$ \eq{lasteq} reduces to
\be D_x  w_\r^{(2)}+D_\r w^{(2)}_x = 0 \ .\label{Dwrho} \ee
Eqs. \eq{COMSTrho2} and \eq{Dwrho} form an integrable set of equations (at second order in perturbation theory) and control the dynamics of the gauge fields on ${\boldsymbol X}_3\times \Real$. Actually, the gauge parameter associated to $w^{(2)}$ can be used to set $w_\r^{(2)}=0$\footnote{Indeed, the system \eq{COMSTrho2}-\eq{Dwrho} can be rewritten as 
\bea  D_x  w_x^{(2)}+\left.D_x \widetilde w^{(2)}_x\right|_{z^-=0} &=& 0\ ,\nonumber\\ 
D_x  w_\r^{(2)}+D_\r w^{(2)}_x = 0\nonumber \ ,\eea 
and while $w_x^{(2)}$ cannot be gauged away (since $\left.D_x \widetilde w^{(2)}_x\right|_{z^-=0}$ is a cohomological element), $D_\r w_x^{(2)}$ is $D_x$-exact. Integrating the first equation in fact gives $w_x^{(2)}=K_x+D_x\e$, where $K_x:=-D^{\ast}_x\left(\left.D_x \widetilde w^{(2)}_x\right|_{z^-=0}\right)$  is the particular solution, and we use the shorthand notation $D^{\ast}_x$ for the homotopy integration along $x^m$ \cite{COMST}. But $D_\r K_x=0$, as $D_\r$ anticommutes with $D^{\ast}_x$ and $D_\r\left.D_x \widetilde w^{(2)}_x\right|_{z^-=0}=0 $, which is, in fact, the consistency condition of the system \eq{COMSTrho2}-\eq{Dwrho}. This implies $D_\r w_x^{(2)} = D_\r D_x \e$, which, substituted in the second equation, enables to peel off one $D_x$ and leave $w^{(2)}_\r = D_\r \e$.}, leaving the simpler system
\bea  D_x w^{(2)}_x &=& i\,e^{\a\gamma}e_\gamma{}^\b\,\partial_{q^{-\a}} \partial_{q^{-\b}}  \left.{\cal J}\left[\widetilde J'\right]\right|_{z^-=0}\ ,  \label{COMST1fin}\\
 D_{\r} w^{(2)}_x &=& 0 \label{COMST2fin} \ ,\eea
determining the coupling of CHSG gauge fields with matter currents and the scale dependence of the gauge fields. Eq. \eq{COMST2fin} can be seen, from the 3D point of view, as a renormalization group equation of a free theory in which all fields evolve according to their classical scaling dimensions.

Eqs. \eq{COMST1fin}-\eq{COMST2fin} are directly obtained from perturbative analysis of our CCHSG reduction. It is interesting to compare these results with those derived in \cite{Misha} from a different starting point, the pull-back of linearized 4D HSG equations on a 3D leaf. There, starting with a doubled set of currents, due to the theory's 4D origin, a system analogue to \eq{COMST1fin}-\eq{COMST2fin} was obtained by means of one specific choice of boundary conditions for bulk fields. We devote Section \ref{App:D} to the detailed comparison between the second-order expansion of our CCHSG system on ${\rm Mink}_3\times \Real$, above studied, and the pull-back on 3D leaves of the linearized HSG equations on the Poincar\'e patch of $AdS_4$ studied in \cite{Misha}.

\subsection{Analysis of the colour gauge field equation}\label{sec:Vsource}

We now turn to examining the source term for the colour gauge field $V$ from \eq{V2}.
To simplify the analysis, we limit ourselves to the $\r=1$ leaf of $\boldsymbol{X}^{(\Real)}_4$ and break the colour gauge group down to $U(N)$ by setting 
\begin{align}
c^{(1)\hat I}\equiv (c^{(1)I}_+,c^{(1)I}_-)=(c^{(1)I},0)\ .
\end{align}
The resulting source terms is given by
\be \overline C^{(1)}\star {C}^{(1)}\star I_{\Real}= \ket{e^{+,I}}\bra{(i/2)}\star \bar c_I^{(1)}\star c^{(1)J} \star \ket{(-i/2)}\bra{e^+_J}\star I_{\Real}  =: \ket{e^{+,I}} \bra{e^+_J}U_I{}^J\star I_{\Real}\ ,\label{Vsource}\ee
As anticipated, the adjoint matrix elements $U_I{}^J$ \eq{Ucol} need to be finite. In order to compute the latter it is convenient to expand the scalar field on a basis of states with well-defined inner product, like the momentum-eigenstate basis $e^{\tfrac{i}2\l y^-}  \star \ket{(-i/2)}=\ket{+;\e;\l}$ \eq{coherent}. Moreover, as seen in Paper I, the momentum-eigenstate basis, diagonalizing $y^+_\a$ (see \eq{coherent}), has also the advantage of manifestly separating positive and negative eigenvalues of the 3D energy generator $T_0=-\frac{1}{4}(\gamma_0)^{\a\b}y^+_\a y^+_\b$ in accordance with the real or imaginary nature of $\l_\a$, since
\be T_0\star \ket{+;\e;\l}  = -\frac{1}{4}(\gamma_0)^{\a\b}\l_\a\l_\b \ket{+;\e;\l} = \frac{\l^2_1+\l^2_2}{4}\,\ket{+;\e;\l} 
\ee
(see Appendix \ref{App:emb} for our realization of $\mso(1,2)$ matrices $\gamma_m$). Thus, we can separate out positive-frequency ($c^{(1)I}_>$) and negative-frequency ($c^{(1)I}_<$) components of the 3D scalar field,
\bea  & \displaystyle  C^{(1)}=c^{(1)I} \star \ket{(-i/2)}\bra{e^+_I} 
= \left[c^{(1)I}_> + c^{(1)I}_< \right]\star \ket{(-i/2)}\bra{e^+_I} \ ,&\\ 
\hspace{-1cm} & \displaystyle \overline C^{(1)}=   \ket{e^{+,I}}\bra{(i/2)} \star  \bar c^{(1)}_I  :=
 \ket{e^{+,I}}\bra{(i/2)} \star \left[\bar c^{(1)}_{I>} + \bar c^{(1)}_{I<} \right] \ , & \eea
where, in terms of real momenta $\ell_\a$, 
\bea & \displaystyle c^{(1)I}_> := \int_{\Real^2} \frac{d^2\ell}{4\pi}\,\widetilde{\phi}^I_>(\ell,x)\,e^{\tfrac{i}2\ell y^-} \ , \qquad c^{(1)I}_< := \int_{\Real^2}  \frac{d^2\ell}{4\pi}\,\widetilde{\phi}^I_<(\ell,x)\,e^{-\tfrac{1}{2}\ell y^-} \ ,& \label{cfreq1}\\ 
 & \displaystyle \bar c^{(1)}_{I>} := \int_{\Real^2}  \frac{d^2\ell}{4\pi}\,\overline{\widetilde{\phi}}_{I>}(\ell,x)\,e^{-\tfrac{i}{2}\ell y^-} \ , \qquad \bar c^{(1)}_{I<} := \int_{\Real^2}  \frac{d^2\ell}{4\pi}\,\overline{\widetilde{\phi}}_{I<}(\ell,x)\,e^{-\tfrac{1}2\ell y^-}\ , & \label{cfreq2}
 \eea
with transforms $\widetilde{\phi}^I_{\lessgtr}(\ell,x)$ and $\widetilde{\bar \phi}_{I \lessgtr }(\ell,x)$ being even functions of $\ell_\a$. 
Note that, working in the non-minimal bosonic model and with a complex 3D scalar, positive and negative frequency components are independent and not connected via hermitian conjugation.

In order to compute \eq{Vsource} we give a concrete star-product realization of the mixed bimodules $C^{(1)}$ and $\overline C^{(1)}$, as 
\bea & \ket{e^{+,I}}\bra{(i/2)}\star \bar c_I^{(1)}\star c^{(1)J} \star \ket{(-i/2)}\bra{e^+_J} & \nonumber\\
& = \ket{e^{+,I}}\bra{e^+_0}\star \ket{(-i/2)}\bra{(i/2)}\star \bar c_I^{(1)}\star c^{(1)J} \star \ket{(-i/2)}\bra{(i/2)}\star\ket{e^+_0}\bra{e^+_J} &\ , \label{Vsource2}
\eea
with $\ket{(-i/2)}\bra{(i/2)}$ given in \eq{realtwist} and $\ket{e^+_0}$ being a colour reference state with non-trivial overlap with $\ket{(\pm i/2)}$, of which we shall soon give one particularly simple realization\footnote{As already commented in Section \ref{sec:currents}, the concrete realization of $C^{(1)}$ used from \eq{Vsource2} onwards to compute the source terme for $V$ is simply a convenient choice. We could have equivalently carried out the entire computation with a different choice for the ``inner'' states, e.g. $C^{(1)}=c^{(1)I} \star \ket{(-i/2)}\bra{(-i/2)}\star\ket{e^+_0}\bra{e^+_I}$. Indeed, this choice and the one used in the main text only differ by a constant. For example, realizing the ``colour'' states in terms of compact $AdS$ generators and choosing, for simplicity, $N_{\rm col}=1$ and $\bra{e^+_I}=\bra{e^+_0}$ to coincide with the compact singleton ground state, such that $\ket{e^+_0}\bra{e^+_0}=4\exp(-4E)$, it is easy to check that
\bea & \ket{(-i/2)}\bra{(-i/2)}\star \ket{e^+_0}\bra{e^+_0} = 4\exp{-4iD}\star 4\exp{-4E} = 8 \exp(-4E-4iD+4M_{02})& \nonumber \\ 
&= 4i\exp{-4iD}\star\k_y\star 4\exp{-4E} = -2 \ket{(-i/2)}\bra{(i/2)}\star \ket{e^+_0}\bra{e^+_0} \ .&\nonumber\eea
}.

According to \eq{cfreq1}-\eq{cfreq2}, the central factor of \eq{Vsource2} splits into the four terms 
\bea &\ket{(-i/2)}\bra{(i/2)}\star \bar c_I^{(1)}\star c^{(1)J} \star \ket{(-i/2)}\bra{(i/2)} \nonumber  &\\ 
&\ket{(-i/2)}\bra{(i/2)}\star \left (\bar c_{I>}^{(1)}+\bar c_{I<}^{(1)}\right)\star \left(c^{(1)J}_>+c^{(1)J}_<\right) \star \ket{(-i/2)}\bra{(i/2)} \ .\label{intfactor}&
\eea
To compute each of the above terms we need the star-product Lemma
\be \d^2_\Comp(y^+)\star \exp(iky^-)\star \d^2_\Comp(y^+) =  \d^2_{\Comp}(y^+)\d^2_{\Comp}(k)\ ,\ee
where, denoting with $\ell$ the momentum used to expand $c^{(1)I}$ and with $\ell'$ the one used in the expansion of $\bar c_I^{(1)}$, $k$ has the following realization in the four sectors $>>$, $><$, $<>$, $<<$:
\bea >\,> &:& \qquad k = \ell-\ell' \ ,\qquad 
>\,< \ : \ \qquad k = i\ell-\ell' \ ,\nonumber \\ 
<\,> &:& \qquad k = \ell+i\ell' \ ,\qquad 
<\,<  \ : \ \qquad k = i\ell+i\ell' \ .
\eea
Coherently with the Hermitian form \eq{2.124} the $\d^2_{\Comp}(k)$ insertion in \eq{intfactor} leads to non-vanishing overlap only between states with both positive or both negative 3D energy. Hence, taking into account \eq{deltaC},
\bea  & \displaystyle \hspace{-4cm}\ket{(-i/2)}\bra{(i/2)}\star \bar c_I^{(1)}\star c^{(1)J} \star \ket{(-i/2)}\bra{(i/2)}  \nonumber \\[7pt]
& \displaystyle  = (4\pi)^2 \d^2_{\Comp}(y^+) \int \frac{d^2\ell \,d^2\ell'}{(4\pi)^2}\,\left[\widetilde{\bar \phi}_{I>}(\ell',x)\,\widetilde{\phi}^J_>(\ell,x)-\widetilde{\bar \phi}_{I<}(\ell',x)\,\widetilde{\phi}^J_<(\ell,x)\right]4\d^2_{\Comp}(\ell-\ell') \nonumber \\[7pt]
&  \displaystyle  = 4\,\d_\Comp^2(y^+) \int d^2\ell \,\left[\widetilde{\bar \phi}_{I>}(\ell,x)\,\widetilde{\phi}^J_>(\ell,x)-\widetilde{\bar \phi}_{I<}(\ell,x)\,\widetilde{\phi}^J_<(\ell,x)\right] =: U_I{}^J \,\d^2_\Comp(y^+) \ ,\label{calTd}\eea
giving rise to a non-local, composite source. By assumption iv) (see Eq. \eq{Ucol}) we require the Fourier transforms of the 3D scalar to be $L^2$-normalizable, in such a way that $U_I{}^J$ is finite, thus forming the entries of a $U(N)$ generator, a Hermitian matrix $U_{IJ}^\ast=U_{JI}$. Summarizing, in terms of the above building blocks, the source term for $V$ amounts to
\be \overline C^{(1)}\star {C}^{(1)}\star I_{\Real} = U_I{}^J \,\ket{e^{+,I}}\bra{e^+_0}\star\d^2_\Comp(y^+)\star \ket{e^+_0}\bra{e^+_J} \star I_{\Real} \ .  \label{summary}\ee

\paragraph{An example with $N_{\rm col}=1$.}

Let us now give one concrete example showing that the remaining star products, while including singular factors, indeed give rise to a regular function. To this end, we choose, for simplicity, the colour group to be $U(1)$, thus having a single compact state, which we take to coincide with the compact singleton ground state, i.e.,
\begin{align}
   \ket{e^+_0}\equiv\ket{1/2;(0)}\ ,
\end{align}
(where $1/2$ is the eigenvalue of the compact $AdS$ energy generator $P_0\equiv E$, see Appendix \ref{App:emb}), in such a way that the colour factor in \eq{summary} can be written as $\ket{e^+_0}\bra{e^+_0}=4\exp(-4E)$ \cite{fibre,2017,COMST} and realized in terms of the same set of oscillators as the non-compact sector. Thus, in this particular example, the internal gauge field $V$ is sourced by
\be \overline C^{(1)}\star {C}^{(1)}\star I_{\Real} = U\,\ket{e^+_0}\bra{e^+_0}\star\d_\Comp^2(y^+)\star \ket{e^+_0}\bra{e^+_0} \star I_{\Real} \ ,  \ee
where the generator $U\in \Real$. In order to compute the star products it is convenient to represent all elements in $\msl(2,\Comp)$-covariant form, i.e., 
\bea  \ket{e^+_0}\bra{e^+_0}&=&4\exp(-4E)= 4\exp(y\s_0 \yb) \ ,\\
\d^2_\Comp(y^+)&=&-\frac{i}{\pi}\,\exp(iy\s_2\yb)   \star \k_y= -\frac{i}{\pi}\,\exp(-4iD)\star\k_y \ ,\\
I_{\Real}&=& -\pi\,dz^{-\a}\wedge dz^-_\a \,\d^2_\Comp(z-\s_2\zb)\ .\eea
Then, using $\k_y\star \exp(-4E)=\exp(4E)\star\k_y$ and the Lemma
\be  \exp(-4E) \star \exp(-4iD)\star\exp(4E)=\frac{\pi}4\,\d^2_\Comp(y-i\s_0\yb) \ , \ee
which can obtained by a computation analogous to the one carried out in Appendix \ref{App:B}, we finally have that, for this simple example, 
\bea \overline C^{(1)}\star {C}^{(1)}\star I_{\Real} &=& 4\pi i \,dz^{-\a}\wedge dz^-_\a \, U\,\d^2_\Comp(y-i\s_0\yb)\star\k_y\star  \,\d_\Comp^2(z-\s_2\zb) \nonumber \\
&=& 2i\,dz^{-\a}\wedge dz^-_\a U\, \d^2_\Comp(y-i\s_0\yb)\star\,\exp(iy(z-\s_2\zb)) \nonumber\\
&=& -\frac{1}{2\pi i}\,dz^{-\a}\wedge dz^-_\a\, U\,\exp\left(\frac12 y\s_0\yb -\frac14(y\s_{02}y+{\rm h.c.})+\frac12 z\s_0\zb\right.\nonumber \\
&& \left.-\frac14(z\s_{02}z+{\rm h.c.})+\frac{i}2 y(z-\s_2\zb)+\frac{i}2 \yb(\zb-\bar \s_2 z)\right)\ ,\eea
which is indeed a regular function of the oscillators.

\subsection{On the role of the CCHSG colour sector}\label{sec:roleV}

As we have shown in the preceding Sections, our CCHSG system includes a CHSG coupled to coloured matter fields via conformal currents of all spins. The current master field is manifestly star-factorized in terms of the 3D zero-form master field $C^{(1)}$ and of its complex conjugate $\overline C^{(1)}$ \eq{expC}-\eq{expCb} --- where $C^{(1)}$ features a 3D massless scalar unfolded master field acting on an intertwining element $\ket{(-i/2)}\bra{e^I}$. The latter in its turn admits a unique oscillator realization (see Sections \ref{sec:currents}, \ref{sec:Vsource} and Appendix \ref{App:B}). 

This construction completely refers to elements of $MpH(4,\Comp)$ realized in terms of $y^\pm_\a$ oscillators, without introducing separately an abstract 3D scalar Hilbert space. However, it requires that the product $C^{(1)}\star \overline C^{(1)}\star I_{\Real}$, from which one extract the conformal currents, be well defined, as well as $\overline C^{(1)}\star C^{(1)}\star I_{\Real}$ constituting the source in the equation for $V$. 

As shown explicitly in Appendix \ref{App:B}, the hybrid bimodule structure of $C^{(1)}$, with the introduction of colour states, helps smoothing the star product $C^{(1)}\star \overline C^{(1)}$ and obtaining a well-defined source term for the CHSG gauge fields. Indeed, were it not for the insertion of the colour ground-state projector, the star product $C^{(1)}\star \overline C^{(1)}$ would diverge, due to the direct vacuum-anti-vacuum clash $\ket{(-i/2)}\bra{(-i/2)}\star \ket{(i/2)}\bra{(i/2)}$. The insertion of colour states helps instead reducing $C^{(1)}\star \overline C^{(1)}$ to the twisted projector \eq{CCbar}-\eq{realtwist}, represented by an analytic delta function symbol in Weyl order, which then has, in its turn, a regular star product with $I_{\Real}$, as we have seen.

\paragraph{Ruled out alternative.}

In what follows, we provide an argument that rules out the possibility of modifying the theory by replacing the compact colour states with non-compact states with vanishing, regularized inner products and truncating the colour gauge field. 

We first observe that the vacuum-anti-vacuum divergence could alternatively be cured by a regularization procedure based on defining all conformal singleton module endomorphisms via an integral presentation \cite{2011,2017,COMST,corfu19}; but such regularization implies orthogonality between states with different $D$-eigenvalue, which would imply $C^{(1)}\star \overline C^{(1)}$ vanishes, i.e., a trivialization of the generating function of conformal currents. 

On the other hand, when building the source term $\overline C^{(1)}\star C^{(1)}\star I_{\Real}$ for $V$, it is the 3D scalars $c^{(1)I}$ and  $\bar c^{(1)}_I$ that avoid the vacuum-anti-vacuum clash, as shown explicitly  in \eq{calTd} by means of their expansion in momentum eigenstate basis. It is interesting to note that, in principle, one could consider expanding the 3D scalar on the Lorentz-tensorial, conformal singleton states \eq{confsing} and realize the latter with a regularized integral presentation \cite{2011,2017,corfu19}, while still using the intertwining Ansatz \eq{expC}-\eq{expCb}. Such a choice would imply that $\overline C^{(1)}\star C^{(1)}=0$, since, as shown in \eq{calTd}, $\overline C^{(1)}\star C^{(1)}$ contains the star product 
\bea & \ket{(i/2)}\bra{(i/2)}\star \bar c_I^{(1)}\star c^{(1)J} \star \ket{(-i/2)}\bra{(-i/2)} &\nonumber \\
&\displaystyle \sim \sum_{m,n\geqslant 0} c_{mn}\ket{(i/2)}\bra{(i(m+1/2))} \star \ket{(-i(n+1/2))}\bra{(-i/2)}& \eea
which, upon regularization, vanishes by orthogonality; while $C^{(1)}\star \overline C^{(1)}\neq 0$. This means that it would be possible to gauge away $V^{(2)}$ while still retaining non-trivial currents. One can check perturbatively that this extends to the next order, as the source for $V^{(3)}$  only features $\overline C^{(2)}\star C^{(1)}$ and $\overline C^{(1)}\star C^{(2)}$, and $C^{(2)}$ can still be solved in terms of an intertwining Ansatz of type \eq{expC}, as at second order \eq{fullCHSG2} still keeps the form $dC^{(2)}+\O\star C^{(2)}=0$. However, at the next perturbative order this state of affairs is broken by the appearance of $W^{(2)}$ deforming Eq. \eq{fullCHSG2} for $C^{(3)}$. Bringing in $z^-$-dependence, this implies that the solution for $C^{(3)}$ will not in general be of the form \eq{expC}, thereby not guaranteeing that $ \bar C\star C$ vanishes at fourth order. This in turn implies that $V^{(n)}$ for $n\geqslant 4$ cannot be gauged away within the ``standard'' perturbative scheme above used, not even by making use of the regular presentation for conformal states. On the other hand, if on a given exact solution  $\bar C\star C=0$ to all orders, then it becomes possible to consistently truncate the CCHSG model by setting $V=0$ while still retaining a non-trivial coupling to conformal scalar matter via the currents. Indeed, from \eq{fullCHSG1}-\eq{fullCHSG3}, it is evident that if $ \bar C\star C=0$ then $V$ can be gauged away, leaving 
\bea &  dW+W\star W + C\star \bar C \star I_{\Real} = 0\ , & \\
& dC+W\star C =0 \ , & \\
& d\bar C-\bar C\star W =0 \ , & \eea
which is now a consistent system due to the fact that the only term remaining in the consistency condition of \eq{fullCHSG2} is
\be dW\star C-W\star dC = -W\star W\star C-C\star \bar C\star C\star I_\Real +W\star W\star C  = 0\ee
where the last equality is due to $\bar C\star C=0$~.

There are however reasons why the regular prescription is not suitable to represent non-compact states. One is that the divergence appearing in the star product $\ket{(i/2)}\bra{(i/2)}\star \ket{(-i/2)}\bra{(- i/2)}$ is related to the divergence of the boundary two-point function at colliding points \cite{Didenko:2012tv}. In order to preserve this property, in this paper we have not used any regularization. Star products of $C$ and $\overline C$, which drive the perturbative expansion, are instead tamed by their expansion in hybrid bimodules of the fractional-spin algebra, thereby in particular letting colour states help to obtain finite sources. The presence of colour states then brings in the additional gauge field $V$, as explained above.

\paragraph{Summary.}
Thus, the fractional spin algebra expansion and the related coupling with the colour gauge field $V$ is crucial in order to write a fully non-linear system of 3D CHSG coupled to conformal scalar in closed form. Our result can thus be thought of as a non-linear completion of the systems proposed in \cite{Misha} and \cite{Nilsson} (which did not make use of colour states nor internal gauge fields). Moreover, being reductions of the same parent theory, 4D HSG and 3D CCHSG are candidate holographic dual theories (confirming the conjecture of \cite{Misha}) within our proposed AKSZ approach --- a conjecture that we plan to test via overlap conditions in a future paper \cite{OLC}. 

\subsection{Manifest $U(N,N)$-covariant formulation and non-abelian anyons}
 
The reduced system \eq{fullCHSG1}--\eq{fullCHSG3} admits a set of boundary conditions, viz., 
\begin{align}
C=C^{\hat I}\star |(-i/2)\rangle\langle e_{\hat I}|\ ,\qquad V=V_{\hat I}{}^{\hat J}\star |e^{\hat I}\rangle\langle e_{\hat J}|\ ,    \qquad 
[y_\alpha,V_{\hat I}{}^{\hat J}]_\star=0=[\bar y_{\dot\alpha},V_{\hat I}{}^{\hat J}]_\star\ ,
\end{align}
which differ from those used on the CCHSG defect, and that leads to a manifestly $U(N,N)$-covariant system, viz., 
\begin{align}
dW+W\star W+ C^{\hat I}\star |(-i/2)\rangle\langle (i/2)|\star I^\xi_\Real\star \overline{C}_{\hat I}={}&0\ ,\label{4.30}\\
(dC^{\hat I}-C^{\hat J}\star V_{\hat J}{}^{\hat I}+W\star C^{\hat I})\star |(-i/2)\rangle={}&0\ ,\\
dV_{\hat I}{}^{\hat J}+V_{\hat I}{}^{\hat K}\star V_{\hat K}{}^{\hat J}+ {\cal M}_{\hat I}{}^{\hat J} I^\xi_\Real={}0 &\ ,
\end{align}
with monodromy matrix
\begin{align}
{\cal M}_{\hat I}{}^{\hat J}:=\langle(i/2)| \star \overline{C}_{\hat I}\star C^{\hat J}\star   |(-i/2)\rangle\ ,
\end{align}
which describes deformations of a family of wave functions  $C^{\hat I}$ and non-abelian connections $V_{\hat I}{}^{\hat J}$ on $S^2_\xi$ as the system is evolved along $\boldsymbol{X}$ by a set of Hamiltonians contained in $W$, which are themselves subject to back-reactions via \eq{4.30}; for a similar interpretation, see \cite{Misha}. We leave this system and in particular its second-quantization within the AKSZ framework, and possible connection to non-abelian anyons \cite{Moore:1991ks,Valenzuela:2016joe}, for future studies.

\section{CCHSG on ${\rm Mink}_3\times \Real$ vs HSG on the Poincar\'e patch}\label{App:D}

In this Section, we compare the second-order expansion of the foliations of CCHSG on ${\rm Mink}_3\times \Real$, studied in Section \ref{sec:rhocurrents}, and the linearized HSG equations on the Poincar\'e patch of $AdS_4$ studied in \cite{Misha}. To this end, we begin in Section \ref{App:D1} by reviewing the results of \cite{Misha} in the bosonic HSG setup, spelled out with our conventions, and by discussing their interpretation. Then, in Section \ref{App:D2} we shall highlight the main differences between our approach and that of \cite{Misha}, discuss them and directly compare the embedding of conformal currents in the two systems.

\subsection{Vasiliev's approach}\label{App:D1}

First of all, both ${\rm Mink}_3\times \Real$ and $AdS_4$ Poincar\'e patch share the same topology, and can be equipped with the same $\mathfrak{so}(2,3)$-valued background connection $\Omega$.
Using Poincar\'e coordinates $x^a=(x^m,\rho)$ in which the 4D metric and Vierbein reads
\begin{align}
   ds_4^2=\frac{dx^2+d\rho^2}{\rho^2}\ ,\qquad e^a = \frac{dx^a}{\rho}\ , 
\end{align}
and the 3D dittos read
\begin{align}
    ds_3^2=dx^2\ ,\qquad e^m=dx^m\ ,
\end{align}
the background $\Omega$ splits in conformal basis as\footnote{In \cite{Misha}, the convention used is $T_{\a\b}=-\frac12 {\rm y}^{-}_\a {\rm y}^-_\b$, where ${\rm y}^\pm_\a=\frac{e^{-i\pi/4}}{\sqrt{2}}y^\pm_\a$, and, consequently, $\exp(\pm i y^+ y^-) = \exp(\mp 2{\rm y}^+{\rm y}^- ) $ and the 3D Poincar\'e-invariant vacuum projector is $4 e^{i y^+y^-}=4 e^{-2 {\rm y}^+{\rm y}^- }$ .} 
\be \O =  i \,\frac{dx^m}{\rho}\,T_m -i\, \frac{d\rho}{\r}\,D   = -\frac{i}{4\r} \left(dx^{\a\b}y^+_\a y^+_\b + d\r\,y^+y^-\right)\ .\label{D1}\ee
Decomposing instead $\Omega$ concerning the direct-product structure of the foliation, one has
\begin{align}
    \Omega=\frac{1}{\rho}e+\O_\rho\ ,\label{d4}
\end{align}
where $e$ is the Dreibein on ${\rm Mink}_3$. 
Comparing with \eq{Omega}, we find that the Vielbein takes the following manifestly $SL(2,\Real)\times SO(1,1)$-covariant form
\be  e^{\a\bd} =\frac{dx^{\a\bd}}{2\r} = \frac{1}{2\r}\left(dx^{\a\b}\e_\b{}^{\bd} - i \e^{\a\bd}d\r\right) \ ,\qquad \e_{\a\bd}:= i(\s_2)_{\a\bd}\ ,\label{vb} \ee
using \eq{gammasigma} to split the 4D coordinates as
\be x^{\a\bd}=  x^a(\s_a)^{\a\bd}= x^m(\gamma_m)^{\a\b}\e_\b{}^{\bd}+x^2(\s_2)^{\a\bd} = x^{\a\b}\e_\b{}^{\bd}-i\r \,\e^{\a\bd} \ .\ee
Decomposing also $d=d_x + d_\r$, where $d_x=dx^m \partial_{m}$ and $d_\rho=d\r \,\partial_\r$, the linearized equation for the twisted-adjoint zero-form on the HSG branch \eq{COMSTHSG2}, viz.,
\be d\Phi + [\O,\Phi]_{\pi} = 0\ , \label{twadj}\ee
splits into 
\bea d_x\Phi + \frac{1}{\r}[e,\Phi]_{\pi} &=& 0 \label{D3} \\
d_\r\Phi + [\O_{\r},\Phi]_{\pi} &=& 0 \ , \label{D4}\eea
where, for notational simplicity, as we remain at linearized level, in this Section we shall identify $\Phi\equiv \check \Phi^{(1)}$, and  $[f,\Phi]_\pi:= f\star \Phi - \Phi\star \pi^\ast_y(f)$. 
Following \cite{Misha}, the twisted-adjoint $\mathfrak{so}(2,3)$-module is sliced into conformal tensors, i.e., into 3D Lorentz tensors with definite $\mso(1,1)_D$-eigenvalues, by introducing the projector $4 e^{-iy^+y^-}=|(-i/2)\rangle\langle (-i/2)|$ on the $\mathfrak{iso}(1,2)$-invariant vacuum state of the first-quantized theory (which obeys $y^+_\a\star e^{-iy^+y^-} =0= e^{-iy^+y^-}\star y^-_\a$) via the change of variables
\be \Phi = \rho \,e^{-iy^+y^-} J\ . \label{D8bis}\ee
Indeed, expanding $J$ in a formal power series in $(y^+,y^-)$, the right-hand side of Eq. \eqref{D8bis} can be rearranged into an object of the form $\phi(y^-)\star |(-i/2)\rangle\langle (-i/2)| \star \bar\phi(y^+)$ where $\phi$ and $\bar\phi$ belong to conformal $\mathfrak{so}(2,3)$-modules for unfolded conformal scalar fields, as will be spelled out in more detail in the next Subsection (see also \cite{2011,COMST,BTZ,corfu19,Iazeolla:2022dal} for analogous examples in compact and non-compact bases).
Substituting \eq{D8bis} into Eq. \eq{D3} yields
\be \left[d_{x}+\frac{i}{4\r}\,dx^{\a\b} \left(\frac{\partial^2}{\partial y^{+\a}\partial y^{+\b}}+ \frac{\partial^2}{\partial y^{-\a}\partial y^{-\b}}\right)\right]J=0 \ .\label{d9a} \ee
Introducing new twistor-space coordinates
\begin{align}
q_\a:=\r^{1/2} y_\a\ ,\qquad \bar q_\a = \r^{1/2} \bar y_a\ ,\qquad \bar y_\a := \e_\a{}^{\ad}\yb_{\ad}= i(\sigma_2 \bar y)_\a\ ,\label{D8}
\end{align}
which are thus complex $SL(2,\Real)$-doublets obeying  $(w_\alpha)^\dagger=\bar w_\alpha$ and $y\yb=-iy^+y^-$, Eq. \eqref{d9a} takes the form
\be \left[d_{x}+\frac{i}{2}\,dx^{\a\b} \frac{\partial^2}{\partial q^{\a}\partial \bar q^{\b}}\right]J = 0 \ .\label{D9} \ee
i.e., the unfolded equations for 3D conserved currents \cite{Mishasiegel,Misha} comprising the conservation laws for primaries and a chain of identifications of descendants, while Eq. \eq{D4} reads 
\be \left[d_{\r}+\frac{1}{2}\,d\r \frac{\partial^2}{\partial q^{\a}\partial \bar q_{\a}}\right]J = 0  \ ,\label{rhoJ} \ee
which relates $\r$-dependence and contractions among boundary spinor indices in $J$.  

As for the linearized adjoint one-form, again simplifying the notation as $A\equiv \check A^{(1)}$ on the HSG branch, we decompose  
\begin{align}
    A=A_x+A_\rho\ ,
\end{align}
and, correspondingly, its equations of motion \eq{COMSTHSG1} decompose as
\begin{align}
    D_x A_x + \S^A(e,e,\Phi)_{xx}=0\ ,\label{d18}\\
    D_\rho A_x+D_x A_\rho+\S^A(e,e,\Phi)_{x\rho}=0\ ,\label{d18b}
\end{align}
where the cocycle $\S^A(e,e,\Phi)$ is given in Eq. \eq{ChEil}
In view of \eq{d4} and \eq{D8}, the $xx$-component of the cocycle read
\bea \S^A(e,e,\Phi)_{xx}  &\sim & \frac{1}{\r^2}(dx^2)^{\a\b}\left(b \frac{\partial^2}{\partial \bar y^\a  \bar y^\b}\Phi(0,\yb,x,\r)+\bar b \frac{\partial^2}{\partial y^\a  y^\b}\Phi(y,0,x,\r)\right) \nonumber\\
& = & (dx^2)^{\a\b}\left(b \frac{\partial^2}{\partial \bar q^\a  \bar q^\b}J(0,\bar q,x,\r)+\bar b \frac{\partial^2}{\partial q^\a  q^\b}J(q,0,x,\r)\right) \ ,\label{D11}
\eea
that is, $A_x$ is sourced by the primaries in $J$. 
Introducing the rescaled, real $SL(2,\Real)$-doublets
\begin{align}
   q^\pm_\alpha := \r^{1/2}y_\alpha^\pm\ ,\qquad p_\alpha^\pm := \r^{-1/2}y_\alpha^\pm\ ,
\end{align}
in terms of which the complex dittos can be written as 
\be q_\a= \frac{e^{i\pi/4}}{\sqrt{2}}\,(q^-_\a - i\r\,p^+_\a) \ , \qquad \bar q_\a= \frac{e^{-i\pi/4}}{\sqrt{2}}\,(q^-_\a + i\r\,p^+_\a)  \ ,\label{d22}\ee
Eq. \eq{d18} reads
\bea  \hspace{-1cm} D_x A_x(q^-,p^+,x,\r)  &\sim  &-\frac{i}{2}\,(dx^2)^{\a\b}\frac{\partial^2}{\partial q^{-\a}  q^{-\b}}\left[-b J\left(0,\frac{e^{-i\pi/4}}{\sqrt{2}}\,(q^- + i\r\,p^+),x,\r\right)\right.\nonumber \\
&& \hspace{3cm} + \ \ \left.\bar b J\left(\frac{e^{i\pi/4}}{\sqrt{2}}\,(q^- - i\r\,p^+),0,x,\r\right)\right]  \ ,\label{D12}
\eea
treating $(q^-_\alpha,p^+_\alpha)$ as the new independent variables.
Keeping these variables fixed in the limit $\r\to 0$, yields (modulo an irrelevant rescaling)
\bea D_x A_x(q^-,p^+,x,0) & \sim & (dx^2)^{\a\b}\frac{\partial^2}{\partial q^{-\a}  q^{-\b}}{\cal T}_{(-)}(q^-,x) \ ,\label{D13}
\eea
where 
\be {\cal T}_{(-)}(q^-,x) := \bar b J(e^{i\pi/4}q^-,0,x,0) -b J(0,e^{-i\pi/4}q^-,x,0) \ee
is a generating function of primary currents. 
Thus, in the Type A ($b=1$) and B ($b=i$) models, every fixed spin-$s$ gauge field curvature is sourced by the combinations $i^{-s}{\cal T}_{(-)\a(2s)}=J_{\a(2s)}+(-1)^{s+1}\bar J_{\a(2s)}$ and $i^{1-s}{\cal T}_{(-)\a(2s)}=J_{\a(2s)}+(-1)^{s}\bar J_{\a(2s)}$, respectively. Extending to primary currents the terminology used for 4D Weyl tensor components\footnote{In analogy with the standard analysis of the spin-$1$ and spin-$2$ cases (see, e.g., \cite{PR,Stephani}), a generalized spin-$s$ primary Weyl tensor $C^{a(s),b(s)}$, $s\geqslant 1$, with spinor components $C_{\a(2s)}$ and $\bar C_{\ad(2s)}$ is said to be purely electric if the invariant $K:=C^{\a(2s)}C_{\a(2s)}$ is such that ${\rm sign}(K)=(-1)^{s}$, and purely magnetic if ${\rm sign}(K)=(-1)^{s+1}$, the alternating sign being related to the number of timelike indices, changing with the spin $s$, raised or lowered in forming Lorentz invariants.}, this conclusion can be summarized by saying that in the Type-A model the gauge fields are sourced by magnetic components of the boundary current and in the Type-B model by electric ones. As a consequence, the Type A and B models admit boundary conditions at $\r=0$ such that ${\cal T}_{(-)}$ vanishes --- thus decoupling, at linearized level, the 3D gauge fields from the matter currents at the boundary --- without implying that the whole $J$ vanishes\footnote{As anticipated in the preamble to this Section, the analysis of  \cite{Misha} actually considers a HSG model with integer and half-integer Lorentz spins, which implies that $\Phi$ consists of four sectors, $\Phi^{ij}$, $i,j=\{0,1\}$, of which $\Phi^{01}$ and $\Phi^{10}$ contain propagating 4D fields. 
The resulting source in the COMST is 
$$ {\cal T}^{ii}(q^-, x,) := \bar b J^{i,1-i}(e^{i\pi/4}q^-,0,x,0) -b J^{1-i,i}(0,e^{-i\pi/4}q^-,x,0) \ , $$
which for $b=1$ or $b=i$ can be set to zero by choosing appropriate boundary conditions at $\r\to 0$. Upon projecting to the bosonic model, one is left with two independent bosonic sectors with Weyl zero forms, respectively, $\Phi_{\pm} = \frac12(\Phi^{01}\pm \Phi^{10})$. Thus, the current $J$ that was introduced in our bosonic setup corresponds to one of the combinations $J_{\pm} = \frac12(J^{01}\pm J^{10})$.}: indeed, imposing
\be {\cal T}_{(-)}=0 \ , \qquad {\rm for } \ \ b=\{0,i\}\ , \label{decoupling} \ee
leaves electric components of the currents in $J$ free to fluctuate in the Type A model, and magnetic components in the Type B, corresponding to the components arising from scalar and spinor singletons squared respectively \cite{fibre}, viz.,
\begin{align}
    \mbox{Type A:}\ &\qquad \left.J\right|_{{\cal T}_{(-)}=0}=\mbox{electric} \ ,\\
    \mbox{Type B:}\ &\qquad \left.J\right|_{{\cal T}_{(-)}=0}=\mbox{magnetic}\ .
\end{align}
The mixed $x\rho$ component \eq{d18b} instead controls the $\rho$-evolution of the gauge fields, and, together with \eq{rhoJ}, should be interpreted as a holographic renormalization group equation. Indeed, performing the changes of variables in Eqs. \eq{D8bis}, \eq{D8} and \eq{d22} yields 
\begin{align}\label{holren}
\S^A(e,e,\Phi)_{x\r} \sim{}&  dx^{\a\b} d\r \left(b \frac{\partial^2}{\partial \bar w^\a  \bar w^\b}J(0,\bar w,x,\r)-\bar b \frac{\partial^2}{\partial w^\a  w^\b}J(w,0,x,\r)\right)\nonumber \\
={}& \frac{i}2\, dx^{\a\b} d\r \frac{\partial^2}{\partial  q^{-\a}  q^{-\b}}\left[b J\left(0,\frac{e^{-i\pi/4}}{\sqrt{2}}\,(q^- + i\r\,p^+),x,\r\right)\right. \nonumber  \\
{}&+ \left.\bar b J\left(\frac{e^{i\pi/4}}{\sqrt{2}}\,(q^- - i\r\,p^+),0,x,\r\right)\right] \ ,& 
\end{align}
%
%
where the relative sign between the two terms on the r.h.s. is because the $\r$-component of the vielbein \eq{vb} is aligned with the dilation operator (in the Poincar\'e patch) which produces an imaginary component in spinor basis.  As a consequence, the source term in \eq{holren} contains the components of $J$ of opposite electric/magnetic nature than those of \eq{D12}. This means that imposing the boundary conditions  \eq{decoupling} sets to zero the source in \eq{D12} while leaving the source in \eq{holren} non-vanishing and free to drive the $\rho$-evolution of the gauge fields.  Thus, with boundary conditions \eq{decoupling}, the Type A and B foliations can be interpreted as holographic RG flows in which conformal scalar and spinor fields source the $\r$-evolution of 3D CHSG fields. 

On the other hand, in the Type A and B models one can also set boundary conditions
\be {\cal T}_{(+)}=0 \ , \qquad {\rm for } \ \ b=\{0,i\}\ , \label{zeroT+}\ee
where
\be {\cal T}_{(+)}(q^-,x) := \bar b J(e^{i\pi/4}q^-,0,x,0) +b J(0,e^{-i\pi/4}q^-,x,0) \ ,\ee
by which the source of the equations governing the scaling with $\r$ of 3D gauge fields vanishes, leaving currents of type 
\begin{align}
    \mbox{Type A:}\ &\qquad \left.J\right|_{{\cal T}_{(+)}=0}=\mbox{magnetic} \ ,\\
    \mbox{Type B:}\ &\qquad \left.J\right|_{{\cal T}_{(-)}=0}=\mbox{electric}\ .
\end{align}
free to couple to the 3D gauge fields. With this type of boundary conditions, the resulting system describes a matter-coupled CHSG at a UV holographic renormalization group fixed point. 

Finally, it is important to recall that the definition of the Type A and B models, which concerns parity-invariance \cite{Sezgin:2003pt}, goes beyond linearized level, as $b=\{1,i\}$  allow for parity-truncated HSG at full non-linear level. In \cite{Misha}, the fact that the decoupling condition \eq{decoupling} can be imposed at linearized level only for the Type A and B models was taken a rational for that, in such models, the decoupling extends to fully non-linear level, i.e., that the corresponding boundary conditions select a free boundary dual of HSG; for a verification of this conjecture to second order, see \cite{Didenko:2017lsn}.

\subsection{Comparison with our results}\label{App:D2}

In comparing the results obtained in Section \ref{sec:CHSG} of the paper with those of \cite{Misha}, first of all, a few general observations are in order:

\begin{itemize}

\item As we have just revisited, the results of \cite{Misha} can be seen as a bottom-up unfolded approach to the holographic duality, which focuses on reinterpretating the bulk linearized HSG equations in terms of a 3D theory of conformal currents sourcing topological gauge fields of all spins on every leaf at constant $\r$. Consistency of the equations beyond the linearized approximation implies that the 3D theory becomes non-linear, with all-order couplings of the currents to the conformal HS gauge fields and a deformation of the current conservation condition.  However, the concrete construction of the fully non-linear CHSG model is not attempted. Moreover, as Vasiliev's construction moves from the 4D bulk, the currents appear as primitive dynamical objects, and there is no notion of their composite structure in terms of 3D conformal scalar (and spinor) fields (although, as recalled in \cite{Misha}, this conclusion can be inferred from the study of the relation between rank-two and rank-one unfolded equations formulated in a generalized spacetime extendend with additional coordinates, so that the conformal group acts geometrically \cite{Misharank}).  

\item  On the other hand, we employ a top-down approach to the holographic correspondence, whereby 4D HSG and 3D CCHSG both result from  two inequivalent consistent reductions of one non-linear parent theory. The two reductions are singled out by two-form expectation values, respectively $I_{\Comp}$ and $I_{\Real}$, captured by second Chern classes, and characterized by related unbroken structure groups. This way we land directly on a fully non-linear CCHSG model of coloured conformal scalar fields interacting with topological fields of all spins via sesquilinear currents. Our construction realizes explicitly the 4D current master fields in terms of the star product of the zero-form master field (containing the 3D conformal scalar master field) and its complex conjugate, in a manifest Flato-Fronsdal-like construction. The colour states also help softening star products appearing at second and higher orders in perturbation theory without invoking regularization prescriptions which would be difficult to incorporate in a quantum theory. The presence of this internal sector in turn involves the addition of a colour gauge field to the whole system, sourced by the conformal matter fields. 
As explained in the Introduction, 4D HSG and such 3D CCHSG are candidate holographic dual to each other, and we shall devote a future paper \cite{OLC} to checking this conjecture via overlap conditions spelled out in terms of invariant functionals of the parent theory.

\end{itemize}


The colour-singlet sector of our CCSHG model, expanded at second order, corresponds to a CHSG coupled to colourless currents, and can thus be compared with the linearized model of \cite{Misha}: 

\begin{itemize}

\item First of all, at the first non-trivial order our CCHSG system is defined by Eqs. \eq{COMST1fin}, \eq{COMST2fin} and the linearized unfolded equation for the (complex) scalar field \eq{C1xR}, to be compared with Eqs. \eq{d18}, \eq{d18b} (accompanied by Eqs. \eq{D11} and \eq{holren} defining their r.h.s.) and the current equations \eq{D9}-\eq{rhoJ}.

\item The latter two equations are effectively incorporated in our system as well, since they are implied, as usual, by the linearized unfolded equation for the complex scalar field, given the composite nature \eq{calJrho} of the current master field ${\cal J} \sim C \star \overline C \star \d^2_{\Comp}(z^-)$. We shall examine this issue in greater detail in the remainder of this Section.

\item Eqs. \eq{COMST1fin} and \eq{d18} (with \eq{D11}) both couple topological gauge fields of all spins with conserved currents. However, there is a crucial difference introduced by the different two-form source of the two models: $I_{\Comp}$ introduces deformations along both holomorphic and anti-holomorphic sectors, resulting in the two terms of \eq{D11} whereas $I_{\Real}$ only deforms along $z^-$ and gives rise to the single terms on the r.h.s. of \eq{COMST1fin}. As recalled in Section \ref{App:D1}, the decoupling of the 3D currents from the topological gauge fields occurs specifically from the interplay of the holomorphic and anti-holomorphic term in $I_{\Comp}$.  In other words, the matter-coupled CHSG of \cite{Misha}, due to its 4D HSG origin, starts life with twice as many primary currents ($\Phi|_{y=0}$ and $\Phi|_{\yb=0}$) as our model, and can thus set to zero certain combinations and retain others. Our CCHSG system is instead always non-linear, and it is not obvious to generalize our Ansatz as to accommodate the conjecture that the dual system to Type A and Type B HSG models is free. For the same reason, our CCHSG model cannot be truncated according to parity. 

\item Moreover, \eq{COMST2fin} and \eq{d18b}-\eq{holren} both describe the behaviour of 3D gauge fields with the radial coordinate/foliation parameter $\r$. However, again the difference of the two-forms  $I_{\Real}$ and $I_{\Comp}$ triggers an important deviation: in our case the scaling of the gauge fields is independent of the matter currents, which also implies that the second-order, one-form connection along $\r$ can be gauged away; while in \eq{d18b} the source term \eq{holren} can only be set to zero for special boundary conditions \eq{zeroT+} in the Type A and in the Type B model. Interestingly, choosing a different homotopy contraction (i.e., not the simplest one along $\vec E_-$) for the integration of our CCHSG system and a different cohomology projection on $\boldsymbol{Z}^{(\Real)}_4$ (i.e., choosing a physical surface different from $z^-=0$), may allow us to keep a non-trivial current source driving the $\r$-evolution of the gauge fields. We hope to return to this issue, and in general to the meaning of different generalized gauges and variable choices in the context of CCHSG and of the holographic duality, in a future work.  

\item Boundary conditions \eq{decoupling} decouple, at linear level, matter and CHSG fields and correspond, from the bulk point of view to Neumann boundary conditions on the gauge fields (as they set the tangential curvature components to zero leave the normal derivatives free to fluctuate). On the other hand, boundary conditions \eq{zeroT+}, leading to a matter coupled CHSG at a renormalization group fixed point, give rise to a system which essentially agrees with \eq{COMST1fin}-\eq{COMST2fin}, and are compatible with inhomogeneous Dirichlet boundary conditions on the CHSG fields. 

\end{itemize}

Given the differences summarized above, it is interesting to compare the two approaches in greater detail to find out, in the remainder of this Section, how a generating function of conformal currents is obtained from the pullback of the 4D HSG to a 3D leaf and from our reduction Ansatz.     

Thus, let us now aim at explicitly connecting the two procedures by building a 4D bulk Weyl zero-form $\Phi$ starting from a colour-singlet, sesquilinear construct of the 3D conformal scalar $c^I$. For notational simplicity, here and in the following of this Section we omit the labels denoting the perturbative order, with the understanding that $C$, $c^I$ and their complex conjugates are all computed at first order. As a consequence, the perturbative expansion of the bulk field $\Phi$ as well as of $A$ over the background $A^{(0)}=\O$, $\Phi^{(0)}=0$ will be understood to start from second order. To complement somehow the treatment offered in Sections \ref{sec:currents}-\ref{sec:rhocurrents}, in this Section we shall build all quantities by means of local data and background gauge functions. 

The gauge function of the Poincar\'e patch is\footnote{$\exp_\star A$ denotes a star-power expansion $\exp_\star A=1+A+\ft12 A\star A+...$ .}\cite{GiombiYin2,Didenko:2012tv,corfu19}
\be L^{} = L_{x} \star L_\r = \exp_\star (ix^mT_m) \star \exp_\star(-i \ln \r\, D)\ , \label{D19} \ee
such that 
\be L^{-1}\star dL^{} = \O = L^{-1}_\r\star e \star L_\r + L^{-1}_\r\star dL_\r \ ,\ee
 where $\O$ is given in \eq{D1}. The factorization of $L^{}$ parallels the decomposition of $\O$ into the 3D Minkowski connection $\O_{x}$ plus the radial part \eq{D1}. Notice in particular that setting $\r=1$ reduces $L^{}$ to $L_{x}$, just like it reduces $\O$ to $\O_{x}$. 
 
Working with $U(N)$-coloured scalar fields, the basic building blocks are thus $L^{}$ and the local data $c^{I\prime}(y^-)$ for a free conformal scalar field on a 3D Minkowski leaf. Having chosen the boundary translation generator to be the $D$-raising operator, a conformal scalar can be obtained from the fibre element
\be C'(y^-) =  c^{I\prime}(y^-)\star\ket{(-i/2)}\bra{e^+_I} \ , \ee
where $c^{I\prime}$ builds on the Poincar\'e invariant vacuum $\ket{(-i/2)}$ and is expanded as in \eq{phider},
\bea c^{I\prime}(y^-)  &=& \sum_{n=0}^\infty \frac{1}{(2n)!}\,\phi^{\prime I,\a(2n)}\,y^-_{\a_1}\ldots y^-_{\a_{2n}} \nonumber\\ &=&  \sum_{n=0}^\infty \frac{(-1)^n}{(2n)!}\,\phi^{\prime I ,\a(2n)}\left.\left(\frac{\partial^2}{\partial u^{+}\partial u^{+}}\right)_{\a(2n)}^n \exp(iu^+ y^-)\right|_{u^+=0} \label{phiprime}\eea
(likewise for the complex conjugate $\overline c\,'$) with constant coefficients $\phi^{\prime I,\a(2n)}$. ${\rm Mink}_3\times\Real$ spacetime dependence is introduced via a left star product with the gauge function $L^{-1}$. The relevant star-product lemmas are
\bea  L^{-1}\star \ket{(-i/2)}\bra{(-i/2)} &=& L_\r^{-1}\star \ket{(-i/2)}\bra{(-i/2)} =  \sqrt{\r} \ket{(-i/2)}\bra{(-i/2)} \nonumber \\
&=& \ket{(-i/2)}\bra{(-i/2)} \star L_\r^{-1} = \ket{(-i/2)}\bra{(-i/2)} \star \pi^\ast(L) \label{L1}\\
L^{-1} \star y^+_\a \star L^{} &= & \frac{1}{\sqrt{\r}}\, y^+_\a =: p^+_\a  \ ,\label{L2} \\
L^{-1} \star y^-_\a \star L^{} &= & \sqrt{\r}\,y^-_\a + \frac{x^m}{\sqrt{\r}}(\gamma_m)_\a{}^\b \, y^+_\b =: q^-_\a + (x p^+)_\a  \label{L3}\eea
(from which also the action of $L_x$ alone can be obtained by setting $\r=1$ and that of $L_\r$ alone by setting $x^m=0$), where $x=x_\a{}^\b=x^m (\gamma_m)_\a{}^\b$. Thus, we obtain the 3D zero-form master field\footnote{Let the vacuum gauge function $L^{}$ be an $Mp(4;\Comp)$ group element and $Y^{L\underline\alpha}:=L^{-1}\star Y^{\underline\alpha}\star L^{}$; since  both $Y^{\underline\alpha}$ and $Y^{L\underline\alpha}$ are canonical, it follows that if  $f(Y^{\underline{\alpha}})$ is a Weyl ordering symbol, then $L^{-1}\star f\star L^{}= f(Y^{L\underline{\a}})$.}
\bea C &=& L^{-1}\star C' = L^{-1}\star c^{\prime I}(y^-) \star L^{} \star L^{-1} \star\ket{(-i/2)}\bra{e^+_I} \nonumber\\
&=& \sqrt{\r}\,c^{\prime I}(q^-+xp^+)\star \ket{(-i/2)}\bra{e^+_I}  \ ,  \label{D26a} \eea
where
\be c^{I\prime}(q^-+xp^+)  =  \sum_{n=0}^\infty \frac{(-1)^n}{(2n)!}\,\phi^{\prime I ,\a(2n)}\left.\left(\frac{\partial^2}{\partial u^{+}\partial u^{+}}\right)_{\a(2n)}^n \exp(iu^+ (q^-+xp^+))\right|_{u^+=0}\ . \label{phiprimerot} \ee
Eq. \eq{D26a} is by construction a solution to the 3D zero-form equation \eq{C1xR} (imposing the Klein-Gordon equation on $\phi(x)$). It should therefore also admit the form  
\be C = \sqrt{\r}\  c^{\prime I}(y^-+xy^+)\star \ket{(-i/2)}\bra{e^+_I} = \sqrt{\r}\  c^I(x,q^-)\star \ket{(-i/2)}\bra{e^+_I} \ ,  \label{D26b} \ee
used in \eq{expCbis}-\eq{phiderw}, with $c^I(x,q^-)$ given by
\be c^{I}(x,q^-) = \sum_{n=0}^\infty \frac{(-1)^n}{(2n)!}\,\phi^{I ,\a(2n)}(x)\left.\left(\frac{\partial^2}{\partial u^{+}\partial u^{+}}\right)_{\a(2n)}^n \exp(iu^+ q^-)\right|_{u^+=0}  \ .\label{phider2}\ee
The relation between the two expressions can be obtained by taking the star products of both \eq{phiprimerot} and \eq{phider2} with the vacuum state $\ket{(-i/2)}\bra{(-i/2)} = 4 \exp(-iy^+ y^-)$ using
\bea & \exp(iu^+(q^-+xp^+))\star \ket{(-i/2)}\bra{(-i/2)} = 4  \exp(iu^+(q^-+xp^+))\star \exp(-iy^+ y^-) & \nonumber \\
& = 4 \exp(-iy^+y^-+2iu^+ q^-+ i u^+xu^+) &\label{D27}\eea
(while the colour bra in \eq{D26b} is a mere spectator). By comparing the two resulting expansions one deduces the Taylor-like expansions
\bea \phi^{I,\a(2k)}(x) &=& \left.\sum_{n=k}^\infty \frac{(-1)^{n-k}}{(2n-2k)!}\,\phi^{\prime\,I,\a(2n)} (\partial_{u^+}^2)^{n-k}_{\a(2n-2k)} e^{iu^+x u^+}\right|_{u^+=0} \nonumber\\
&=& \sum_{n=k}^\infty \frac{(2i)^{n-k}}{(2n-2k)!!}\,\phi^{\prime\,I,\a(2n)}\,x^{n-k}_{\a(2n-2k)} \ , \label{D30}\eea
consistently, in particular, with $\phi^{\prime\,I,\a(2k)}=\phi^{I,\a(2k)}(0)$, as it should be since $C=L^{-1}\star C'$. 

By construction, the fields in $C$ ($\overline C$) form a left (right) $\mathfrak{so}(1,2)_{M_{mn}}\oplus\mso(1,1)_D$-module. It is thus natural to consider $C'\star \overline C'$ as a candidate local datum for the 3D conformal current module, sesquilinear in the (complex) scalar field.
Indeed, as we have studied in Section \ref{sec:CHSG}, in our approach 3D conserved currents are extracted from 
\be \left. {\cal J}\right|_{z^-=0} =  \frac{1}{\r}\,\left. C\star\overline C\star \pi\delta^2_\Comp(z^-) \right|_{z^-=0} \ , \label{cJ0comp}
\ee
with the adjoint element $C\star\overline C=L^{-1}\star C'\star \overline C'\star L$
(see \eq{calJrho}, \eq{calJ0rho} and \eq{COMST1fin}). Having established the relation \eq{D26b}, the computation of \eq{cJ0comp} proceeds as in Section \ref{sec:rhocurrents}. 

On the other hand, in the approach of \cite{Misha} the current master field $J$ is extracted directly from the bulk linearized 4D HSG twisted-adjoint field $\Phi$ as in \eq{D8bis}. Thus, it is natural to guess that the local datum $\Phi'$ of the latter must contain the same core element $C'\star \overline C'$. Taking into account that $\k_y$ (defined in Weyl ordering by \eq{kappay}) maps fibre adjoint and twisted-adjoint modules into one another (see, e.g., \cite{Didenko:2009td}), one concludes that 
\be \Phi'= {\cal N}\,C'\star \overline C'\star \k_y \sim  c^{I\prime}(y^-)\star\ket{(-i/2)}\bra{(-i/2)}\star\overline c\,'_I(y^+)  \label{D20}\ee
where ${\cal N}$ is a normalization, irrelevant for our considerations. An Ansatz like \eq{D20} corresponds to a fibre element $\Phi'$ expanded over endomorphisms $\ket{-i(m+1/2)}\bra{-i(n+1/2)}$ of the conformal singleton module of Paper I, also considered in \cite{2011,BTZ,corfu19}. Thus,
\be \Phi = {\cal N}\, L^{-1}\star C'\star \overline C'\star L \star \k_y = {\cal N}\, L^{-1}\star C'\star \overline C'\star \k_y\star \pi^\ast_y(L)= {\cal N}\, C\star \overline C \star \k_y \ .\label{PhifromC}\ee
Indeed, if $C$ and $\overline C$ satisfy the 3D equation \eq{C1xR}, then it follows that $C\star\overline C\star\k_y$ satisfies \eq{D3}-\eq{D4}. In other words, in the light of the results obtained in this paper, it is possible to ``reverse'' the path followed in \cite{Misha} and summarized in Section \ref{App:D1}, and to build, in an explicit star-factorized fashion, 4D gauge field curvatures from a 3D conformal scalar. The holographic duality at linearized level is this way manifestly encoded into the construction of $\Phi$ in terms of $C$, and takes place at any fixed $\r$ as it is, in fact, $\r$-independent (the $\rho$-dependence, with the scaling regulated by \eq{D4} is, in fact, entirely carried by the gauge function $L_\r$). Indeed, by relating directly bulk fields (and conformal currents, via \eq{D8bis}) to the product of two conformal scalar modules, Eq. \eq{PhifromC} encapsulates the Flato-Fronsdal theorem \cite{FF}. 

The current master field is obtained from $\Phi$ via \eq{D8bis}, i.e.,
\be J = \frac{1}{\r}\,e^{iy^+y^-} \Phi \ , \ee
and, in particular, conserved currents are extracted from the projections 
\be \left.J\right|_{\yb=0} = \frac{1}{\r}\,\left.\Phi\right|_{\yb=0} \label{J0comp} \ee
(equivalently, $\left.\Phi\right|_{y=0}$), as can be seen from the r.h.s. \eq{D11} of \eq{d18}. 

Even once the structure \eq{PhifromC} is assumed, the two generating functions for primary currents, given in \eq{cJ0comp} and \eq{J0comp}, are proportional to each other only at the boundary, i.e.,
\be  \left. {\cal J}\right|_{z^-=0}  \propto \left.J\right|_{\yb=0, \r=0}\ ,\ee
as we shall now demonstrate by first examining ${\cal J}$ and then $J$.

Concerning ${\cal J}$, even though $ \left. {\cal J}\right|_{z^-=0} $ was already computed in Section \ref{sec:rhocurrents}, we shall repeat the computation here using the gauge function and the form \eq{D26a} for $C$. 
Using \eq{D26a}-\eq{phiprimerot}, we compute the building block $\Phi\star\k_y$ with $\Phi$ of the form \eq{PhifromC}, neglecting irrelevant normalization factors 
\be \Phi\star \k_y =  C\star\overline C = \r  \, {\cal F} \ , \label{rhocalF} \ee
with
\bea  {\cal F} := \sum_{m,n=0}^\infty \frac{(-1)^{m+n}}{(2m)!(2n)!}\,\phi^{\prime I ,\a(2n)}\bar\phi^{\prime I ,\b(2n)}\left(\partial^2_{u^+}\right)_{\a(2m)}^m\left(\partial^2_{v^+}\right)_{\b(2n)}^n F(x,\r,y^+,y^-;u^+,v^+) \ , \label{calF}\eea 
where
\bea & F(x,\r,y^+,y^-;u^+,v^+) := \exp(iu^+(q^-+xp^+))\star \ket{(-i/2)}\bra{(i/2)}\star \exp(iv^+(q^-+xp^+)) &  \nonumber \\
& \ \ \ =  8\pi \exp \left[i(u^++v^+)q^-  +iu^+ x u^+ -i v^+ x v^+  \right]\d_{\Comp}^2\left(y^+-\sqrt{\r}(u^+-v^+)\right) \ .& \label{D31}\eea
In terms of ${\cal F}$, 
\be {\cal J}={\cal F}\star \pi\d^2_{\Comp}(z^-)\ , \qquad J = e^{iy^+y^-}{\cal F}\star \k_y \ .\ee
To compute the star products with the Klein operators, we can use the lemmas
\bea & \displaystyle f(y^+,y^-)\star \d^2_{\Comp}(z^-)= e^{iy^+ z^-}\int\frac{d^2 \xi^+}{(2\pi)^2}\, e^{iz^-\xi^+} f(\xi^+,y^-) \ , &\label{fkz}\\
& \displaystyle f(y^+,y^-)\star \k_y  =  2i\int\frac{d^2 \xi^-}{2\pi} \,e^{-i(y^++iy^-)\xi^-} f(y^++i\xi^-,y^-+\xi^-) \ ,& \label{fky}\eea
the latter of which can be obtained by rewriting 
\be \k_y = 2\pi \d_{\Comp}^2(y) = 4\pi \,i\, \d^2_{\Comp}(y^++iy^-) \ , \ee
as can be easily checked recalling \eq{ypmreal} and the properties of the complex delta function. Taking the star product of \eq{D31} with $\d^2_{\Comp}(z^-)$ we obtain 
\be  F \ \star\ \pi\d^2_{\Comp}(z^-)=4 \exp\left[iy^+z^-+i(u^++v^+)q^--i\sqrt{\r}(u^+-v^+)z^- + iu^+xu^+-iv^+xv^+\right] \ . \ee
Thus, apart from constant factors, irrelevant for this discussion,  
\bea   \left. {\cal J}\right|_{z^-=0} & \propto &  \sum_{m,n=0}^\infty \frac{(-1)^{m+n}}{(2m)!(2n)!}\,\phi^{\prime I ,\a(2n)}\bar\phi^{\prime I ,\b(2n)}\left(\partial^2_{u^+}\right)_{\a(2m)}^m\left(\partial^2_{v^+}\right)_{\b(2n)}^n\nonumber \\
&& \hspace{2cm} \exp\left[i(u^++v^+)q^- + iu^+xu^+-iv^+xv^+\right]\ ,\label{D45} \eea
which is identical to \eq{calJ0rho} except that in \eq{D45} we are using $x$-independent coefficients $\phi^{\prime I ,\a(2n)}$ and $\bar\phi^{\prime I ,\b(2n)}$ and the $x$-dependence is included in the exponent (as a result of directly building the master fields via gauge functions), while in \eq{calJ0rho} we encoded the $x$-dependence in the scalar expansion coefficients.
Indeed, as a verification,  we can check that identical conserved currents are extracted from the form \eq{D45}. Using the expansion \eq{Jexprho}, it is easy to see that the $y^-$-independent element of \eq{D45} coincides with the spin-$0$ ``current''
\bea {\cal J}_0 &=& \left.{\cal J}\right|_{z^-=0=y^-}= \sum_{m,n}\frac{(-1)^{m+n}}{(2m)!\,(2n)!}\,\phi^{\prime\,I,\a(2m)}\bar\phi'_I{}^{\b(2n)} \left(\partial^2_{u^+}\right)_{\a(2m)}^m\left(\partial^2_{v^+}\right)_{\b(2n)}^n \nonumber\\
&& \ \ \ \ \ \ \left.\exp\left[iu^+xu^+-iv^+xv^+\right]\right|_{u^+=0=v^+}= \nonumber\\
&=& \sum_{m,n}\frac{(2i)^{m}(-2i)^n}{(2m)!\,(2n)!}\,\phi^{\prime\,I,\a(2m)}\bar\phi'_I{}^{\b(2n)} x^m_{\a(2m)}x^n_{\a(2n)} = \phi^I(x)\overline\phi_I(x)\ , \label{J0}\eea
where in the last equality \eq{D30} was used, relating ${\cal J}_0$ to the scalar mass term, as expected. Generally, the spin-$s$ coefficient of \eq{D45} (i.e., the coefficient of the $2s$-th power of $q^-$ in the expansion) is
\bea {\cal J}_{s,\a(2s)} &=& \sum_{m,n}\frac{(-1)^{m+n}}{(2m)!\,(2n)!}\,\phi^{\prime\,I,\a(2m)}\bar\phi'_I{}^{\b(2n)} \left(\partial^2_{u^+}\right)_{\a(2m)}^m\left(\partial^2_{v^+}\right)_{\b(2n)}^n \nonumber\\
&& \ \ \ \ \ \ \left.(-1)^s (u^++v^+)^{2s}_{\a(2s)}\exp\left[iu^+xu^+-iv^+xv^+\right]\right|_{u^+=0=v^+}\nonumber\\
&=& \sum_{k=0}^s {2s \choose 2k}(-1)^s\sum_{m,n}\frac{(-1)^{m+n}}{(2m)!\,(2n)!}\,\phi^{\prime\,I,\a(2m)}\bar\phi'_I{}^{\b(2n)} \left(\partial^2_{u^+}\right)_{\a(2m)}^m\left(\partial^2_{v^+}\right)_{\b(2n)}^n \nonumber\\
&& \ \ \ \ \ \ \left.i^k \left[(\partial_x)^{k}_{\a(2k)}\exp\left(iu^+xu^+\right)\right](-i)^{s-k} \left[(\partial_x)^{s-k}_{\a(2s-2k)}\exp\left(-iv^+xv^+\right)\right]\right|_{u^+=0=v^+}\nonumber\\
&=& i^s\sum_{k=0}^s {2s \choose 2k}(-1)^k (\partial_x)^{k}_{\a(2k)}\phi^I(x)(\partial_x)^{s-k}_{\a(2s-2k)}\overline\phi_I(x)\ ,\eea
in agreement with \eq{4.59}-\eq{currents} (as well as \eq{currents2}, obviously, as the currents keep the same form at any $\r$).   

In order to build $J$, starting from the building block \eq{rhocalF},
we instead take the star product of ${\cal F}$ with $\k_y$. With the help of \eq{fky}, we can first of all compute
\be F \ \star\ \k_{y}\propto 2i\exp\left[-iy^+y^-+2i\sqrt{\r}\,u^+y^- -2\,\sqrt{\r}v^+y^+ + iu^+xu^+-iv^+xv^+ -2\r u^+ v^+\right] \ .\label{D49}\ee
Extracting the generating function for primary currents by projecting to $\yb=0$ and using \eq{ypmreal}, we have
%
%
%
\be \left. F \ \star\ \k_{y}\right|_{\yb=0}\propto \exp\left[\sqrt{2}e^{i\pi/4}\,(u^+-v^+) q + iu^+xu^+-iv^+xv^+ -2\r u^+ v^+\right] \ .\label{D51} \ee
Thus, in terms of the rescaled  $\tilde q:= \sqrt{2}\, e^{-i\pi/4}q=2q^-|_{\yb=0}$ we can write 
\bea \left. J\right|_{\yb=0} &=& \left. {\cal F}\star \k_y\right|_{\yb=0} \propto \sum_{m,n=0}^\infty \frac{(-1)^{m+n}}{(2m)!(2n)!}\,\phi^{\prime I ,\a(2n)}\bar\phi^{\prime I ,\b(2n)}\left(\partial^2_{u^+}\right)_{\a(2m)}^m\left(\partial^2_{v^+}\right)_{\b(2n)}^n\nonumber \\
&& \hspace{1cm} \exp\left[i(u^+-v^+)\tilde q + iu^+xu^+-iv^+xv^+-2\r \, u^+ v^+\right]  \label{Jyb0}\eea
which is identical to \eq{D45} \emph{in the limit $\r\to 0$} (keeping $\tilde q$ fixed and observing that the generating functional is invariant under the sign change $v^+ \to -v^+$ in the bosonic theory). 

To summarize, apart from constant normalization factors, the generating function of conserved currents in our model coincides with that of \cite{Misha} in the limit $\r \to 0$ as stated in \eq{cJ0comp}.

The last term in the exponent in \eq{Jyb0} is relevant in ensuring that the full unfolded current module, containing the primary currents as well as their descendants\footnote{Incidentally, we note that Eq. \eq{D49} confirms that the factorized Weyl zero-form \eq{PhifromC}, here assumed to connect explicitly our results with those of \cite{Misha}, indeed has the expected form \eq{D8bis} in Weyl order, since, as from \eq{rhocalF} and \eq{calF}, $\Phi=\r{\cal F}\star \k_y=\rho \, e^{-iy^+y^-}J$, where $J$ is given in \eq{fullJ}.},
\bea  & \displaystyle  J \propto  \left. \sum_{m,n}\frac{(-1)^{m+n}}{(2m)!\,(2n)!}\,\phi^{\prime\,I,\a(2m)}\bar\phi'_I{}^{\b(2n)} \left(\partial^2_{u^+}\right)_{\a(2m)}^m\left(\partial^2_{v^+}\right)_{\b(2n)}^n \right.&\nonumber\\
&\displaystyle   \left. \exp\left[\sqrt{2}\,e^{i\pi/4}u^+ (q-i\bar q)- \sqrt{2}\,e^{i\pi/4}v^+ (q+i\bar q) + iu^+xu^+-iv^+xv^+ - 2\r\, u^+v^+\right]\right|_{u^+=0=v^+} \ , &\label{fullJ} \eea
satisfies Eq. \eq{rhoJ}. Let us assess the role of that term on the primary currents. Encoding, for simplicity, the $x^m$-dependence back into the expansion coefficients as done in Section \ref{sec:rhocurrents}, the generating function \eq{D51} can be rearranged as
\bea & \displaystyle \left.  J\right|_{\yb=0}  \propto  \left. \sum_{m,n}\frac{(-1)^{m+n}}{(2m)!\,(2n)!}\,\phi^{I,\a(2m)}(x)\bar\phi'_I{}^{\b(2n)} (x)\left(\partial^2_{u^+}\right)_{\a(2m)}^m\left(\partial^2_{v^+}\right)_{\b(2n)}^n \right.&\nonumber\\
&\displaystyle   \left. \exp\left[i( u^+-v^+) \tilde q -2\r \, u^+ v^+  \right]\right|_{u^+=0=v^+} \ . &  \eea
By virtue of the last term in the exponent, away from the boundary the primary currents get trace corrections involving all scalar modes weighted by powers of $\r^2$. The resulting, so-corrected currents are also conserved once their surviving terms at $\r=0$ are conserved. For example, the spin-$0$ boundary ``current'' \eq{J0} becomes
\bea  \displaystyle J_0 &=&  \left.  J\right|_{\yb=0,\,q=0 }\propto   \sum_{n=0}^{\infty}\frac{\r^{2n}}{(2n)!}\,\phi^{I,\a(2n)}(x)\bar\phi_I{}_{\a(2n)} (x) \nonumber \\
&=& \sum_{n=0}^{\infty} \frac{\r^{2n}}{(2n)!}\, \partial^{\a(2n)}\phi^I(x)\partial_{\a(2n)}\bar\phi_I(x)\ .\label{Hdscalar}\eea
This is to be contrasted with the primary current generating function in our model $ \left. {\cal J}\right|_{z^-=0}$, which gives rise to the same, uncorrected currents on every leaf at any fixed $\r$. 

Summarizing: our $\r$-foliated CCHSG model at second order contains at any $\r$ the same conditions on conformal currents obtained in \cite{Misha} from pulling back on a 3D leaf the linearized 4D HSG equations on the Poincar\'e patch. Moreover, we build all currents explicitly in terms of 3D matter fields. However, in the two models the conserved currents are extracted in a different way: in our case, the perturbative expansion of the fully non-linear  CCHSG system naturally leads to the generating function $ \left. {\cal J}\right|_{z^-=0}=\r^{-1} \left. C\star\overline C\star \pi\d^2_{\Comp}(z^-)\right|_{z^-=0}$, which agrees with $ \left. J\right|_{\yb=0}=\r^{-1}\left.\Phi\right|_{\yb=0}$ of \cite{Misha} only at the boundary  $\r=0$ (and for the bosonic theory). Anywhere else, the currents in the two models differ by higher-derivative corrections (which do not ruin conservation): the primary currents in our model are the usual ones on any leaf; whereas (assuming that the bulk HSG Weyl zero-form has the composite structure \eq{PhifromC}, which seems very natural) those in \cite{Misha} away from the boundary get higher-derivative corrections, as shown in \eq{Hdscalar}.

\section{Conclusions of Part II}\label{sec:conclusions}

This paper completes the analysis initiated in \cite{paperI}, which represents a first step towards reformulating the study of HS/CFT dualities within the framework of the AKSZ formalism, and more particularly of the Frobenius-Chern-Simons (FCS) formulation of Vasiliev's 4D HSG. 
The main result of these two joint papers is that the flatness condition for the superconnection of the FCS model, valued in a fractional-spin algebra, can be taken as a parent field equations for both 4D HSG and for a novel 3D non-linear theory of coloured conformal matter fields coupled to conformal higher-spin gauge fields and colour gauge fields, alias, CCHSG. This common origin relates the two theories directly and provides a rationale for the holographic duality, to be derived from AKSZ partition functions on cylinders with dual boundary conditions, which we shall study in detail in a future paper \cite{OLC}.

While in Paper I we mostly focused on the construction of the parent model and on singling out the dynamical two-form cohomology of relevance in singling out HSG and CCHSG reductions, the present work explored the details of the two reductions. As for the 4D defect, we began the exploration of the FSG model --- describing the coupling of coloured singletons to HSG and colour gauge fields --- which can be truncated to Vasiliev's HSG. Most of this paper was instead devoted to building and studying the 3D CCHSG defect. The CCHSG model can be thought of as a fully non-linear completion of the models of CHSG coupled to conformal matter \cite{Misha,Nilsson} obtained by introducing topological, colour-like gauge fields that couple non-locally to the conformal matter fields. Such extra, internal states are naturally incorporated via the expansion of the parent-theory superconnection in terms of a fractional-spin algebra. While leaving the conformal HSG fields colourless --- as they are sourced by colour-singlet currents, sesquilinear in the 3D matter fields --- such colour states are helpful in smoothing star products of the fundamental master-fields in the model without requiring any regularization. We have also carried on a detailed comparison of the results here obtained with those of \cite{Misha}. As already stressed in \cite{Misha}, conjecturing a matter-coupled CHSG as holographic dual of 4D HSG is not in contradiction with the Maldacena-Zhiboedov theorem \cite{MZh}, since the coupling to a gauge theory violates at least one of the hypotheses that the theorem is based on. 
If supported by appropriate holographic tests --- which we intend to carry out at the level of the parent theory via overlap conditions (see comments in Section 1 and in the Conclusions of \cite{paperI}) --- the results of this paper would thus refine Vasiliev's HS holography proposal \cite{Misha} (itself a refinement of the holography proposal of \cite{Sezgin:2002rt,KP}). 

There are many questions and future research directions that the results of this paper suggest. Let us begin by mentioning several open issues related to the systems here studied. 

First, one may think of the assumption that the superconnection is in the fractional-spin algebra with its $\mathfrak{u}(N,N)$-component, which is a boundary condition in the fibre directions, as an HSG analog of introducing Dirichlet branes and anti-branes into string theory and supergravity. It would be interesting to make this view more precise, though we stress the stand-alone character of the construction presented here. As shown, the fractional-spin algebra expansion also leads to the identification of HSG as a further reduction of a more general 4D FSG theory, the accurate study of which we postpone to future work. The introduction of the fractional-spin degrees of freedom also makes it possible to modify the boundary conditions on the dynamical two-form leading to a manifestly $U(N,N)$-covariant formulation that shares some similarity with a non-abelian anyon on $Z$-space, which we hope to clarify.

In this paper, we have limited ourselves to constructing a bosonic CCHSG system, which we conjecture to be dual to the bosonic truncation of 4D HSG. As remarked in Footnote \ref{Fermfoot}, however, our construction can be extended to the full supersymmetric system including fermionic matter fields, which we intend to do in a future publication. 

Furthermore, as in this paper, we had many other aspects to focus on, for the sake of simplicity we have studied the perturbative expansion of our CCHSG only in the simplest $\boldsymbol{Z}$-space gauge and by using the unshifted homotopy contraction along the vector field $\vec E_- := z^{-\a}\partial_{z^{-\a}}$ and cohomology projection on $z^-=0$. Besides the fact that choosing shifted homotopy contractions improves the locality properties of the interactions (as studied, in the HSG context, by many works, see e.g. \cite{Boulanger:2015ova,Vasiliev:2017cae,Gelfond:2018vmi,Didenko:2018fgx,Didenko:2019xzz,Gelfond:2019tac,COMST,Didenko:2020bxd,Vasiliev:2022med}), it would be interesting to understand whether anything may change, in the holographic relation between the two theories, when changing homotopy contraction and cohomology projection. 

Our CCHSG reduction Ansatz may also be generalized to accommodate the minimal-model projection \cite{Konstein:1989ij,review99,Sezgin:2003pt}, and it would be interesting to study whether it can be extended to a more general expectation value for the two-form which allows for a decoupling of gauge fields from currents along the expectations of \cite{Misha} (see Section \ref{App:D} for more details). Indeed, a possibility that we have begun investigating in this direction is a generalization of our CCHSG reduction Ansatz that enables introducing deformations along both $z^-$ and $z^+$, implying a doubling of conserved currents for every spin: the standard conformal ones obtained in this paper would be accompanied by twin currents built not in terms of the standard perturbative 3D conformal scalar, but on its ``dual'' --- essentially the fundamental solution to the 3D Klein-Gordon equation --- introduced via $y^+$ oscillators acting on the Poincar\'e non-invariant lowest-weight state $\ket{(i/2)}$. We hope to complete this investigation in a future work.  

In general, it would also be interesting to explore further the expansion of the colour sector of our CCHSG model, to understand the details of its coupling to the conformal matter sector in spacetime. As we have seen in Section \ref{sec:Vsource} on a generic 3D leaf embedded in the Poincar\'e patch, the coupling takes place, at the first non-trivial order, via a non-local, scalar-matter composite source term realizing a dynamically generated colour matrix, which raises the question whether it may be possible to reduce it to a local, Chern-Simons-like coupling in the IR limit. 

Of course, the most important open issue is now the actual check of the proposed duality within the AKSZ formalism. The common origin of HSG and CCHSG as reductions of one parent system, living in one higher dimension, provides support to establishing the holographic duality in terms of AKSZ partition functions: the two dual theories are embedded as two boundaries of a cylinder at the parent theory level, resulting in the definition of an entangled vacuum state.  The information of the
holographic correspondence is thus all contained in such an entangled vacuum, as the latter obeys topological and local overlap conditions that effectively relate algebraic features and observable quantities of the two dual theories. We plan on reporting on these results in a separate work, currently in progress \cite{OLC}. As a preliminary check, in Paper I we have evaluated a parent-theory Chern class, previously computed on 4D HSG \cite{Boulanger:2015kfa,COMST}, on the CCHSG reduction: the result ensures that the theory is non-trivial and is compatible with the expected holographic duality between the two theories. 

Finally, it is interesting to generalize our system to include a non-trivial zero-form vacuum expectation value, inducing a Wigner-deformation of the non-commutative parent geometry. The latter admits reduction to fractional-spin deformations of the 4D HSG and corresponding massive deformations of the 3D CCHSG model. We hope to report on this matter in a future work \cite{wip}.

\paragraph{Acknowledgements.}
We would like to thank the Referees for insightful questions and suggestions that helped us improve our paper. We have benefitted from discussions with L. Andrianopoli, R. Aros, M. Bianchi, N. Boulanger, S. Deger, V. E. Didenko, J. Lang, S. Lysov, Y. Neiman, B.E.W. Nilsson, C. Reyes, E. Skvortsov, D. Sorokin, M. Trigiante, M. Tsulaia, M. Valenzuela, B. Vallilo, M. A. Vasiliev and J. Zanelli. PS is grateful for the support during various stages of the project of the Centro de Ciencias Exactas at Universidad del Bio-Bio; the Centro de Estudios Cientificos at Universidad San Sebastian; the Service de Physique de l’Univers, Champs et Gravitation at
Universit\'e de Mons; the Department of Mathematics at Bogazici University; and the Quantum Gravity Unit of the Okinawa Institute of Science and Technology. FD and PS would like to thank the Institute of Mathematics of the Czech Academy of Sciences for hospitality during the final stage of this project. The work of PS is partially supported by the funding from the European Research Council (ERC) under Grant No. 101002551 and by the Tubitak Bideb-2221 fellowship program. The work of FD is supported by {\sc Beca Doctorado nacional} 
(ANID) 2021 Scholarship No. 21211335, ANID/ACT210100 Anillo Grant ``{\sc Holography and its applications to High Energy Physics, Quantum Gravity and Condensed Matter Systems}'' and FONDECYT Regular grant No. 1210500.

\begin{appendix}

\section{Bases and oscillator realizations of $\mso(2,3)$}\label{App:emb}

\subsection{Compact vs. conformal basis of $\mathfrak{so}(2,3)$}

In the conventions of \cite{fibre}, the $\mso(2,3)$ generators $M_{AB}=-M_{BA}$, $A,B=0',0,1,2,3$, are taken to obey 
\be [M_{AB},M_{CD}]_\star =\ 4i\y_{[C|[B}M_{A]|D]}\ ,\qquad
(M_{AB})^\dagger\ =\ M_{AB}\ ,\label{sogena}\ee
where $\eta_{AB}={\rm diag}(--+++)$.
The generators of the Lorentz subalgebra $\mso(1,3)$ are taken to be $M_{ab}$, $a,b=0,1,2,3$; the transvections 
\begin{align}
    P_a:=M_{0'a}~
\end{align}
in units where the cosmological constant  $\L=-3$, obey 
\be 
[M_{ab},P_c]_\star\ =\ 2i\y_{c[b}P_{a]}\ ,\qquad [P_a,P_b]_\star\ =\
i M_{ab}\ ,\label{sogenb}\ee
where $\eta_{ab}={\rm diag}(-+++)$.

In order to exhibit the maximal compact subalgebra $\mso(2)_E\oplus\mso(3)_M$ generated by the energy generator $E=M_{0'0}=P_0$ and the spatial rotation generators $M_{rs}$ with $r,s=1,2,3$, we arrange the remaining generators into energy-raising and lowering operators
\be L^\pm_r =M_{0r}\mp iM_{0'r}\ =\ M_{0r}\mp iP_r\ ,\label{Lplusminus}\ee
leading to the following $E$-graded decomposition of the
commutation rules \eq{sogena}:
\begin{align} 
\label{el} 
[E,L^{\pm}_r]_\star  ={}& \pm L^{\pm}_r\ , && [L^-_r,L^+_s]_\star = 2iM_{rs}+2\d_{rs}E \ ,\\
[M_{rs},M_{tu}]_\star ={}& 4i\d_{[t|[s}M_{r]|u]} \ ,&& [M_{rs},L^\pm_t]_\star =
2i\d_{t[s}L^\pm_{r]}\ .\label{ml}
\end{align}
The generators $(E,M_{rs},L^\pm_r)$ are referred to as the \emph{compact basis}, or \emph{compact split} of $\mso(2,3)$. Representations in which $E$ is bounded from below and above, respectively, referred to as lowest- and highest-weight representations, arise from specific functions in the enveloping algebra of $\mso(2,3)$ modulo ideals. In particular, the ultra-short unitary irreducible singleton and anti-singleton representations arise by factoring out the ideal generated by
\begin{align}\label{A.7}
    V_{AB} :={}& \frac12 M_{(A}{}^C \star M_{CB)}+\frac15 \eta_{AB}C_2 = 0~,\qquad V_{ABCD} := M_{(AB}\star M_{CD)} = 0~,
\end{align}
implying the Casimir constraint \cite{fibre}
\begin{align}
    C_2 := \frac12 M^{AB}\star M_{AB} = -\frac54~.
\end{align}
Equivalently, the states forming the (anti-)singleton representation can be obtained from the one-sided star-product action 
\be {\cal D}^\pm(\pm 1/2) := {\rm Env}(\mso(2,3))\star {\cal P}_{\pm 1/2|\pm 1/2} \ee
of the enveloping algebra of $\mso(2,3)$ on the projectors
\begin{align} \label{eq:vac proj}
{\cal P}_{\pm 1/2|\pm 1/2} \ = 4e^{\mp 4E} \ , \qquad {\cal P}_{\pm 1/2|\pm 1/2} \star {\cal P}_{\pm 1/2|\pm 1/2} ={\cal P}_{\pm 1/2|\pm 1/2}~, 
\end{align}
which are the images of the Wigner-Ville map applied to the projectors onto the singleton lowest-weight $(+)$ and anti-singleton highest-weight $(-)$ states $|\pm 1/2,(0)\rangle$, viz.,
\begin{align}
    \cP_{\pm 1/2|\pm 1/2} = | \pm 1/2,(0)\rangle \langle \pm 1/2,(0)| ~.\label{compsing2}
\end{align}
Such projectors carry quantum numbers of the compact subalgebra such that
\begin{align} E\star\cP_{\pm 1/2|\pm 1/2} \ ={}& \ \cP_{\pm 1/2|\pm 1/2} \star E \ = \ \pm\frac12 \cP_{\pm 1/2|\pm 1/2}  \ , \\
 M_{rs}\star\cP_{\pm 1/2|\pm 1/2} \ ={}& \ 0 \ = \ \cP_{\pm 1/2|\pm 1/2}\star M_{rs}  \ , \end{align}
and their lowest- and highest-weight properties are manifested by 
\bea  L^\mp_r\star\cP_{\pm 1/2|\pm 1/2} \ = \ 0 \ = \ \cP_{\pm 1/2|\pm 1/2}\star L^\pm_r \ . \eea
The Lie algebra also admits a \emph{conformal basis} $(D,M_{mn},T_m, K_m)$, viz.,
\begin{align}\label{dt}
[D,T_m] =  iT_m\ ,\quad [D,K_m] = -iK_m \ ,\quad 
[K_m,T_n] = 2i(\eta_{mn}D-M_{mn}) \ , 
\end{align}
\vspace{-1cm}
\begin{align}
[M_{mn},M_{pq}] = 4i\eta_{[p|[n}M_{m]|q]} \ ,\quad   [M_{mn},T_p] = 2i\eta_{p[n}T_{m]}\ , \quad [M_{mn},K_p] = 2i\eta_{p[n}K_{m]}\ .\label{mt}
\end{align}
which is 3-graded with respect to the dilation operator
$D$ of the non-compact subalgebra $\mso(1,1)_D\oplus\mso(1,2)_{M_{mn}}$, and exhibits the translations $T_m$ and special conformal transformations $K_m$ of 3D conformal Minkowski spacetime. Embedding the boundary conformal algebra in such a way that all its generators are hermitian, the dilation generator $D$ can be identified with any spacelike transvection. 
%
For oscillator realizations, and with our conventions on van der Waerden symbols as in \eqref{A.35}, it is convenient to identify the (boundary) dilation generator as
\begin{align}\label{DasP2}
   D=P_2~, 
\end{align}
and thus the (boundary) Lorentz generators $M_{mn}$, $m,n=0,1,3$, and $D$-raising and $D$-lowering combinations\footnote{Of course, the identification \eq{DasP2} is purely a convenient choice, and we could have rather embedded both compact and conformal slicings by introducing a normalized frame $(L_i^a,L^a)$ obeying 
\begin{align}
L^a L_a=\epsilon~, \qquad L_i^a L_a =0~,\qquad L^a_i L_{ja}=\eta_{ij}=(+,+,-\epsilon)~,\nonumber
\end{align}
and letting 
\begin{align}
K:=L^a P_a~,\qquad K^\pm_i := (\e L^ b M_{ab} \mp \sqrt{\epsilon} P_a)L^a_i~,\qquad M_{ij}:=\e L^a_i L^b_j M_{ab}~, \nonumber
\end{align}
where $K$ is referred to as the principal Cartan generator, and the compact and conformal bases arise for $\epsilon = -1$ and $\epsilon = 1$, respectively; for example see \cite{Sezgin:2005pv,cosmo}. The specific realizations above used thus correspond to the particular choices 
\bea
   & \epsilon = -1~:\qquad  L^a=(1,0,0,0)~, \qquad  K = P_0 = E\ , &\nonumber\\ 
   & \epsilon = 1~:\qquad   L^a=(0,0,1,0)~, \qquad K= P_2 = D \ .&  \nonumber
\eea}. 
\be T_m \ = \ M_{m2}-P_m \ ,\qquad K_m \ = \ M_{m2}+P_m \ .\label{TmKm}\ee
The construction of lowest/highest-vector modules induced from $\mso(1,1)\oplus\mso(1,2)$-modules proceeds in a completely parallel fashion to the compact-basis case, with the only difference that, in order to compensate for the non-compact nature of $D$, an extra factor of $i$ enters the metaplectic realization of the lowest/highest weight state projectors. Indeed, the conformal analogue of \eq{compsing2} is the realization of the conformal (anti-)singleton highest-weight (lowest-weight) projector 
\be |\pm i/2,(0)\rangle\langle \pm i/2,(0)| \ \equiv \ \cP_{\pm i/2|\pm i/2 } \ = \ 4e^{\pm 4 iD}  \ , \ee
satisfying
\bea & D\star\cP_{\pm i/2|\pm i/2 } \ = \ \pm \frac{i}{2}\cP_{\pm i/2|\pm i/2 } = \cP_{\pm i/2|\pm i/2 }\star D\ , & \\
 & M_{mn}\star \cP_{\pm i/2|\pm i/2} \ = \ 0  \ = \ \cP_{\pm i/2|\pm i/2} \star M_{mn}  \ ,& \eea
and respectively annihilated by $K_m$ from the left (and $T_m$ from the right) 
\bea  K_m\star \cP_{i/2|i/2} \ = \  0 \ = \ \cP_{i/2|i/2} \star T_m\ , \eea
and $T_m$ from the left (and $K_m$ from the right),
\bea  T_m\star\cP_{-i/2|-i/2} \  = \ 0 \ = \ \cP_{-i/2|-i/2}\star K_m \ , \label{A.23}\eea
where we note that $\pi^\ast(K_m)=T_m$.  
In the body of the paper we have frequently used the shorthand notation 
\be \ket{(\pm i/2)} := \ket{\pm i/2; (0)} \ . \ee
All states created via the one-sided action of the enveloping algebra of $\mso(2,3)$ on $ 4e^{-4iD}$ ($ 4e^{4iD}$) are $\mso(1,2)$-tensors of left $D$-eigenvalue $-i(2s+1)/2$ ($i(2s+1)/2$) and rank $s$, $s=0,1,2,...$, corresponding to states $\ket{-i(2s+1)/2;(s)}$ ($\ket{i(2s+1)/2;(s)}$), and give rise to the \emph{conformal (anti-)singleton} representation\footnote{The reason why we refer to ${\cal T}^-(-i/2)$ as conformal singleton, instead of anti-singleton --- reversing the convention used in compact basis --- is because we conventionally choose to realize the 3D Minkowski translation $T_m$ as $D$-raising operator, which, due to \eq{A.23}, singles out $\ket{(-i/2)}$ as 3D Poincar\'e invariant vacuum, ``breaking the symmetry'' in the definition od conformal singleton and anti-singleton.} of $\mso(2,3)$,
\be  {\cal T}^\pm(\pm i/2):={\rm Env}(\mso(2,3))\star \cP_{\pm i/2|\pm i/2 } \ . \label{confsing}\ee
States in ${\cal T}^\pm(\pm i/2)$ are bounded from below ($+$) and above ($-$) in the eigenvalue $i\D$ of $D$. Note that the $\pi$-map exchanges highest- and lowest-weight modules, i.e., reverses the sign of $\D$.

\subsection{Spinor conventions and oscillator realizations of $\mathfrak{so}(2,3)$}

In terms of the Majorana oscillators $Y_{\underline\a}$ satisfying the commutation relations $[Y^{\underline{\a}},Y^{\underline{\b}}]_\star = 2iC^{\underline{\a\b}}$, the realization of the generators of $\mso(2,3)$ is taken to be
\be M_{AB}~=~ -\ft18  (\C_{AB})_{\underline{\a\b}}\,Y^{\underline\a}\star Y^{\underline\b}\ ,\label{MAB}
\ee
using real Dirac matrices $(\Gamma_{A})^{\underline{\alpha\beta}}$ obeying $(\C_A)_{\underline\a}{}^{\underline\b}(\C_B C)_{\underline{\b\c}}=
\eta_{AB}C_{\underline{\a\c}}+(\C_{AB} C)_{\underline{\a\c}}$.   
Going to a Weyl basis $Y_{(W)}^{\underline\alpha} = (y^{\alpha},\bar{y}^{\dot\alpha})$ that diagonalizes  $\Gamma^5:=i\Gamma^{0123}$, the Dirac matrices decompose as follows 
\begin{align}
C_{\underline{\alpha\beta}}={}&\left(
\begin{array}{cc}
    \epsilon_{\alpha\beta} & 0 \\
     0 & \epsilon_{\dot\alpha\dot\beta}\end{array}\right)~, && \left(\Gamma^5_{(W)}\right)_{\underline\alpha}{}^{\underline{\beta}} = \left( 
\begin{array}{cc}
\delta_{\alpha}^\beta & 0 \\ 
0 & -\delta_{\dot\alpha}^{\dot\beta}%
\end{array}
\right)~, \\ \left(\Gamma_{(W)}^{0'}\right)_{\underline\alpha}{}^{\underline\beta} ={}& \left( 
\begin{array}{cc}
i\delta_{\alpha}^\beta & 0 \\ 
0 & -i\delta_{\dot\alpha}^{\dot\beta}%
\end{array}
\right)~, && \left( \Gamma ^{a}_{(W)}\right)_{\underline{\alpha }}^{\ \ \underline{\beta }
}=\left( 
\begin{array}{cc}
0 & -i\left( \sigma ^{a}\right) _{\alpha }^{\ \ \dot{\beta}} \\ 
i\left( \bar{\sigma}^{a}\right) _{\dot{\alpha}}^{\ \ \beta } & 0%
\end{array}
\right) \ , \\ 
\left( \Gamma^{0'a}_{(W)}\right) _{\underline{\alpha }}^{\ \ \underline{\beta }
}={}&\left( 
\begin{array}{cc}
0 & \left( \sigma ^{a}\right) _{\alpha }^{\ \ \dot{\beta}} \\ 
\left( \bar{\sigma}^{a}\right) _{\dot{\alpha}}^{\ \ \beta } & 0%
\end{array}
\right) \ , && \left( \Gamma^{ab}_{(W)}\right)_{\underline{\alpha }}^{\ \ \underline{\beta}} =\left( 
\begin{array}{cc}
\left( \sigma ^{ab}\right) _{\alpha }^{\ \ \dot{\beta}} & 0 \\ 0 &
\left( \bar{\sigma}^{ab}\right) _{\dot{\alpha}}^{\ \ \dot\beta } %
\end{array}
\right) \ ,
\end{align}
one has
\be
 M_{ab}\ =\ -\frac18 \left[~ (\s_{ab})^{\a\b}y_\a\star y_\b+
 (\sb_{ab})^{\ad\bd}\bar y_{\ad}\star \yb_{\bd}~\right]\ ,\qquad P_{a}\ =\
 \frac{1}4 (\s_a)^{\a\bd}y_\a \star \yb_{\bd}\ ,\label{mab}
 \ee
where the van der Waerden symbols obey
 \bea
 & (\s^{a})_{\a}{}^{\ad}(\sb^{b})_{\ad}{}^{\b} ={} \y^{ab}\d_{\a}^{\b}\
 +\ (\s^{ab})_{\a}{}^{\b} \ , \qquad
 (\sb^{a})_{\ad}{}^{\a}(\s^{b})_{\a}{}^{\bd}~=~\y^{ab}\d^{\bd}_{\ad}\
 +\ (\sb^{ab})_{\ad}{}^{\bd} \ ,& \label{so4a}\\
 &\ft12 \e_{abcd}(\s^{cd})_{\a\b} = {} i (\s_{ab})_{\a\b}\ ,\qquad  \ft12
 \e_{abcd}(\sb^{cd})_{\ad\bd}~=~ -i (\sb_{ab})_{\ad\bd}\ , &\label{so4b}
\\ & \e^{\a\b}\e_{\c\d} \ ={} \ 2 \d^{\a\b}_{\c\d} \ , \qquad 
\e^{\a\b}\e_{\a\c} \ = \ \d^\b_\c \ , & \\ 
 &(\s^a)_{\a\bd})^\dagger={}
(\sb^a)_{\ad\b} ~=~ (\s^a)_{\b\ad} \ ,\qquad  ((\s^{ab})_{\a\b})^\dagger\ =\ (\sb^{ab})_{\ad\bd} \ , \qquad  (\e_{\a\b})^\dagger \ = \ \e_{\ad\bd} \ . &
\eea
and two-component spinor indices are raised and lowered according to the
conventions $A^\a=\epsilon^{\a\b}A_\b$ and $A_\a=A^\b\epsilon_{\b\a}$. 
In the text we frequently employ the implicit-index notation for contracted indices, in which case we always juxtapose  spinors and spinor-tensors from left to right according to so-called NorthWest-SouthEast rule, e.g.,
\begin{align}\label{implicit}
VW
  &:=
  V^\a W_\a
  =
  -WV\ , \qquad  VABW:=
  V^\a A_\a{}^{\b} B_\b^{\phantom{\b}\gamma} W_\gamma \ .
\end{align}

In terms of this oscillator basis, the $\mso(2,3)$-valued connection
 \be
  \O = -i \left(\frac12 \omega^{ab} M_{ab}+e^a P_a\right) ~:=~ \frac1{2i}
 \left(\frac12 \omega^{\a\b}~y_\a \star y_\b
 +  e^{\a\dot\b}~y_\a \star {\bar y}_{\dot\b}+\frac12 \bar{\omega}^{\dot\a\dot\b}~{\bar y}_{\dot\a}\star {\bar y}_{\dot\b}\right)\
 .\label{Omega}
 \ee
The van der Waerden symbols are realized as 
\begin{align}\label{A.35}
    \epsilon_{\alpha\beta} = i\left(\sigma^2\right)_{\alpha\beta}~,\qquad 
    \left(\sigma^a\right)_{\alpha}{}^{\dot\alpha} = \left(-i \sigma^2,-i\sigma^r\sigma^2\right)_{\alpha}{}^{\dot\alpha}~,\qquad \left(\bar{\sigma}^a\right)_{\dot\alpha}{}^{\alpha} = \left(-i \sigma^2,i\sigma^2\sigma^r\right)_{\alpha}{}^{\dot\alpha}~,
\end{align}
where $\s^r$, $r=1,2,3$, are the Pauli matrices.

Every slicing of the $\mso(2,3)$-algebra, like the compact or the conformal basis, has a corresponding grading generator ($E$ and $D$, in the examples above shown) and adapted choice of oscillator basis. Indeed, the $\Gamma_{AB}$ matrix that selects the grading Cartan generator --- $\Gamma_{0'0}$ in the compact case, $\Gamma_{0'2}$ in the conformal one according to the realization \eq{DasP2} --- can be used to define projectors inducing a split of the symplectic coordinates $Y_{\ua}$ into canonical pairs $Y_{\ua}^\pm$. In the non-compact case the latter can be extracted as
\be \widetilde Y_{\ua}^\pm \ = \ \sqrt{2}\, \Pi^\pm_{\ua}{}^{\underline{\b}}Y_{\underline{\b}} \ = \  \frac1{\sqrt{2}}\left(\d_{\ua}{}^{\underline{\b}}\pm \Gamma_{0'2\ua}{}^{\underline{\b}}\right)Y_{\underline{\b}} \ , \label{Ypmdef}\ee
with commutation relations
\be [\widetilde Y_{\ua}^\e,\widetilde Y_{\underline{\b}}^{\e'}]_\star \ = \ 4i\d^{\e,-\e'}\Pi_{\underline{\a\b}}^{\e} \ , \qquad \e,\e'=\pm \ ,\ee
where $\Pi^\pm$ are projectors (see \cite{2011} for the general construction of adapted oscillator bases) and the factor of $\sqrt{2}$ in the definition \eq{Ypmdef} has been added for convenience. More explicitly, according to the realization \eq{A.35} of the van der Waerden symbols, the independent canonical pairs are 
\bea \widetilde y^{\pm}_\a  =  \frac1{\sqrt{2}}\left(y\pm\s_2\yb\right)_\a =\frac1{\sqrt{2}}(y_\a\mp i\yb_{\dot \a})\ , \label{ypms}\eea
satisfying the commutation relations
\be [\widetilde y^\e_\a,\widetilde y^{\e'}_\b]_\star  =  2i\e_{\a\b}\delta^{\e,-\e'} \ .\label{commypm}\ee
Clearly, these oscillators are not real, $(\widetilde y^\pm_\a)^\dagger=\pm i\widetilde y^\pm_\a$. A real pair can be easily defined as $y^\pm_\a:=\exp(\pm i\pi/4)\widetilde y^\pm_\a$, i.e.,
\bea  y^{\pm}_\a  =  \frac{e^{\pm i\pi/4}}{\sqrt{2}}\left(y\pm\s_2\yb\right)_\a =\frac{e^{\pm i\pi/4}}{\sqrt{2}}(y_\a\mp i\yb_{\dot \a})\ , \qquad (y^{\pm}_\a)^\dagger=y^{\pm}_\a \ ,\label{ypmreal}\eea
which definition leaves the commutation relations \eq{commypm} unmodified\footnote{To our knowledge, this basis for the oscillator realization of the 3D conformal group was first used in \cite{F&L}.},
\be [y^\e_\a, y^{\e'}_\b]_\star  =  2i\e_{\a\b}\delta^{\e,-\e'} \ .\label{commypmreal}\ee
As follows from \eq{piyyb}, 
\be \pi^\ast_y(y^\pm_\a) = \mp i y^{\mp}_\a \ , \qquad \bar\pi^\ast_{\yb}(y^\pm_\a) = \pm i y^{\mp}_\a\ .\ee
It is possible to fix the relation between the generators of the 3D conformal group in vectorial basis, viz., $(D,M_{mn},K_m,T_m)$, and spinorial basis, viz., $(D,M_{\a\b},K_{\a\b},T_{\a\b})$, as 
\bea   T_{\a\b}=(\gamma_m)_{\a\b}T^m \ , \qquad K_{\a\b}=(\gamma_m)_{\a\b}K^m\ , \qquad  M_{\a\b}  =  -\frac12(\gamma_{mn})_{\a\b}M^{mn} \ ,
\eea
where $\gamma_{mn}=\frac12[\gamma_m,\gamma_n]$, and
\be T_m = -\frac12 (\gamma_m)_{\a\b}T^{\a\b}\ , \qquad K_m = -\frac12 (\gamma_m)_{\a\b}K^{\a\b} \ , \qquad M_{mn}  =  -\frac12(\gamma_{mn})_{\a\b}M^{\a\b} \ , \ee
where, having selected $P_2$ as transvection generator along the direction of foliation, it is natural to define $\e_{\a\bd}:=i(\s_2)_{\a\bd}$ as the element that breaks $AdS_4$-covariance, and thus, clearly,
\bea (\gamma_m)_{\a\b} & := & (\s_m)_{\a}{}^{\bd}\e_{\bd \b} = i(\s_{2m})_{\a\b} \nonumber \\ & = &  \left\{\left(\begin{array}{cc}-1 & 0 \\ 0 & -1 \end{array}\right), \left(\begin{array}{cc}0 & 1 \\ 1 & 0 \end{array}\right), \left(\begin{array}{cc} 1 & 0 \\ 0 & -1 \end{array}\right)\right\} = (\s_m)_{\a\bd} ,\label{gammasigma}\eea
all real, as expected from boundary Lorentz algebra $\mso(1,2)\sim \msp(2,\Real)$ generators, and satisfying
\be (\gamma_{mn})_{\a\b} \ = \ \epsilon_{mn}{}^r(\gamma_r)_{\a\b} \ ,\qquad \{\gamma_m,\gamma_n\}=2\eta_{mn} \ . \ee
with $\e_{023}=1$. In these conventions, the spinor realization of the conformal group generators is
\be T_{\alpha\beta}=\frac12\,y^+_\alpha y^+_\beta\ , \qquad K_{\alpha\beta}=-\frac12\,y^-_\alpha y^-_\beta \label{PKapp}\ee
\be M_{\alpha\beta}=\frac12\,y^+_{(\alpha} y^-_{\beta)}\ ,\qquad D= \frac14 \,y^{+\alpha} y^-_{\alpha} \ .\label{MDapp}\ee

Analogously, one can define the combinations
\bea  z^{\pm}_\a  =  \frac{e^{\pm i\pi/4}}{\sqrt{2}}\left(z\pm\s_2\zb\right)_\a =\frac{e^{\pm i\pi/4}}{\sqrt{2}}(z_\a\mp i\zb_{\dot \a})\ , \qquad (z^{\pm}_\a)^\dagger=-z^{\pm}_\a \ ,\label{zpmreal}\eea
satisfying the commutation relations
\be [z^\e_\a, z^{\e'}_\b]_\star \ = \ -2i\e_{\a\b}\delta^{\e,-\e'} \ .\ee
where we note that the extra sign in the reality conditions, making $z^\pm_\a$ purely imaginary, is a direct consequence of the reality conditions \eq{realz}. 

The split $\widetilde Y_{\ua}^\pm =  \frac1{2}\left(\d_{\ua}{}^{\underline{\b}}\pm i\Gamma_{0'0\ua}{}^{\underline{\b}}\right)Y_{\underline{\b}} $ yields canonical coordinates in compact basis, leading to the definition of the $SU(2)$ creation/annihilation doublets
\be
     a^{\dagger i} = \frac12 \delta^{i\a}\left(y - i \sigma_0\bar{y}\right)_\a ,\qquad a_i = (a^{+i})^\dagger~,\qquad [a_i,a^{\dagger j}]_\star = \delta_i^j~,
\ee
where we have defined the mixed, intertwining symbol $\delta^{i\a}=(\sigma^{0})^{i\a}$, i.e.,
\be
     a^{\dagger 1} = \frac12 \left(y - i \sigma_0\bar{y}\right)_1\ , \qquad a^{\dagger 2} = \frac12 \left(y - i \sigma_0\bar{y}\right)_2  \ .
\ee

We recall here also the definition of \emph{non-compact singleton}, consisting of $L^2$-normalizable states
\begin{align}
&{\cal S}^{(\sigma;\epsilon)}(-i\xi/2)\ni |\xi;\epsilon;\sigma;\phi\rangle:=\int_{\zeta_\epsilon\Real^2} \frac{d^2\lambda}{4\pi} \phi^{(\epsilon)}_{\sigma}(\lambda)|\xi;\epsilon;\lambda\rangle\ ,\quad\phi^{(\epsilon)}_{\sigma}(-\lambda)=(-1)^{\sigma+1} \phi^{(\epsilon)}_{\sigma}(\lambda)\ ,\label{mom}\\
&\zeta_{\epsilon} := e^{i(1-\epsilon)\pi/4}\ ,\quad \phi^{(\epsilon)}_{\sigma}\in L^2(\zeta_\epsilon\Real^2)\ ,\quad \Pi^{(\sigma)}_K\star|\xi;\epsilon;\sigma';\phi\rangle=\delta_{\sigma,\sigma'}|\xi;\epsilon;\sigma;\phi\rangle\ ,
\end{align}
expanded over \emph{momentum eigenstates}\footnote{In an abuse of nomenclature, we refer here collectively to both $y^+$- and $y^-$-eigenstates as ``momentum eigenstates'', though, in view of the commutation relations \eq{commypmreal}, we should refer to them as momentum and coordinate eigenstates.} 
\begin{align}\label{coherent}
|\xi;\epsilon;\lambda\rangle:=\exp(\frac{i}2 \lambda y^{-\xi})\star |(-i\xi/2)\rangle\ ,  \quad (y^\xi_\alpha-\lambda_\alpha)|\xi;\epsilon;\lambda\rangle=0\ ,\quad \lambda\in \zeta_\epsilon\Real^2\ ,
\end{align}
defined in \cite{paperI} (which we refer to for the definitions of all the labels in the notation, as well as for the projector $\Pi^{(\sigma)}_K$) and used in this paper for the expansion of the CCHSG zero-form in Section \ref{sec:Vsource}. The regularized trace defined in Paper I equips the non-compact singleton module with the Hermitian form 
\begin{align}\label{2.124}
\langle \xi;\epsilon;\lambda|\xi';\epsilon';\lambda'\rangle= 4\d_{\e\e'}\left(\d_{\xi\xi'}4\pi \d^2_\Comp(\l'+\bar\l)+\d_{\xi,-\xi'}e^{i\bar\lambda \lambda'/2}\right)\ ,
\end{align}
which, recalling \eq{deltaC}, is positive on the positive-energy states ($\e=1$, i.e., $\l\in \Real$) and negative on negative-energy ones ($\e=-1$, i.e., $\l\in i\Real$), and with respect to which positive-energy and negative-energy states are orthogonal.

\section{Delta function from triple product of projectors}\label{App:B}

In this Appendix we give a concrete realization of the intertwining zero-form master field $C^{(1)}$  in terms of star-product projectors, and we explicitly compute $C^{(1)}\star \overline C^{(1)}$ to obtain the real ``twisted projector'' (in the terminology of \cite{2017,COMST,meta}) \eq{realtwist}. To justify the latter equation in the simplest possible set-up, we shall limit our considerations to the $U(1)$ colour group case, with an expansion of type \eq{expC} involving a single compact state, i.e.,
\be C^{(1)} = c^{(1)} \star \ket{(-i/2)}\bra{e_0^+}  \ , \label{B1}\ee
that we take to be the compact singleton ground state, i.e. $\ket{e_0^+}=\ket{1/2;(0)}$ (see \eq{eq:vac proj}-\eq{compsing2}).
Thus, such intertwining operator can be realized by means of the star product of an external, non-compact projector and an internal, compact one,
\be  \ket{(-i/2)}\bra{e^+_0} = \ket{(-i/2)}\bra{(-i/2)}\star \ket{e_0^+}\bra{e_0^+} = 4e^{-4iD}\star 4e^{-4E} \ . \label{B2}\ee
Then,
\be C^{(1)}\star \overline C^{(1)} = c^{(1)} \star 4e^{-4iD}\star 4e^{-4E} \star 4 e^{4iD} \star \bar c^{(1)} \ ,  \ee
which is manifestly real, a property shared by the core element $4e^{-4iD}\star 4e^{-4E} \star 4 e^{4iD}$. Now, the bra-ket notation suggests that 
\be 4e^{-4iD}\star 4e^{-4E} \star 4 e^{4iD} \propto \ket{(-i/2)}\bra{(i/2)} \label{B3}\ee
which is a twisted projector. The latter can be realized via the product $4e^{-4iD}\star\k_y$ or $\k_y\star 4e^{4iD}$, but these are manifestly imaginary. Thus, we expect that
\be \ket{(-i/2)}\bra{(i/2)} = \gamma 4e^{-4iD} \star \k_y = 4\pi i\gamma\d^2(y^+) \ , \label{B5}\ee
where, in order to match the reality properties, $\gamma$ has to be imaginary. 

However, it may look somewhat surprising that the triple product of regular elements, albeit projectors, gives rise to a delta function (considering that each separate star product gives rise to a projector, and not to $\k_y$). As we shall see, this happens because the star product of one compact and one non-compact projector gives rise to a new Gaussian projector with a more complicated quadratic form matrix. A further star product with another non-compact projector, whose $D$-eigenvalue has opposite sign with respect to the first, results in a Gaussian integral with singular quadratic form, giving rise to a delta density (or analytic delta function) \cite{meta}.  

Let us begin from the first star product, \eq{B2}, letting $s=\pm 1$ encode the sign of the first exponential,
\be e^{-4isD}\star e^{-4E}=\frac{1}{1+s^2}\,e^{-4E-4iwD+4wM_{02}} = \frac{1}{1+s^2}\, e^{\frac12 YMY} \ , \ee
where $w:=\frac{2s}{1+s^2}$ and the matrix
\be M:= \Gamma_{0'0}+w(i\Gamma_{0'2}+\Gamma _{02})  \ , \qquad M^2=-1 \ , \quad \det M = 1\ ,\ee
which implies that $e^{-4E-4iwD+4wM_{01}}$ is itself a star-product projector up to a normalization. Let us now study the remaining star product with $\exp(-4is'D)$ ($s'=\pm 1$), i.e.,
\be e^{-4isD}\star e^{-4E}\star e^{-4is'D} = \frac{1}{1+s^2}\,e^{\frac12 YMY} \star e^{\frac12 s'YK' Y}  \ , \qquad K':=i\Gamma_{0'2} \ ,\quad K^{\prime 2} = -1\ ,\ee
and note that 
\be  M=A+wK' \ , \qquad A:= \Gamma_{0'0}+w\Gamma_{02} \ , \quad \{A,K'\}=0 \ .\ee
Then, 
\be e^{\frac12 YMY} \star e^{\frac12 s'YK' Y} = e^{\frac12 s'YK' Y}  \int \frac{d^4V}{(2\pi)^2}e^{\frac12 V(M+s'K')V+(iY+s'YK')V} \ .\label{finalint}\ee
The matrix of the quadratic form at the exponent is now
\be M':=M+s'K'=A+(w+s')K' \ , \qquad M^{\prime 2} =-1-s^{\prime 2}-2s'w \ , \quad \det M' = (1+s^{\prime 2}+2s'w)^2 \ , \ee
i.e., $M^{\prime -1}=-\frac1{\zeta}M'$, where $\zeta:=1+s^{\prime 2}+2s'w$. Notice that, for $s=-s'$, $M'=\Gamma_{0'0}+s\Gamma_{02}=\Gamma_{0'0}(1+s\Gamma_{0'2})$, which squares to zero.  Summarizing, the result of the integration in \eq{finalint}, viz.,
\be e^{\frac12 YMY} \star e^{\frac12 s'YK' Y} = \frac1{\sqrt{\zeta^2}} \, e^{\frac12 s'YK' Y -\frac1{2\zeta}(iY+s'YK')M'(-s'K'Y+iY)}    \label{x}\ee
can be expanded as $s=1+\e_s$, $s'=-1+\e_{s'}$, $ss'+1=\e$, leading to 
%
%
%
\be e^{-4iD}\star e^{-4E}\star e^{4iD}  =   \frac{1}{4\e} \exp \left[iy\s_2 \yb -\frac{1}{\e} \big(y\s_{02}y+\yb\bar \s_{02}\yb+2y\s_0\yb\big)\right] \ .\label{B13}\ee
up to irrelevant terms. But in the limit $\e\to 0$ this expression is a two-dimensional analytic delta-sequence, approaching the complex delta function with defining scaling property\footnote{The phase-preserving scaling, $\delta_\Comp(e^{i\varphi}x)=e^{-i\varphi}\delta_\Comp(x)$ in the one-dimensional case, more properly defines a delta density. Such a complex generalization of the (multi-dimensional) Dirac delta function finds its rationale when viewed as the symbol of an element of the complex \emph{holomorphic} metaplectic group $Mp(2n,\Comp)$ \cite{meta,paperI}. In this context, an analytic continuation of Gaussian symbols is sufficient to establish a correspondence between the two branches of the metaplectic double covering and the two sheets of the Riemann surface ${\cal S}^{2n}$ of the square root. See Appendix B in \cite{meta} for the details of the construction and the motivations for such an extension within the higher-spin context.} \eq{deltaC}. More precisely (see \cite{meta} for details on such complex delta sequences), given a $2n$-dimensional symmetric matrix $R_{IJ}$ such that $R_I{}^J R_{J}{}^K=\d_I{}^K$, 
\be \lim_{\e\to 0} \frac{1}{\e^n}\,\exp\left(\frac{i}{2\e}XRX\right) = (2\pi)^n \,\d^{2n}_\Comp(X) \ ,  \ee
independently of the direction of the limit $\e\to 0$ in $\Comp$. Thus rewriting the r.h.s. of \eq{B13} as  
\be e^{-4iD}\star e^{-4E}\star e^{4iD} = \exp(iy\s_2\yb)\,\frac1{4\e}\exp\left[-\frac1{\e}(y-\yb\bar\s_2)\s_{02}(y+\s_2\yb) \right] \ ,\ee
taking the limit and recalling the definition \eq{ypmreal}, one finds
\be e^{-4iD}\star e^{-4E}\star e^{4iD} = \frac{\pi}2 \,\d^2_\Comp(2y^+)= \frac{\pi}{8}\,\d^2_\Comp(y^+)  \ ,\ee
which is indeed manifestly real. Reinstating the normalization of the projectors we can thus conclude that, for the specific realization \eq{B2} of the intertwining element $\ket{(-i/2)}\bra{e^+_0}$,
\bea & \ket{(-i/2)}\bra{(-i/2)}\star \ket{e^+_0}\bra{e^+_0} \star\ket{(-i/2)}\bra{(-i/2)} = 4e^{-4iD}\star 4e^{-4E}\star 4e^{4iD}& \nonumber \\
& = 8\pi\,\d^2_\Comp(y^+)= 2\ket{(-i/2)}\bra{(i/2)}  \ ,\label{B18} \eea
i.e., referring back to \eq{B5}, $\gamma=-i$. The extra factor of $2$ that appears in \eq{B18} with respect to the simple bra-ket evaluation of $\ket{(-i/2)}\bra{e^+_0} \star \ket{e^+_0}\bra{(-i/2)}=\ket{(-i/2)}\bra{(-i/2)}$ is due to the non-trivial overlap between the states $\ket{e^+_0}$ and $\ket{(\pm i/2)}$. As anticipated in Section \ref{sec:currents}, the state $\ket{(-i/2)}\bra{e^+_0}$ is in fact unique modulo normalizations, that can be absorbed in $c^{(1)}$.

This result also shows explicitly how the hybrid bimodule structure of $C^{(1)}$ and the introduction of colour states has the effect of smoothing star products of 3D master fields, giving rise to a well-defined generating function for currents (see Section \ref{sec:currents}). Indeed, had $C^{(1)}$ been a pure conformal bimodule, i.e., had no colour states been there, the building block $C^{(1)}$ \eq{B1} would have featured a ``naked'' conformal anti-vacuum state on the right, $C^{(1)}=c^{(1)}\star \ket{(-i/2)}\bra{(-i/2)}$ and the conformal current building block $C^{(1)}\star\overline C^{(1)} =c^{(1)}\star \ket{(-i/2)}\bra{(-i/2)} \star \ket{(i/2)}\bra{(i/2)}\star\bar c^{(1)}  $ would thus have featured a direct clash of vacuum and anti-vacuum state projectors, which, as can be seen from the above computation, simply diverges as $\e^{-1}$. It is possible to regularize such products and achieve orthogonality between non-compact states with different $D$-eigenvalue, as studied in \cite{2011,2017,COMST,corfu19}, but then orthogonality would make $C^{(1)}\star \overline C^{(1)}$ vanish.

\end{appendix}

\providecommand{\href}[2]{#2}\begingroup\raggedright\endgroup

\end{document}